\catcode`\@=11
\def\citen#1{\if@filesw \immediate\write \@auxout {\string\citation{#1}}\fi%
\@tempcntb\m@ne \let\@h@ld\relax \def\@citea{}%
\@for \@citeb:=#1\do {\@ifundefined {b@\@citeb}%
    {\@h@ld\@citea\@tempcntb\m@ne{\bf ?}%
    \@warning {Citation `\@citeb ' on page \thepage \space undefined}}%
    {\@tempcnta\@tempcntb \advance\@tempcnta\@ne
    \setbox\z@\hbox\bgroup\ifcat0\csname b@\@citeb \endcsname \relax
    \egroup \@tempcntb\number\csname b@\@citeb \endcsname \relax
    \else \egroup \@tempcntb\m@ne \fi \ifnum\@tempcnta=\@tempcntb
    \ifx\@h@ld\relax \edef \@h@ld{\@citea\csname b@\@citeb\endcsname}%
    \else \edef\@h@ld{\hbox{--}\penalty\@highpenalty
    \csname b@\@citeb\endcsname}\fi
    \else \@h@ld\@citea\csname b@\@citeb \endcsname \let\@h@ld\relax \fi}%
\def\@citea{,\penalty\@highpenalty\hskip.13em plus.13em minus.13em}}\@h@ld}
\def\@citex[#1]#2{\@cite{\citen{#2}}{#1}}%
\def\@cite#1#2{\leavevmode\unskip\ifnum\lastpenalty=\z@\penalty\@highpenalty\fi%
  \ [{\multiply\@highpenalty 3 #1%
  \if@tempswa,\penalty\@highpenalty\ #2\fi}]}   %
\makeatother 

\catcode`\@=12

\def\A             {{\rm A}}
\def\AG            {A_G}
\def\abihom        {alternating bi-homo\-mor\-phism}

\def\alg           {algebra}

\def\AS            {S^{\sss(A)}}
\def\atype         {automorphism type}
\def\auta          {\mbox{Aut$(\cala)$}}
\def\auto          {automorphism}
\def\bashi         {boundary block}

\def\BB            {\Beta_\lambda}
\newcommand\bbb[1] {\rp\Beta_{\rp #1}}
\def\BBB           {\rp{\rm B}_\lambdab}
\newcommand\bbbJ[1]{\rp\Beta_{\J\rp #1}}
\newcommand\bbbJo[1]{\rp\Beta_{\Jo\rp #1_\circ}}
\newcommand\bbD[1] {\odtyp\Beta_{#1}}

\newcommand\bbN[1] {\evtyp\Beta_{#1}}

\def\bc            {boundary condition}
\def\Bc            {Boundary condition}
\def\be            {\begin{equation}}
\def\Be            {{\rm B}_\lambda^{(\bfe)}}
\def\bearl         {\begin{array}{l}}
\def\bearll        {\begin{array}{ll}}
\def\bearlll       {\begin{array}{lll}}
\def\Bet           {{\rm b}}
\def\Beta          {{\rm B}}

\def\Betab         {{\rp\Beta}}
\def\bfe           {{\bf1}}
\def\Bg            {{\rm B}_\lambda^{(\gg)}}
\def\Bgrho         {{\rm B}_\lambda^{(\gg_\rho)}}
\def\bihom         {bi-homo\-mor\-phism}
\def\bq            {{q}}

\def\bR            {{r}}
\def\Br            {{r'}}
\def\BR            {\rp{\cal R}}
\def\bs            {{s}}
\def\bS            {\rp S}
\def\caD           {{\rm o}}
\def\cala          {{\mathfrak A}}
\def\calap         {{\rp\cala}}
\def\calapp        {{\cala'}}
\def\calb          {{\cal B}}
\def\calc          {{\cal C}}
\def\calh          {{\cal H}}
\def\calhb         {\rp{\cal H}}
\newcommand\calhbJ[1]{\rp{\cal H}_{\J\rp{#1}}}
\def\calhl         {{\cal H}_\lambda}
\def\calhlp        {{\cal H}_\lambdap}
\def\calhv         {{\cal H}_\vac}
\def\caln          {{\cal N}}
\def\calm          {{\cal M}}

\def\cals          {{\cal S}}

\def\calu          {{\cal U}}

\def\calup         {\calu\oei}

\def\caN           {{\rm e}}
\def\cblock        {chiral block}
\def\cc            {{\sigm_{\rm c}}}
\def\ccc           {{\sigm_{\rm c}^*}}
\def\cdoT          {\,{\cdot}\,}
\def\cft           {conformal field theory}
\def\Cft           {Conformal field theory}
\def\cfts          {conformal field theories}
\newcommand\cg[2]  {\mbox{$\calc^{(#1;#2)}$}}
\def\chii          {\raisebox{.15em}{$\chi$}}

\def\chira         {chiral algebra}
\def\cirC          {\,{\circ}\,}
\newcommand\Cl[1]  {{\rm C}^{B\,(\rp #1)}}
\def\cla           {classifying algebra}
\def\clA           {\mbox{$\calc(\cala)$}}
\def\clAb          {\mbox{$\calc(\calap)$}}
\def\clAe          {\mbox{$\calc^{(\bfe)}\!(\calap)$}}
\def\clAg          {\mbox{$\calc^{(\gg)}\!(\calap)$}}
\def\clAgg         {\mbox{$\calc^{(\gg)}\!(\calapp)$}}
\def\clAm          {\mbox{$\odtyp\calc(\calap)$}}
\def\class         {classification}
\def\clo           {{\cal O}}
\def\complex       {{\dl C}}
\def\cocon         {coset construction}
\def\con           {conformal }
\def\Con           {Conformal }
\def\corfu         {correlation function}

\def\ctype         {Chan\hy Paton type}

\newcommand\CY[2]  {\tC_{{#1},{#2}}}
\def\df            {\,{:=}\,}
\def\dim           {dimension}
\def\Dim           {{\rm dim}\,}
\def\dl            {\mathbb }
\newcommand\dmatrix[9]{\left(\begin{array}{ccc}
                   #1&#2&#3\\#4&#5&#6\\#7&#8&#9\end{array}\right)}
\newcommand\Dmatrix[3]{\dmatrix {\!\!\!#1\one_{d_1}\!\!\!}00
                   0{\!#2\one_{d_2}\!}0 00{\!\!\!#3\one_{d_3}\!\!\!}}
\newcommand\dN[3]  {\odtyp{\rm N}_{#1,#2}^{\;\ \ #3}}

\def\dS            {\odtyP S}
\def\DS            {S^{\sss(D)}}
\def\DSs           {S^{{\sss(D)}*}}
\def\dsty          {\displaystyle}

\newcommand\e[1]   {{{\rm E}(#1)}}
\def\ee            {\end{equation}}
\def\eE            {{\rm e}}
\def\eear          {\end{array}}
\def\ekz           {\mbox{$\frac k2$}}

\def\eps           {{\psi'}}
\def\epss          {{\psi''}}
\def\eq            {\,{=}\,}
\newcommand\era[1] {(\ref{#1})$_{\!\sss A}$}
\newcommand\erA[2] {(\ref{#1#2})$_{\!\sss A}$}
\newcommand\erf[1] {(\ref{#1})}
\newcommand\erF[1] {(\ref{#1})}
\newcommand\Erf[2] {(\ref{#1#2})}
\newcommand\ErF[2] {(\ref{#1#2})}
\def\evtyp         {{}_{}^{\caN\sss\!}}
\def\evtyP         {{}_{}^{\caN\sss\!\!}}
\def\evodtyp       {{}_{}^{\caN\caD\sss\!\!}}
\def\F             {{\cal F}}
\def\fc            {{f_{\rm c}}}
\def\feta          {{(f{,}\psi)}}
\def\fetab         {{(\rp f{,}\psi)}}
\def\FFO           {E}
\def\findim        {finite-dimensional}
\def\FO            {{\cal E}}

\newcommand\Frac[2]{\mbox{\large$\frac{#1}{#2}$}}

\def\furu          {fusion rule}
\def\futnote#1     {\footnote{~#1}\ }
\def\g             {\mbox{$\mathfrak g$}}
\def\G             {{\rm G}}
\def\ga            {\gamma}
\def\gb            {\mbox{$\bar\mathfrak g$}}

\def\geps          {{(g{,}\eps)}}
\def\gepsb         {{(\rp g{,}\eps)}}
\def\gepSb         {{[\rp g{,}\eps]}}
\def\gf            {{g_f^{}}}
\def\gF            {{g_f^{\phantom|}}}
\def\gfs           {{g_f^*}}
\def\gg            {{g}}
\def\GG            {{\sf G}}
\def\GGG           {{\rm H}}

\def\gka           {\alpha}
\def\gkb           {\beta}

\def\Gmap          {\mbox{$\Gamma$}}

\def\gs            {{\gg^\star}}
\def\Gs            {{\cal G}}
\def\GS            {{G^*}}
\def\Gss           {{\cal G}''}
\def\gt            {\succ}
\def\GT            {\mbox{$\Gamma_{\!B}$}}
\def\Gtot          {\mbox{$\Gamma_{\!Z}$}}
\def\Hat           {}

\def\hepp          {{(h{,}\epss)}}
\def\heppb         {{(\rp h{,}\epss)}}

\newcommand\hN[3]  {\hat{\rm N}_{#1,#2}^{\;\;\ #3}}
\newcommand\HN[3]  {\tilde{\rm N}_{#1,#2}^{\;\;\ #3}}
\newcommand\HNe[1] {\tilde{\rm N}_{#1}}

\def\Hp            {\Gs\oei}
\def\hr            {{\Hat r}}
\def\Hr            {{\tilde r}}
\def\hs            {{\Hat s}}
\def\hS            {{\hat S}}
\def\Hs            {{\tilde s}}
\newcommand\hsp[1] {\mbox{\hspace{#1 em}}}
\def\hT            {{\Hat t}}  
\def\Ht            {{\tilde t}}
\def\hu            {{\Hat u}}
\def\Hu            {{\tilde u}}
\def\hy            {$\mbox{-\hspace{-.66 mm}-}$}
\def\I             {\cite{fuSc11}}
\def\ib            {\ii }
\def\id            {\mbox{\sl id}}
\def\ID            {\{\vacb\}}

\def\ii            {{\rm i}}
\def\II            {{\rm I}}

\def\iN            {\,{\in}\,}

\def\Infdim        {Infinite-dimensional }
\def\inv           {invariance}

\def\irrep         {irreducible representation}
\def\J             {{\rm J}}
\def\Jg            {\J_g}
\def\Jgp           {\J_g'}
\def\jgso          {\mbox{$J_{\rm GSO}$}}
\def\JJ            {{}_{\J}}

\def\JJL           {{}_{\JL}}
\def\JJo           {{}_{\Jo\!}}

\def\JK            {{\rm K}}
\def\Jl            {\J_\lambda}
\def\JL            {{\rm L}}
\def\Jo            {{{\rm J}_\circ}}
\def\Jop           {{{\rm J}'_\circ}}
\def\Jopp          {{{\rm J}''_\circ}}

\def\Kg            {\JK_\gg}
\def\Kggp          {\JK_{\gg\gg'}}
\def\Kgp           {\JK_{\gg'}}
\def\Kh            {\hat\JK_g}
\def\Khg           {\hat\JK_{\hat g}}
\def\Khgp          {\hat\JK_{\hat g'}}
\def\Khp           {\hat\JK_{g'}}
\def\kma           {Kac\hy Moody algebra}
\def\kzc           {Knizhnik\hy Zamolodchikov connection}
\renewcommand\L[1] {N_{#1}}
\long\def\labl#1   {\label{#1}\ee}
\long\def\Labl#1#2 {\label{#1#2}\ee}
\def\lambdab       {{\rp\lambda}}
\def\lambdaB       {{[\lambdab]}}
\def\Lambdab       {{[\lambdab,\psu]}}

\def\lambdabg      {{\lambda'}}
\def\lambdabo      {{\lambdab_\circ}}

\def\lambdaBO      {{[\lambdabo]}}
\def\LambdaBo      {{[\lambdabo,\psu]}}
\def\lambdabop     {{\lambdab^{\!+}_{\circ}}}
\def\lambdabp      {{\lambdab^{\!+}_{\phantom i}}}

\def\lambdag       {{\lambda'}}
\def\lambdao       {{\lambda_\circ}}
\def\lambdaO       {\lambda}
\def\lambdap       {{\lambda^{\!+}_{\phantom i}}}
\def\lb            {{\rp l}}

\def\lie           {Lie algebra}
\def\liefont       {\mathfrak }
\def\llb           {\mbox{\large(}}
\def\Llb           {\mbox{\Large(}}
\def\lrb           {\mbox{\large)}}
\def\Lrb           {\mbox{\Large)}}
\def\M             {I}
\def\Mb            {\rp\M}

\def\Me            {\M_{1/2}}
\def\Meb           {\Mb_{1/2}}
\def\Mf            {\M_{\rm f}}
\def\Mfb           {\Mb_{\rm f}}
\def\mi            {\,{-}\,}
\def\mimo          {minimal model}
\newcommand\Mo[3]  {{\rm M}^{#1,#2}_{\ \ \ \ \ \ \ \ \ \ #3}}
\def\Mo            {\M_\circ}
\def\Mob           {\rp\Mo}
\def\modinv        {modular invarian}

\def\mub           {{\rp\mu}}
\def\mubg          {{\mu'}}
\def\mug           {{\mu'}}
\newcommand\N[3]   {\rp{\rm N}_{\rp #1,\rp #2}^{\ \ \ \rp #3}}
\def\n             {\beta_\circ}
\def\nb            {\mbox{\small$\!\!\sqrt2$}}
\def\nbm           {\mbox{\small$\!\!-\sqrt2$}}
\newcommand\ND[3]  {\evtyp{\rm N}_{#1,#2}^{\;\;\ #3}}
\newcommand\NDn[3] {\evtyp{\rm N}_{#1,#2,#3}}
\def\nE            {\,{\ne}\,}
\newcommand\Ng[3]  {{\rm N}_{#1,#2}^{(\gg)\;#3}}
\newcommand\NJ[3]  {\rp{\rm N}_{\J\rp #1,\rp #2}^{\,\ \ \ \ \rp #3}}
\newcommand\NJJ[3] {\rp{\rm N}_{\rp #1,\J\rp #2}^{\,\ \ \ \ \rp #3}}
\newcommand\NJJJ[3]{\rp{\rm N}_{\rp #1,\rp #2}^{\ \ \J\rp #3}}
\def\nl            {{n_\lambda^{}}}
\def\NM            {{\rm M}}

\def\nn            {$N\,{=}\,2$ }
\newcommand\Nn[3]  {\rp{\rm N}_{\rp #1,\rp #2,\rp #3}}

\newcommand\norm[2]{{\cal N}_{\rp #1,#2}}
\newcommand\Norm[2]{{\cal N}_{#1}}
\newcommand\Ntot[3]{{\rm M}_{#1,#2}^{\;\ \ #3}}
\def\nubg          {{\nu'}}
\def\nug           {{\nu'}}
\newcommand\nxt[1] {\\\raisebox{.12em}{\rule{.35em}{.35em}}\hsp{.6}#1}

\def\oei           {'}

\def\odtyp         {{}_{}^{\caD\sss\!}}
\def\odtyP         {{}_{}^{\caD\sss\!\!}}
\def\Odtyp         {{}_{\phantom I}^{\caD\sss}\!}
\def\odevtyp       {{}_{}^{\caD\caN\sss\!\!}}

\def\Om            {{\rm o}}
\def\omm           {\varpi}
\def\ommdual       {\omm^\star}

\def\one           {\mbox{\small $1\!\!$}1}
\def\onedim        {one-dimen\-sional}
\newcommand\oNn[3] {\pNn{#2}{#1}{#3}}
\newcommand\oS[2]  {\breve S_{\rp{#1},\rp{#2}}}

\def\opluS         {\,{\oplus}\,}
\def\ordg          {{|G'|}}
\def\ot            {\raisebox{.07em}{$\scriptstyle\otimes$}}
\def\oT            {\,\ot\,}
\def\otimeS        {\,{\otimes}\,}

\def\parfu         {partition function}
\def\Phig          {\Phi^{(\gg)}}
\def\Phigp         {\Phi^{(\gg')}}
\def\phio          {{\varphi_\circ}}
\newcommand\pho[1] {\phi_{#1,\tilde{#1}}}

\def\phu           {{\hat\varphi}}

\def\pl            {\,{+}\,}

\newcommand\pN[3]  {\breve{\rm N}_{\rp #1,\rp #2}^{\ \ \ \rp #3}}
\newcommand\pNn[3] {\breve{\rm N}_{\rp #1,\rp #2,\rp #3}}

\def\po            {p_\circ}

\def\PSI           {{}^\Psi\!}
\def\psio          {{\psi_\circ}}
\def\psiop         {{\psi'_\circ}}
\def\psiopp        {{\psi''_\circ}}

\def\psu           {{\hat\psi}}
\def\Psu           {{\hat\J}}
\def\psuo          {{\psu_\circ}}
\def\psup          {{\psu\oei}}
\def\q             {quantum }

\def\qo            {q_\circ}
\newcommand\rank[1]{\mbox{rank}(#1)}

\newcommand\Rc[3]  {{\rm R}^{#1}_{#2;#3}}
\def\rationals     {{\dl Q}}
\def\rchi          {\raisebox{.15em}{$\rp\chi$}}
\def\rep           {representation}
\def\resp          {respectively}
\def\rhob          {{\rp\rho}}
\def\rhoB          {{[\rhob]}}
\def\RhoB          {{[\rhob,\psu_\rho]}}
\def\RHoB          {{[\rhob,\psu]}}

\def\RhoBe         {{[\rhob_1,\psu_1]}}
\def\rhobg         {\rhog}

\def\RhoBz         {{[\rhob_2,\psu_2]}}

\def\rhod          {{\rho_3^{}}}
\def\rhoe          {{\rho_1^{}}}
\def\rhog          {{\rho'}}
\def\rhov          {{\rho_4^{}}}
\def\rhoz          {{\rho_2^{}}}

\def\ro            {{\rm r}}
\def\rp            {\bar}
\def\rpa           {{\rp\alpha}}
\def\rpb           {{\rp\beta}}
\def\rpc           {{\rp\gamma}}

\newcommand\Sb[2]  {\rp S_{\rp{#1},\rp{#2}}}
\newcommand\SD[2]  {\evtyp S_{#1,#2}}
\newcommand\SDo    {\evtyP S}
\newcommand\sect[1]{\section{#1}\setcounter{equation}{0}}
\def\si            {{\rm s}}
\def\sigm          {\sigma}
\def\sigmab        {{\rp\sigma}}

\def\SigmaBp       {{[\sigmab,\psu_\sigma]\oei}}

\def\sigs          {{\sigm^*}}
\newcommand\Sj[2]  {\rp S_{\J\rp{#1},\rp{#2}}}
\def\SJ            {S^\J}
\def\sltwo         {\mbox{$\mathfrak{sl}(2)$}}
\def\so            {{\rm so}}

\def\SO            {{\rm SO}}

\def\Spin          {{\rm Spin}}
\def\ss            {{\rm s}}
\def\sss           {\scriptscriptstyle}
\def\ssty          {\scriptstyle}
\def\stt           {string theory}

\def\SU            {{\rm SU}}

\newcommand\sumb[1]{\sum_{\rp #1\in\Mb}}
\newcommand\sumbo[1]{\sum_{\ssty\rp #1 \atop Q_\Gs(#1)=0}}
\newcommand\sume[1]{\sum_{#1\in\Me}}
\newcommand\sumeb[1]{\sum_{\rp #1\in\Meb}}

\newcommand\sumfb[1]{\sum_{\rp #1\in\Mfb}}

\newcommand\sumofb[1]{\sum_{\rp #1\in\Mob\cup\Mfb}}

\newcommand\Sumpsipsu[1]{\sum_{\ssty\psi\in\cals_{#1}^* \atop \psi\gt\psu}}

\def\sutwo         {\mbox{$\liefont{sl}(2)$}}
\def\sym           {symmetry}
\def\syms          {sym\-me\-tries}
\def\Tau           {\Theta}
\def\tBeta         {{\Tilde\Beta}}
\def\tC            {C^B}

\def\tf            {\Tau_\gf}
\def\tg            {\Tau_\gg}
\def\tgp           {\Tau_{\gg'}}

\def\Tilde         {\tilde}
\def\timeS         {\,{\times}\,}

\def\tims          {\oT}

\newcommand\tNl[3] {\Tilde{\rm N}_{#1,#2,#3}^{}}
\newcommand\tNm[3] {\Tilde{\rm N}_{#1,#2}^{\;\;\ #3}}
\newcommand\tNn[3] {\Tilde{\rm N}_{#1,#2}^{\ \ #3}}
\def\tomm          {\Tau_{\omm}}
\def\tPhi          {\Tilde\Phi}

\def\trfo          {transformation}

\def\tS            {\Tilde S}

\def\tto           {\to\hsp{-1.42}\to}

\def\twodim        {two-dimensional}

\def\uone          {\mbox{$\liefont{u}(1)$}}
\def\uonE          {{\liefont{u}(1)}}
\def\univ          {{\cal C}^\infty}
\def\untw          {^{\sss(1)}}

\def\ustab         {untwisted stabilizer}
\def\vac           {\Omega}
\def\vacb          {{\rp\vac}}

\def\voa           {vertex operator algebra}
\def\V             {{\cal V}}
\def\vir           {\mbox{${\mathfrak V}${\sl ir}}}

\def\vphi          {\varphi}
\def\Vpsu          {\V_{\psu_\lambda}}
\def\Vpsup         {\V_{\psu^+_\lambda}}
\def\vv            {{\rm v}}
\def\vvir          {v_{\rm Vir}}
\def\wrt           {with respect to }
\def\wrtt          {with respect to the }
\def\wzwm          {WZW model}
\def\wzwt          {WZW theory}
\def\wzwts         {WZW theories}

\newcommand\YS[2]  {\tS_{{#1},{#2}}}
\def\zet           {{\dl Z}}

\def\zetay         {\zeta_Y}
\def\zetplus       {{\dl Z}_{>0}}
\def\zetpluso      {{\dl Z}_{\ge0}}
\def\zg            {{{<}{g}{>}}}
\def\zgs           {{{G'}^*_{}}}

\documentclass[12pt]{article} \usepackage{amssymb,amsfonts}
\def\theequation{{\thesection.\arabic{equation}}}
\begin{document}
\setcounter{section}6


\begin{flushright}  {~} \\[-1cm] {\sf hep-th/9908025} \\[1mm]
{\sf ETH-TH/99-04} \\[1 mm]
{\sf August 1999} \end{flushright}
 
\begin{center} \vskip 22mm
{\Large\bf SYMMETRY BREAKING BOUNDARIES}\\[4mm]
{\Large\bf II.\ MORE STRUCTURES; EXAMPLES}\\[22mm]
{\large J\"urgen Fuchs} \ and \ {\large Christoph Schweigert}\\[5mm]
Institut f\"ur Theoretische Physik \\
ETH H\"onggerberg \\[.2em] CH -- 8093~~Z\"urich
\end{center}
\vskip 26mm
\begin{quote}{\bf Abstract}\\[1mm]
Various structural properties of the space of symmetry breaking boundary
conditions that preserve an orbifold subalgebra are established. To each
such boundary condition we associate its automorphism type. We show that
correlation functions in the presence of such boundary conditions are
expressible in terms of twisted boundary blocks which obey twisted Ward
identities. The subset of boundary conditions that share the same
automorphism type is controlled by a classifying algebra, whose 
structure constants are shown to be traces on spaces of chiral blocks.
T-duality on boundary conditions is not a one-to-one map in general.
\\
These structures are illustrated in a number of examples. 
Several applications, including the construction
of non-BPS boundary conditions in string theory, are exhibited.
\end{quote}
\newpage


\section*{Introduction}

The study of conformally invariant \bc s in \twodim\ \cft\
is of considerable interest both
for applications in condensed matter physics and in string theory.
In such applications typically only the (super-)conformal symmetry needs to be 
preserved by the boundaries, while the rest of the chiral bulk symmetries 
$\cala$ may be broken.

In \I\ we have studied conformally invariant
\bc s for an arbitrary \cft\ that preserve a (consistent) sub\alg\ $\calap$ 
of $\cala$, such that 
  $$   \calap = \cala^G_{}  $$ 
is the sub\alg\ that is fixed under a finite abelian group $G$ of 
automorphisms of $\cala$. We have shown that such \bc s are 
governed by a {\em \cla\/} \clAb, in the sense that the reflection coefficients
\cite{lewe3,prss3} -- the data that characterize the \bc\ --
are precisely the \onedim\ \irrep s of \clAb.

This paper is a continuation to \I. 
Our numbering of sections will be consecutive, i.e.\ 
by sections 1 to 6 we refer to those of the preceding paper, while here we
will start with section 7; accordingly equation numbers with section label 
${\le}\,6$ refer to formulas in \I.\,%
 \futnote{A few of those formulas are, however, reproduced in the appendix
of the present paper. This is indicated by an additional subscript `$A$' of
the equation number.}
 A very brief summary of the pertinent results of \I\ is as follows.
The \alg\ \clAb\ is a commutative associative semisimple algebra. Thus its
regular \rep\ is fully reducible, and the structure constants are expressible
through the corresponding diagonalizing matrix $\tS$ by an analogue of the
Verlinde formula. This matrix $\tS$, in turn, can be expressed in terms of
various quantities that are already known from the chiral \cft\ associated
to $\calap$ (see formula \erA tS).

The \cft\ with chiral \alg\ $\calap$ can be obtained from the $\cala$-theory
as an {\em orbifold\/} by the group $G$; conversely, the original 
$\cala$-theory is recovered from the $\calap$-theory as an integer spin 
{\em simple current extension\/}, with the group $\Gs$ of simple currents
being the character group of the orbifold group, $\Gs\eq G^*$. Denoting
the labels for the primary fields of the $\cala$-theory by $\lambda$ and those
for the primary $\calap$-fields by $\lambdab$,
a natural basis of \clAb\ is labelled by pairs $(\lambdab,\varphi)$, where 
$\lambdab$ refers to a $\calap$-primary in the untwisted sector and
$\varphi\iN\cals_\lambda^*$ is a group character, while the boundary conditions
are labelled by pairs $\RhoB$ consisting of an arbitrary primary label
$\rhob$ of the $\calap$-theory and a character $\psu_\rho\iN\calu_\lambda^*$.
Here the stabilizer $\cals_\lambda$ and the untwisted stabilizer 
$\calu_\lambda$ are subgroups of the simple current group $\Gs\eq\GS$; 
$\cals_\lambda$ consists of all simple currents in $\Gs$ that leave
$\lambdab$ fixed, and $\calu_\lambda$ is the subgroup of $\cals_\lambda$ on
which a certain \abihom\ $F_\lambda$, which can be defined through the 
modular properties of one-point chiral blocks on the torus, is trivial.
For the precise meaning of these terms we refer to \I\ (in particular appendix 
A) and to \cite{fusS6}. 
Quite generally, the number of basis elements of \clAb\ -- or equivalently, 
the number of independent {\em boundary blocks\/}, i.e.\ chiral blocks
for one-point \corfu s of bulk fields on the disk -- and the number of \bc s
have to be equal. It is a rather non-trivial result of the analysis that
this equality indeed holds for the two sets of labels, so that in particular 
the matrix $\tS\,{\equiv}\, \tS_{(\lambdab,\varphi),\RhoB}$ is a square matrix.

The results established in \I\ clearly demonstrate an unexpectedly nice
behavior of the space of conformally invariant \bc s.
In the present paper we show that indeed this space is endowed with
even more structure. The following section \ref{t.0} deals with the
implementation of the orbifold group on the \rep\ spaces of the chiral \alg\ 
$\cala$; this discussion does not yet involve \bc s at all, but the results
will be needed in the sequel.  Sections \ref{t.1} -- \ref{t.6} are devoted to 
the discussion of several additional generic features of symmetry breaking 
\bc s.  We start in section \ref{t.1} by analyzing the notion of 
{\em automorphism type}, which in the present setting is a derived 
concept that arises as a direct consequence of the general
structure and does not need be introduced by
hand. Next we show that \bc s of definite \atype\ can be naturally
formulated with the help of {\em twisted boundary blocks\/}, which satisfy
twisted Ward identities (section \ref{t.2}). Furthermore, to the \bc s of 
fixed \atype\ one can associate their own \cla, which is an invariant sub\alg\
of the total \cla\ \clAb\ and whose structure constants can be understood 
in terms of suitable traces on \cblock s; this is done in section \ref{t.3}.
It is also shown that the individual \cla\ for \atype\ $g$ only depends
on the \auto\ $g$, but not on the specific orbifold subalgebra $\calap$, 
i.e.\ not on the group $G$ containing $g$.

Afterwards, in section \ref{t.4}, we turn to a detailed study of the 
dependence of the \cla\ on the chosen torus \parfu, which leads to the 
concept of T-{\em duality\/} of \bc s. First we show that only the `difference'
between an automorphism characterizing the torus partition function and the 
\atype\ of a \bc\ is observable, and then we discuss aspects of T-duality 
among (families of) \bc s for fixed choice of the torus \parfu.
We emphasize that T-duality on \bc s is not a one-to-one map, in general.
In section \ref{t.5} we establish an action of the orbifold group $G$ 
on the space of boundary conditions, which implies a certain
`homogeneity' among the \bc s for fixed $\Gs$-orbit $\rhoB$.
Finally we introduce in section \ref{t.6} the concept of a 
{\em universal\/} classifying algebra, which governs {\em all\/}
conformally invariant \bc s at the same time, and discuss the possibility
to obtain this \alg\ by a suitable projective limit.

In sections \ref{t.7} and \ref{t.8} we turn our attention to a specific
class of \bc s to which we refer as involutary, namely those where the
orbifold group is $\zet_2$. We first address some general features and then,
in section \ref{t.8}, analyze several classes of examples that are of 
particular interest. Afterwards, in section \ref{t.9}
we provide several classes of examples with more complicated orbifold groups,
in which for instance untwisted stabilizer subgroups occur that are proper
subgroups of the full stabilizers. Finally, some of the pertinent formulae 
from \I\ that will be needed in the sequel are collected in appendix A.

\sect{The action of the orbifold group on $\cala$-modules} \label{t.0}

The elements $g$ of the orbifold group $G$ are \auto s of the chiral \alg\
$\cala$. In the sequel we will have to deal with various subgroups of $G$
and their properties. We first observe that every \auto\ $\gg$ of $\cala$ 
can be implemented on the physical $\cala$-modules $\calh_\lambda$ by maps
  \be  \tg\,{\equiv}\,\tg^{(\lambda)}:\quad
  \calhl \to \calh_{\gs\!\lambda} \,, \Labl tg
which obey the $\gg$-twisted intertwining property
   \be  \tg\,Y = \gg(Y)\,\tg \qquad\mbox{for all }\; Y\iN\cala \,,  \Labl tq
and the maps $\tg$ are defined 
by this property up to a scalar multiple. (For a concrete 
realization of these maps in \wzwts, see \cite{bifs}.) In general, such an 
implementation $\tg^{(\lambda)}$ maps a given space $\calh_\lambda$ to some other 
$\cala$-module $\calh_{\gs\!\lambda}$, thereby organizing the primary fields of
the $\cala$-theory into orbits, much like the simple current group 
$\Gs\,{\cong}\,G^*$ organizes \cite{scya6} the $\calap$-primaries into orbits. 

To every $\cala$-primary $\lambda$ we can associate the {\em stabilizer\/}
  \be  S_\lambda := \{ g\iN G \,|\, \gs\lambda\eq\lambda \} \,,  \ee
which is a subgroup of $G$ whose elements constitute 
{\em endo\/}morphisms $\calh_\lambda\,{\to}\,\calh_\lambda$. 
Stabilizers of $\cala$-primaries on the same $G$-orbit are identical
(in the more general case of non-abelian $G$, they are conjugate subgroups);
the vacuum has a maximal stabilizer, $S_\vac\eq G$. Also,
$\cala$-modules on the same $G$-orbit are isomorphic as $\calap$-modules.
The endomorphisms $\tg$ for $\gg\iN S_\lambda$ provide us with 
an action of $S_\lambda$ on $\calh_\lambda$ which is, in general, only 
projective, and hence determines a two-cocycle $\FO_\lambda$ of $S_\lambda$
or, more precisely (in agreement with the fact that the maps $\tg$ are
defined only up to normalization), the cohomology class of $\FO_\lambda$.
We denote by $U_\lambda$ the subgroup of $S_\lambda$ that 
corresponds to the regular elements of the 
associated twisted group \alg\ $\complex_{\FO_\lambda}S_\lambda$, i.e.
  \be  U_\lambda := \{ g\iN S_\lambda \,|\, \FO_\lambda(g,g')\eq
  \FO_\lambda(g',g)\ {\rm for\;all}\ g'\iN S_\lambda \} \,.  \ee

The scalar factors in the definition of the implementers $\Tau_\gg$ 
can be chosen in such a way that the maps $\tg$ with $\gg\iN U_\lambda$
provide us with a honest \rep\ of $U_\lambda$
on $\calh_\lambda$. It follows that the $\cala$-modules $\calh_\lambda$ 
can be decomposed as
  \be \calh_\lambda^{} \cong \bigoplus_{\Psu\in U_\lambda^*} V_{\Psu}
  \otimes \calhb_{\lambdab,\Psu} \,,  \Labl VH
where the spaces $V_\Psu$ are projective $S_\lambda$-modules and the spaces
$\calhb_{\lambdab,\Psu}$ are $\calap$-modules. We make the mild technical 
assumption that all these modules $V_\Psu$ and $\calhb_{\lambdab,\Psu}$ are 
{\em irreducible\/}; this holds true in all known examples, and is rigorously 
proven for the vacuum $\vac$ \cite{dolm5,doMa2} as well as \cite{doMa3} for 
other $\cala$-modules, including twisted sectors.
In the case of the vacuum, no multiplicities appear in this decomposition;
thus the action of $G$ is genuine and we have $U_\vac\eq S_\vac\eq G$.

Since by construction $\gg$ leaves the sub\alg\ $\calap$ of $\cala$ fixed,
the maps $\tg$ are {\em ordinary\/} intertwiners for $\calap$; hence in 
the decomposition \Erf VH they act solely on the degeneracy space $V_\Psu$.
Moreover, by the general properties of twisted group \alg s (compare 
appendix B of \I), all the spaces $V_\Psu$ have the same dimension 
$\sqrt{|S_\lambda|/|U_\lambda|}$, and the basis of the twisted group \alg\ 
$\complex_{\FO_\lambda}\!S_\lambda$ can be chosen in such a way that every 
$\gg\iN U_\lambda$ is implemented as a diagonal matrix acting on $V_\Psu$.

The result \Erf VH should be compared to the similar decomposition \erF{deco}
that arises from the simple current point of view, i.e.
  \be  \calh_\lambda^{} \equiv \calh_\Lambdab =
  \bigoplus_{\J\in\Gs/\cals_\lambda} \V_\psu \otimes \calhb_{\J\lambdab} \,.
  \labl{Deco}
Here the spaces $\V_\psu$ are projective $\cals_\lambda$-modules,
with corresponding cocycle $\F_\lambda$ of $\cals_\lambda$, while the spaces
$\calhb_{\J\lambdab}$ are $\calap$-modules; by assumption, the latter modules
are irreducible (this assumption is indeed satisfied for all cases we know of).
 Simple current theory \cite{mose4,scya,scya6,fusS6} shows that in the
decomposition \erf{Deco} isomorphic $\calap$-modules appear precisely
as a consequence of fixed point resolution; therefore the multiplicity of 
$\calhb_{\J\lambdab}$ in this decomposition is given by $|\calu_\lambda^*|$. 
On the other hand, as seen above, elements of the orbifold
group $G$ that are not in the stabilizer $S_\lambda$ relate isomorphic 
$\calap$-modules. Thus we can identify the groups $\calu_\lambda^*$ 
and $G/S_\lambda$. To make this manifest, we dualize the exact 
sequence $0\to \calu_\lambda\to\Gs$ and complete it to an exact sequence
  \be 0 \to S_\lambda \to G=\Gs^* \to \calu_\lambda^* \to 0 \,.  \Labl01
Conversely, in the decomposition of a $\cala$-module $\calh_\lambda$ there
appear $|U_\lambda^*|$ many irreducible $\calap$-modules. In simple current 
language, the number of irreducibles is just the length of the orbit, and 
hence again we can dualize $0\to U_\lambda \to G$ and complete it to
  \be  0 \to \cals_\lambda \to \Gs=G^* \to U_\lambda^* \to 0 \,.  \Labl00
As a consequence, the cardinalities of the respective subgroups of $G$ and
$\Gs$ are related by
  \be |\cals_\lambda|\, |U_\lambda| = |\Gs|= |G| = |S_\lambda|\,|\calu_\lambda|
  \,,  \Labl09
and hence in particular the \dim s of the degeneracy spaces in
the decompositions \Erf VH and \erf{Deco} coincide:
  \be  (\Dim\V_\psu)^2 = {|\cals_\lambda|}\,/\,{|\calu_\lambda|} 
  = {|S_\lambda|}\,/\,{|U_\lambda|} = (\Dim V_\Psu)^2 \,.  \ee
We abbreviate these dimensions by
  \be  d_\lambda := \Dim\V_\psu = \Dim V_\Psu  \,.  \ee

There is also a manifest relationship between the groups $\cals_\lambda$ and
$G/U_\lambda$. First we realize that the implementation of $G$ on the 
whole $G$-orbit of $\lambda$ provides us with a two-cocycle of $G$ with 
values in U(1) whose restriction to $S_\lambda\,{\times}\,S_\lambda$
coincides with $\FO_\lambda$; we again denote this cocycle by the symbol
$\FO_\lambda$. Given such a cocycle, its {\em commutator cocycle\/} 
$\FFO_\lambda$, which is defined by
  \be  \FFO_\lambda(g,g') := \FO_\lambda(g,g') / \FO_\lambda(g',g)  \ee
for all $g,g'\iN G$, constitutes \I\ 
a {\em \bihom\/} on $G\,{\times}\,G$ which is {\em alternating\/}
in the sense that $\FFO_\lambda(g',g)\eq\FFO_\lambda(g,g')^*$.
Now let us characterize for every $\J\iN\cals_\lambda$ an element $h_\J$ 
of $G$ by the property that
  \be  \FFO_\lambda(h_\J,g) = \J(g) \quad\ {\rm for\ all}\;\ g\iN G \,;  
  \Labl90
such a group element $h_\J$ exists because, owing to the exactness of the 
sequence \Erf00,
for every $\J\iN\cals_\lambda$ we have $\J(g)\eq1$ for all $g\iN U_\lambda$,
and \Erf90 characterizes $h_\J$ uniquely up to an element of $U_\lambda$.
Furthermore, as a consequence of the character property of $\FFO_\lambda$ in 
the first argument, we have
  \be  \FFO_\lambda(h_{\J\J'},g) = \FFO_\lambda(h_\J h_{\J'},g)  \ee
for all $g\iN G$, which tells us that $h_\J h_{\J'}\eq h_{\J\J'}$ modulo 
$U_\lambda$. It follows that the mapping
  \be  \J \,\leftrightarrow\, h_\J\,U_\lambda  \ee
constitutes an isomorphism between $\cals_\lambda$ and $G/U_\lambda$.
It is worth noting that this isomorphism is logically independent from
the isomorphism between $G^*\!{/}\cals_\lambda$ and $U_\lambda^*$ that exists 
according to the sequence \Erf00.

Again this result has an obvious dual analogue. To work this out we recall 
from appendix A of \I\ that the commutator cocycle $F_\lambda$ of 
$\F_\lambda$, defined by $F_\lambda(\J,\JL)\eq\F_\lambda(\J,\JL)/
\F_\lambda(\JL,\J)$ for $\J,\JL\iN\cals_\lambda$, possesses a natural 
extension. $F_\lambda$ is an \abihom\ on $\cals_\lambda\,{\times}\,
\cals_\lambda$, while its extension is a \bihom\ on $\Gs\,{\times}\,
\cals_\lambda$; by imposing the alternating property it can be further 
extended to a \bihom\ on $\Gs\,{\times}\,\Gs$, still to be denoted by 
$F_\lambda$. By the character property of $F_\lambda$ in the second 
argument, we can then associate to every $\gg\iN S_\lambda$ an element
$\Kg\,{\equiv}\,\Kg^{(\lambda)}$ of $\Gs$ by stipulating that
  \be  F_\lambda(\Kg,\JL) = \gg(\JL) \quad\ {\rm for\ all}\;\
  \JL\iN\Gs \,,  \Labl Kg
which determines $\Kg$ uniquely up to elements of $\calu_\lambda$.
The character property of $F_\lambda$ in the first argument implies
  \be  F_\lambda(\Kggp,\JL)
  = F_\lambda(\Kg\Kgp,\JL) \,,  \ee
so that $\Kg\Kgp\eq\Kggp$ modulo $\calu_\lambda$. Thus
 the map
  \be  \gg \,\leftrightarrow\, \Kg\,\calu_\lambda  \Labl gK
between $S_\lambda$ and $\Gs/\calu_\lambda$ is an isomorphism.

For later reference we also mention another isomorphism that is similar to
\Erf gK. Namely, we now consider a character $\hat g$ of $\cals_\lambda$
rather than a character $g$ of $\Gs$. Then the requirement that
  \be  F_\lambda(\Khg,\JL) = \gg(\JL) \quad\ {\rm for\ all}\;\
  \JL\iN\cals_\lambda  \Labl Kh
determines $\Khg\iN\cals_\lambda$ uniquely modulo $\calu_\lambda$, and
  \be  \hat g \,\leftrightarrow\, \Khg\,\calu_\lambda  \Labl KH
is an isomorphism between those elements of $\cals_\lambda^*$ which are the
identity on $\calu_\lambda$ (and hence can be regarded as restrictions of
elements of $S_\lambda$ to $\cals_\lambda$) and $\cals_\lambda/\calu_\lambda$.
Furthermore, combining the prescriptions above we learn that the restrictions
of the relevant \bihom s to the stabilizer groups are closely related.
Indeed, denoting by $\hat g$ the restriction of a given element $g$ of
$S_\lambda$ to $\cals_\lambda$, we have
  \be  \FFO_\lambdaO(\gg,\gg') = \Khgp(\gg) = \hat g(\Khgp)
  = F_\lambdaO(\Khg,\Khgp) \Labl EF
with $\gg,\gg'\iN S_\lambda$ and $\Khg,\Khgp\iN\cals_\lambda$. 

\sect{Automorphism types} \label{t.1}
 
We have demonstrated in \I\ that the reflection 
coefficients $\Rc\RhoB{(\lambdab,\varphi)}\vacb$, i.e.\ the operator product 
coefficients in the expansion \era{oben2} of a bulk field approaching
the boundary, are equal to \onedim\ \irrep s $R_\RhoB$ of the \cla\ \clAb, 
evaluated at the basis element $\tPhi_{(\lambdab,\varphi)}$ of \clAb. Thus
they are given by (see formula \erF R) 
  \be  \Rc\RhoB{(\lambdab,\varphi)}\vacb
  = R^{}_\RhoB (\tPhi_{(\lambdab,\varphi)})
  = \tS_{(\lambdab,\varphi),\RhoB}/ \tS_{\vacb,\RhoB} \,.  \Labl71
Moreover, it follows from the sum rule \erA sr and the fact that 
\clAb\ is semisimple, that the reflection coefficients even provide {\em all\/} 
inequivalent irreducible \clAb-\rep s. The isomorphism classes of
irreducible \clAb-\rep s are in one-to-one correspondence with the
conformally invariant boundary conditions that preserve the orbifold 
subalgebra $\calap\eq\cala^G$ of the chiral algebra $\cala$.
In this section we discuss some implications of this basic result.

Let us associate to each \bc\ $\rho\,{\equiv}\,\RhoB$ the collection of 
all monodromy charges $Q_\J(\rho)$, $\J\iN\Gs$, of $\rhob$. The 
monodromy charges do not depend on the choice of a representative of 
the $\Gs$-orbit $\rhoB$, and via the prescription
  \be  \gg_\rho(\J) := \exp(2\pi\ii Q_\J(\rho))  \Labl gg
for all $\J\iN\Gs$, they furnish a character 
  \be  \gg_\rho^{}
  \in{\Gs}^*  \ee
of the simple current group.
(This $\Gs$-character should not be confused with $\psu_\rho$, which is a 
character of the subgroup $\calu_\rho\,{\subseteq}\,\Gs$.)
The group ${\Gs}^*\eq{(\GS)}^*_{}$ can be naturally 
identified with the orbifold group, ${\Gs}^*\,{\equiv}\,G$,
and hence the quantity $\gg_\rho$ can be regarded as an element of $G$. 

To proceed, we observe that because of the simple current symmetry
   \be  \tS_{\J(\lambdab,\varphi),\RHoB}
   = \gg_\rho(\J)\, \tS_{(\lambdab,\varphi),\RHoB}  \Labl sq
that was established in formula \ErF SQ, we have 
  \be  \bearll
  R_\RhoB (\tPhi_{\J(\lambdab,\varphi)}) \!\!\!
  &= \Frac{\tS_{\J(\lambdab,\varphi),\RhoB}}{\tS_{\vacb,\RhoB}}
   = \gg_\rho(\J) \, \Frac{\tS_{(\lambdab,\varphi),\RhoB}}{\tS_{\vacb,\RhoB}}
   = \gg_\rho(\J) \cdot R_\RhoB(\tPhi_{(\lambdab,\varphi)})
  \eear \labl3
for every simple current $\J\iN\Gs$. Now the reflection coefficients 
constitute the main ingredient in the relation between the boundary 
{\em blocks\/} $\tBeta_{(\lambdab,\varphi)}$ (defined in formula \era{tBeta})
and the boundary {\em states\/} $\calb_\RHoB$. When using the notation 
introduced in \Erf gg, the precise relationship, established in formula 
\ErF89, reads
  \be  \calb_\RHoB
  = \bigoplus_{\ssty\lambdab \atop \gg^{}_\lambda\equiv1}
  \bigoplus_{\varphi\in\cals_\lambda^*}
  \Rc\RHoB{(\lambdab,\varphi)}\vacb\, \langle\Psi^{\RHoB\,\RHoB}_\vacb\rangle\,
  \tBeta_{(\lambdab,\varphi)} \,.  \labl{89'}
Thus the observation \erf3
tells us that the boundary blocks $\tBeta_{(\lambdab,\varphi)}$
for primary fields $\lambdab$ of the $\calap$-theory that lie on one and the 
same $\Gs$-orbit contribute to the boundary states $\calb_\RHoB$ with a
fixed relative phase, which is determined by the element $\gg_\rho$
of the orbifold group. Put differently, in the presence of the \bc\
$\rho\,{\equiv}\,\RHoB$ the reflection of a bulk field at the boundary is 
{\em twisted\/} by the action of the group element $\gg_\rho\iN G$. 

This observation suggests that, in the terminology of 
\cite{fuSc6}, the orbifold group element $\gg_\rho$ provides us with 
the {\em automorphism type\/}\,%
 \futnote{The corresponding term in \cite{reSC} is the {\em gluing \auto\/}.
The information contained in a \bc\ that goes beyond the \atype\ 
was referred to as the {\em\ctype\/} in \cite{fuSc6}.
Thinking in analogy with the general analysis of modular invariant
partition functions on the torus, it may seem to be more suggestive 
to take the \furu\ \auto\ $\gs$ as a starting point for the description of
the \atype\ \cite{fuSc6}. However, several different \atype s may give rise to
one and the same permutation $\gs$. For instance, in the case of \wzwts,
\auto s of the underlying \findim\ compact simple \lie\ \gb\ 
provide us with an \atype\ $\gg$, but whenever that \auto\ of \gb\ is 
inner, the associated map $\gs$ is just the identity.
Another example is given by the inner \auto s of the rational free boson
theories whose fixed point \alg s correspond to the boson theory compactified
at an integral multiple of the original radius;
this will be discussed in subsection \ref{t.91}.}
of the \bc\ $\rho$. To establish that this is indeed the case, we insert the 
expression \Erf71 for the reflection coefficients and the explicit values of 
the one-point correlators of the boundary vacuum fields 
$\Psi^{\RHoB\,\RHoB}_\vacb$ into formula \erf{89'}, so as to arrive at
  \be  \calb_\RHoB
  = \bigoplus_{\ssty\lambdab \atop \gg^{}_\lambda\equiv1}
  \bigoplus_{\varphi\in\cals_\lambda^*}
  \tS_{(\lambdab,\varphi),\RHoB}\, \tBeta_{(\lambdab,\varphi)}  \,.  \Labl07
Next we split the summation over all untwisted $\lambdab$ into a
summation over $\Gs$-orbits and one within orbits, and the summation over
$\cals_\lambda^*$ into one over $\calu_\lambda^*$ and one over
$\cals_\lambda^*/\calu_\lambda^*$. To this end we choose (once and for all)
arbitrarily a set $\{\lambdabo\}$ of representatives of the set 
of $\Gs$-orbits and a set $\{\phio\}$ of representatives of the classes of 
$\cals_\lambda^*/\calu_\lambda^*$; more precisely, the symbol $\lambdabo$
will refer to the chosen representative of the orbit $[\lambdab]$, and
$\phio\iN\cals_\lambdaO^*$ to the chosen representative of the class in
$\cals_\lambda^*/\calu_\lambda^*$ that restricts to $\phu\iN\calu_\lambda^*$,
i.e.\ satisfies $\phio_{|\calu_\lambdaO}{=}\,\phu$. Then \Erf07 becomes
  \be  \bearll  \calb_\RHoB \!\!
  &= \dsty\bigoplus_{\ssty\lambdaBO \atop \gg^{}_\lambdaO\equiv1}
  \bigoplus_{\phu\in\calu_\lambdaO^*} \bigoplus_{\J\in\Gs/\calu_\lambdaO}
  \tS_{\J(\lambdabo,\phio),\RHoB}\, \tBeta_{\J(\lambdabo,\phio)}
  \\{}\\[-.8em]
  &= \dsty\bigoplus_{\ssty\lambdaBO \atop \gg^{}_\lambdaO\equiv1}
  \bigoplus_{\phu\in\calu_\lambdaO^*} \tS_{(\lambdabo,\phio),\RHoB}
  \bigoplus_{\J\in\Gs/\calu_\lambdaO}
  \gg_\rho(\J)\, \tBeta_{\J(\lambdabo,\phio)}
  \,.  \eear \Labl21
Here in the second line we have used the simple current relation \Erf sq,
as well as the fact that by this identity the matrix element
$\tS_{(\lambdab,\varphi),\RHoB}$ vanishes when $\gg_\rho(\J)\nE1$ for any
$\J\iN\calu_\lambda$. 
The latter observation shows
that only those \bashi s $\tBeta_{(\lambdab,\varphi)}$ contribute
to the boundary state $\calb_\RHoB$ for which the character $\gg_\rho$ is 
equal to one on the whole untwisted stabilizer $\calu_\lambda\,{\subseteq}\,\Gs$, 
which in turn implies that $\gg_\rho\iN G$ is actually an element of
$S_\lambda\,{\subseteq}\,G$, i.e.\
  \be  \gg_\rho\in S_\lambda \,.  \Labl rS
Thus as a character of $\Gs$ the function 
$\gg_\rho$ factorizes to a character of $\Gs/\calu_\lambda$; 
according to the sequence \Erf01, the latter group can be identified with
$S_\lambda^*$. As we will see later, this result is perfectly natural.

We now concentrate on the $\J$-summation for fixed values of $\lambdabo$
and $\phio$. To proceed, we need a few further tools. First, it turns out 
to be useful to introduce for every 
$\gg\iN S_\lambda$ and every $J\iN\Gs$ the endomorphism
  \be  O_{\gg,\J}
  := d_\lambdaO^{-1/2}\!\! \sum_{\JL\in\cals_\lambdaO/\calu_\lambdaO}
  \gg(\JL)\, \clo_{\JJ\JJL\phio}  \Labl Og
of $\V_\psu$, where $\clo_\psi$ are the endomorphisms defined by \era{clo}.
Inserting that formula for $\clo_\psi$ and interchanging the order of 
summations, the maps $O_{\gg,\J}$ can also be written in the form
  \be  \bearll  O_{\gg,\J} \!\!
  &= d_\lambdaO^{-2} \dsty\sum_{\JL\in\cals_\lambdaO/\calu_\lambdaO}
  \JJ\phio(\JL)^*\,R_\phu(\JL)\! \sum_{\JL'\in\cals_\lambdaO/\calu_\lambdaO}\!
  \gg(\JL')\,F_\lambdaO(\JL,\JL')^*
  \\{}\\[-.8em]
  &= \dsty\sum_{\JL\in\cals_\lambdaO/\calu_\lambdaO}\!
  \JJ\phio(\JL)^*\,R_\phu(\JL) \cdot \delta_{\JL,\Kh}
   = \JJ\phio(\Kh)^*\,R_\phu(\Kh) 
  \,, \eear  \Labl23
where $\Kh\iN\cals_\lambdaO$ is as defined by formula \Erf Kh, with
$\hat g\iN\cals_\lambdaO^*$ given by $\hat g\eq\gg^{}_{|\cals_\lambdaO}$.
(Because of $\gg\iN S_\lambda$, the character $\hat g\iN\cals_\lambdaO^*$
is the identity on $\calu_\lambdaO\,{\subseteq}\,\cals_\lambdaO$, as required in
\Erf Kh. Also recall that $\Kh$ is defined
only up to elements of $\calu_\lambda$; but the result \Erf23 for $O_{\gg,\J}$ 
is independent of the choice of representative.) 

With the help of formula \Erf23 one checks that
  \be  \bearll  O_{\gg,\J\JL} \!\!\!
  &= \gg(\JL)^*\, O_{\gg,\J} \quad\ {\rm for\ all}\;\ \JL\iN\cals_\lambda
  \,, \eear  \Labl(i
and that
  \be  \bearll  O_{\gg,\J}\,O_{\gg',\J} \!\!
  &= \F_\lambdaO(\Kh,\Khp)\,O_{\gg\gg',\J}
   = \FO_\lambdaO(\gg,\gg')\,O_{\gg\gg',\J}
  \eear  \Labl ii
for all $\gg,\gg'\iN S_\lambda$, i.e.\ the endomorphisms $O_{\gg,\J}$ 
with $\gg\iN S_\lambda$ furnish a projective \rep\ of the stabilizer 
$S_\lambda$ (in the last equality we have used the identity \Erf EF). 
Let us also see to which extent these results depend on the choice of 
representative $\phio$. Any other representative is of the form $\phio\psi$ 
with $\psi\iN\cals_\lambda^*$ and $\psi_{|\calu_\lambda}\eq\id$; thus
upon choosing a different representative
the endomorphisms $O_{\gg,\J}$ get replaced by
  \be  \tilde O_{\gg,\J} = \psi(\Kh)^*\, O_{\gg,\J} \,.  \ee
On the other hand, the two-cocycle that characterizes the relevant \rep\ of
$S_\lambda$ does not depend on this choice. Indeed, as a consequence of
$\hat\JK_{\gg\gg'}\eq\Kh\Khp$ modulo $\calu_\lambda$
the maps $\tilde O_{\gg,\J}$ satisfy 
  \be  \tilde O_{\gg,\J}\, \tilde O_{\gg',\J}
  = \psi(\Kh\Khp)^*\, O_{\gg,\J}O_{\gg',\J}
  = \F_\lambdaO(\Kh,\Khp)\, \tilde O_{\gg\gg',\J} \,.  \ee

Next we also choose a set $\{\Jo\}$ of representatives for the classes in
$\Gs/\cals_\lambdaO$ and write, for every $g\iN S_\lambdaO$,
  \be  \tg \equiv \tg^{(\lambdaO)} := \bigoplus_{\Jo\in\Gs/\cals_\lambdaO}
  \gg(\Jo)\, \llb O_{\gg,\Jo} \oT\id\, \lrb \circ P_{\Jo\lambdabo}  \,,  \Labl tG
where $O_{\gg,\Jo}$ is defined in \Erf Og and $P_\mub$ is the projector 
from the $\cala$-module $\calh_\lambda$ to its isotypical component of
type $\calhb_\mub$. Owing to the identity \Erf(i, $\tg$ is in fact 
independent of the choice of representatives $\Jo$ (whereas $O_{\gg,\Jo}$ 
does again depend on that choice), and from \Erf ii it follows that
  \be  \tg\,\tgp
  = \F_\lambdaO(\Kh,\Khp)\, \Tau_{\gg\gg'}  \Labl TG
for all $\gg,\gg'\iN S_\lambda$.
Let us also note that when specializing to the $\cala$-vacuum sector, 
where $U_\vac\eq S_\vac\eq G$, we simply have $\Kh\eq\bfe$ and hence 
$O_{\gg,\J}\eq\id$ for all $\J\iN\Gs$, so that 
$\tg^{(\vac)}\eq\bigoplus_{\J\in\Gs}\gg(\J)P_\J$.

With these results at hand, we can now address the $\J$-summation that
appears in formula \Erf21. 
Inserting the definition \era{tBeta} of the \bashi s $\tBeta$, we find that
  \be \bearll  \dsty\bigoplus_{\J\in\Gs/\calu_\lambdaO}
  \gg(\J)\, \tBeta_{\J(\lambdabo,\phio)} \!\!
  &= \dsty\bigoplus_{\Jo\in\Gs/\cals_\lambdaO} \gg(\Jo)
  \bigoplus_{\JL\in\cals_\lambdaO/\calu_\lambdaO} \gg(\JL)\,
  \tBeta_{(\Jo\lambdabo,\JJo\JJL\phio)}
  \\{}\\[-.8em]
  &= \Norm\lambda\phu\,d_\lambdaO^{-2} 
  \dsty\bigoplus_{\Jo\in\Gs/\cals_\lambdaO} \gg(\Jo)
  \bigoplus_{\JL\in\cals_\lambdaO/\calu_\lambdaO} \gg(\JL)\,
  \\{}\\[-.9em] &\qquad\qquad
   \n\,{\circ}\, \Llb\! \dsty\bigoplus_{\JL'\in\cals_\lambdaO/\calu_\lambdaO}
  \JJo\JJL\phio(\JL')^*\,R_\phu(\JL')\oT\id\, \Lrb \oT \bbbJo\lambda
  \,,  \eear \ee
where $\n{:}\;\Vpsu{\otimes}\Vpsup\,{\to}\,\complex$ is the non-degenerate 
linear form defined in \erF n. Performing the $\JL$-summation, this becomes
  \be  \bearll  \dsty\bigoplus_{\J\in\Gs/\calu_\lambdaO} \!
  \gg(\J)\, \tBeta_{\J(\lambdabo,\phio)} \!\!\!
  &= \Norm\lambda\phu \dsty\bigoplus_{\Jo\in\Gs/\cals_\lambdaO} \gg(\Jo)\,
  \n\,{\circ}\, \llb
  \JJo\phio(\Kh)^*\,R_\phu(\Kh)\oT\id\, \lrb \oT \bbbJo\lambda
   = \Norm\lambda\phu\, \Bg
  \eear \Labl bG
with
  \be  \Bg := \Beta_\lambda^{} \circ (\tg \oT \id\,) \,.  \Labl Bg
Here $\Beta_\lambda\eq\Be$ are the boundary blocks of the 
$\cala$-theory, which are given by the expression \Erf BJ, and we have 
used that via the identity \erA2d they can be written in the form
  \be  \bearll  \Beta_\lambda \equiv \Beta_{[\lambdaO,\phu]} \!\!
  &= \Norm\lambdaO\phu^{-1} \dsty\bigoplus_{\ssty\varphi\in\cals_\lambdaO^*
  \atop\varphi\gt\phu} \bigoplus_{\J\in\Gs/\cals_\lambdaO}
  \tBeta_{(\J\lambdabo,\varphi)}
   = \Norm\lambdaO\phu^{-1} \dsty\bigoplus_{\J\in\Gs/\cals_\lambdaO}
  \bigoplus_{\JL\in\cals_\lambdaO/\calu_\lambdaO} 
  \tBeta_{(\J\lambdabo,\JJL\phio)}
  \\{}\\[-.8em]
  &= d_\lambdaO^{-1/2} \dsty\bigoplus_{\J\in\Gs/\cals_\lambdaO} 
  \n\,{\circ}\, \llb\!\! \sum_{\JL\in\cals_\lambdaO/\calu_\lambdaO}\!
  \clo_{\JJL\phio} \oT\id\,\lrb \oT \bbbJ\lambda 
   = \dsty\bigoplus_{\J\in\Gs/\cals_\lambdaO} \n \oT \bbbJ\lambda
  \,.  \eear \Labl::
At this point it is worth realizing that the twisted intertwining property
\Erf tq of $\tg$ is formulated independently of the sub\alg\ $\calap$, and
hence $\tg$ only depends on the \auto\ $g$ itself, but not on the
particular orbifold group $G$ containing $g$ we are considering.
By the result \Erf Bg, this independence on the choice of $G$ then 
holds for the quantities $\Bg$, too.

Inserting the result \Erf bG into formula \Erf21, we finally see that
  \be  \calb_\rho 
  = \bigoplus_{\ssty\lambdaBO \atop \gg^{}_\lambdaO\equiv1} \Norm\lambda\phu
  \bigoplus_{\phu\in\calu_\lambdaO^*} \tS_{(\lambdabo,\phio),\rho}\,
  \Bgrho \,.  \labl{Bgrho}
Thus, in summary, the boundary state $\calb_\rho$ can be entirely constructed
from the information contained in the boundary blocks $\Beta_{[\lambdaO,\phu]}$ 
together with the action of $S_\lambda$ they carry and in the character 
$g_\rho$ of $G$. As we will see, this implies that $g_\rho$ indeed constitutes
the {\em automorphism type\/} of the \bc\ $\rho$. 
We have also seen that only those \bashi s $\tBeta_{(\lambdab,\varphi)}$ 
contribute to the boundary state $\calb_\rho$ for which the stabilizer 
$S_\lambda$ contains $\gg_\rho$ (which, incidentally, shows that the
factorization to a character of $\Gs/\calu_\lambda$ is a rather natural 
property of the elements of $S_\lambda$). This should be regarded as a
{\em selection rule\/} on the possible boundary
blocks that show up in the boundary state; the concrete form of this
selection rule is completely determined by the \atype\ of the \bc.

Our derivation also demonstrates that in the case of our interest
one can associate an automorphism of the \chira\ $\cala$ to {\em every\/} 
boundary condition. This comes as a {\em result\/} of our analysis and does 
not have to be put in as an assumption.
In contrast, when the sub\alg\ $\calap$ that is preserved by a \bc\ is
not an orbifold sub\alg\ $\cala^G$, then the \bc\ need not necessarily
possess an automorphism type. Indeed, as
we will see in section \ref{t.84}, most conformally invariant boundary
conditions of the $\zet_2$-orbifold of a free boson, compactified at
a rational radius squared, do not posses an automorphism type.

\sect{Twisted blocks and twisted Ward identities}\label{t.2}

In the discussion of \auto\ types above we have introduced, 
for every primary label $\lambda\,{\equiv}\,\Lambdab$ of the $\cala$-theory 
and for every element $\gg\iN S_\lambda$ of the orbifold group that
stabilizes $\lambda$, the map $\tg$ \Erf tG
as well as the linear form $\Bg\eq\Beta_\lambda\,{\circ}\,(\Tau_g\ot\id\,)$
\Erf Bg on the tensor product $\cala$-module $\calhl\otimeS\calhlp$. We will 
refer to the latter linear forms as $\gg$-{\em twisted boundary blocks\/}.
As demonstrated in the previous section, every boundary state $\calb_\rho$
is a linear combination of those twisted boundary blocks 
$\Bgrho$ for which the twist $\gg_\rho$ lies in $S_\lambda$.

What still remains to be established is that the notation for the maps $\tg$ 
that was introduced in \Erf tG is in agreement with the use of the same
notation in section 7, i.e.\ that these maps satisfy the 
$\gg$-twisted intertwining property \Erf tq. To address this issue, we
need some information about the \rep\ $R_\lambda$ of $\cala$ on the 
subspaces in the decomposition \erf{Deco} of a $\cala$-module
$\calhl\,{\equiv}\,\calh_\LambdaBo$ into irreducible $\calap$-modules. 
While a complete description of the action of the vertex operator \alg\ 
that is obtained by a simple current extension is not yet available, 
the known results naturally suggest the following structure.
On the $\calap$-module $\V_\psu{\otimes}\calhb_{\Jo\lambdabo}\,{\subseteq}\,
\calhl$ an element $Y_\J(z)\,{\equiv}\,Y(\rp v_\J;z)$ of $\cala$ with 
$\rp v_\J\iN\calhb_\J\,{\subseteq}\,\calh_\vac$ is represented by
  \be  R_\LambdaBo(Y_\J) = R_\phu(\J') \oT \bigoplus_{\Jo\in\Gs/\cals_\lambda}
  \BR_{\Jo\lambdabo}(\rp Y_\J) \,.  \ee
Here $\J'$ is defined by $\J\eq\J'\Jop$ with $\Jop$ the 
representative for a class in $\Gs/\cals_\lambdaO$ (see before
formula \Erf tG), and $\BR_{\Jo\lambdabo}(\rp Y_\J)$ is a map from 
$\calhb_{\Jo\lambdabo}$ to $\calhb_{\Jop\Jo\lambdabo}$. To proceed we use
the commutation properties
  \be  R_\phu(\J')\,R_\phu(\Kh) \!\!
  = F_\lambda(\J',\Kh)\, R_\phu(\Kh)\,R_\phu(\J')
  = \gg(\J')^*\, R_\phu(\Kh)\,R_\phu(\J')  \Labl'1
and 
  \be  \bigoplus_{\Jo\in\Gs/\cals_\lambda}\!\! \llb \id\oT
  \BR_{\Jo\lambdabo}(\rp Y_\J) \lrb \circ P_{\Jopp\lambdabo}
  = \BR_{\Jopp\lambdabo}\!(\rp Y_\J)
  = P_{\Jop\Jopp\lambdabo} \circ \bigoplus_{\Jo\in\Gs/\cals_\lambda}\!\!
  \llb \id\oT \BR_{\Jo\lambdabo}\!(\rp Y_\J) \lrb  \Labl'2
as well as\,%
 \futnote{This relation is needed for compatibility with the fact that 
while the isomorphism class of the projective $\cals_\lambda$-\rep\
$R_\phu$ is the same for all values of $\lambdab$ within the class
$[\lambdabo]$, its explicit realization can depend on $\lambdab$, and
this dependence should precisely be characterized by $F_\lambda$.}
  \be  \llb R_\phu(\J')\oT\id \,\lrb \circ P_{\Jop\Jopp\lambdabo}
  = F_\lambda(\Jop,\Kh)\, P_{\Jop\Jopp\lambdabo} \circ
  \llb R_\phu(\J')\oT\id \,\lrb \,.  \Labl'3
Then we obtain
  \be  \bearll  R_\lambda(Y_\J) \circ \tg^{(\lambdaO)} \!\!\!
  &= \gg(\J')^*\,F_\lambda(\Jop,\Kh)\!
  \dsty\bigoplus_{\Jopp\in\Gs/\cals_\lambda}\!\!
  \gg(\Jopp)\,{}^{}_\Jopp\phio(\Kh)^*\, \llb R_\phu(\Kh)\oT\id \,\lrb
  \circ P_{\Jop\Jopp\lambdabo} \circ R_\lambda(Y_\J)
  \\{}\\[-.8em]
  &= \gg(\J')^*\,\gg(\Jop)^*\, \dsty\bigoplus_{\Jopp\in\Gs/\cals_\lambda}\!
  \gg(\Jopp)\,{}^{}_\Jopp\phio(\Kh)^*\, \llb R_\phu(\Kh)\oT\id \,\lrb
  \circ P_{\Jopp\lambdabo} \circ R_\lambda(Y_\J)
  \\{}\\[-.8em]
  &= \gg(\J)^* \cdot \tg^{(\lambdaO)} \circ R_\lambdao(Y_\J)
   \,\equiv \tg^{(\lambdaO)} \circ R_\lambda(\gg^{-1}(Y_\J))
  \,.  \eear \Labl tQ
Here in the last expression the element $\gg$ of the orbifold group is
to be regarded as an automorphism of the chiral algebra $\cala$, while
in the intermediate steps it is interpreted as
a character of the simple current group $\Gs\eq G^*$.
Formula \Erf tQ reproduces the twisted intertwiner property \Erf tq 
and hence is the desired result.

We also remark that according to section 7 the twisted intertwiners carry
a projective \rep\ that is characterized by the cohomology class of the
cocycle $\FO_\lambda$ or, equivalently, by the commutator cocycle 
$\FFO_\lambda$. On the other hand, according to relation \Erf TG the
concrete realization \Erf tG of the twisted intertwiners can be
characterized by the commutator cocycle $F_\lambda$. This is compatible
because of the identity \Erf EF.

Now for every field $Y(z)\eq\sum_{n\in\zet}Y_n z^{-n-\Delta_Y}$ of
conformal weight $\Delta_Y$ in the chiral algebra $\cala$, the ordinary
boundary block $\Be$  of the $\cala$-theory obeys the Ward identity 
appropriate for a two-point block on ${\dl P}^1$. That is,
  \be  \Be \circ \llb R_\lambda(Y_n) \oT\bfe + \zetay\, \bfe \oT
  R_\lambda(Y_{-n}) \lrb = 0  \Labl16
with $\zetay\eq(-1)^{\Delta_Y-1}$. 
When combined with the definition \Erf Bg of the twisted boundary blocks,
the twisted intertwiner property \Erf tQ therefore allows us to 
write (suppressing from now on the \rep\ symbol $R_\lambda$)
  \be  \bearll
  \Bg \circ (Y_n\oT\bfe) \!\!
  &= \Beta_\lambda \circ (\tg\oT\id) \cirC (Y_n \oT\bfe) 
   = \Beta_\lambda \circ (\gg(Y_n)\,\oT\bfe) \circ (\tg\oT\id) 
  \\{}\\[-.8em]
  &= - \zetay\, \Beta_\lambdab \circ (\bfe\oT \gg(Y_{-n})) \circ (\tg\oT\id)
   = - \zetay\, \Bg \circ (\bfe\oT \gg(Y_{-n}))
  \,.  \eear \ee
Thus the twisted boundary states satisfy a {\em twisted Ward identity\/}
  \be  \Bg \circ
  \llb Y_n\oT\bfe + \zetay\, \bfe\oT \gg(Y_{-n}) \lrb = 0 \, . \labl{tww}

It is worth noting that we have defined the twisted Ward identities 
\erf{tww} through an automorphism of $\cala$ that only acts on the 
second factor of the tensor product. One could easily generalize this 
by considering the action of two different automorphisms 
$g_L^{}$ and $g_R^{}$ on the two factors, according to
  \be  \Bg \circ
  \llb g_L^{}(Y_n)\oT\bfe + \zetay\, \bfe\oT g_R^{}(Y_{-n}) \lrb = 0 \, . \ee
But obviously what matters is only the combination $g_L^{-1} g_R^{}$.
Thus we can describe the space of automorphism types either as the coset 
space $(G\,{\times}\,G)/G$ or as the group $G$. In fact, the map 
$(g_L^{},g_R^{})\,{\mapsto}\,g_L^{-1}g_R^{}$ provides a natural bijection 
between the two sets which, in the case when $G$ is a Lie group, is even an 
isomorphism of smooth manifolds. (For theories of free bosons, the description 
in terms of $(G\,{\times}\,G)/G$ has been established in \cite{grgu}.)

  \sect{The classifying algebra $\calc^{(\gg)}$ for automorphism type $g$}
\label{t.3}

  \subsection{Individual classifying algebras for fixed automorphism type}

We now restrict our attention to the collection of \bc s that possess some
fixed \atype\ $\gg$. According to the results of section 8 the corresponding
boundary states can all be written as linear combinations of the twisted 
boundary blocks \Erf Bg with fixed $\gg$. This suggests to study analogous
elements of the \cla; accordingly we introduce for
every $\lambda$ with $\gg\iN S_\lambda$ the linear combination
  \be  \Phig_\lambda := \Frac{|\calu_\lambda|}{|\Gs|} \sum_{\J\in\Gs/
  \calu_\lambda} \gg(\J)^*\, \tPhi_{\J(\lambdabo,\psio)}  \labl{Phig}
of basis elements of the \cla\ \clAb, where by $\lambdabo$ and $\psio$ 
(which satisfies $\psio\,{\gt}\,\psuo$ for $\lambda\,{\equiv}\,[\lambdabo,
\psuo]$\,) are the representatives introduced in the paragraph before formula
\Erf21. To proceed, we also note the relation
  \be  \sum_{\J\in\Gs/\calu_\lambda} \gg(\J)^*\, \tS_{\J(\lambdab,\psi),\rho}
  = (|\Gs|/|\calu_\lambda|)\, \delta_{\gg,\gg_\rho}\,
  \tS_{(\lambdab,\psi),\rho}  \Labl9o
that follows for every $\gg\iN S_\lambda$ with the help of the simple 
current symmetry \Erf sq. (When $\gg$ is not in $S_\lambda$, then according
to the remarks after \Erf21 this expression vanishes.)

Using the Verlinde-like formula that expresses the structure constants of
\clAb\ in terms of the diagonalizing matrix $\tS$, the result \Erf9o
allows us to compute the product of two elements of \clAb\ of the form
\erf{Phig} as
  \be  \bearll  \Phig_\lambda \star \Phigp_{\lambda'} \!\!
  &= \delta_{\gg,\gg'} \dsty\sum_{\lambda''}
  \Ng\lambda{\lambda'}{\lambda''} \Phig_{\lambda''}
  \eear \ee
with\,%
 \futnote{Note that this expression is only defined when $g\iN S_\lambda
\cap S_{\lambda'}\cap S_{\lambda''}$.}
  \be  \Ng\lambda{\lambda'}{\lambda''} \equiv
  \Ng\lambda{\lambda'}{\lambda''}\!(\calap) := 
  \sum_{\ssty \rho \atop \ssty \gg_\rho=\gg}
  \tS_{(\lambdabo,\psio),\rho}^{} \tS_{(\lambdabo',\psiop),\rho}^{}
  \tS_{(\lambdabo'',\psiopp),\rho}^* (\tS_{\vacb,\rho})^{-1}_{}
  \,.  \Labl Ng
This means that the elements $\Phig_\lambda$ for all $\lambda$ with 
$\gg\iN S_\lambda$ span not only a subalgebra, but even an {\em ideal\/} 
of the \cla\ \clAb. We call this ideal of \clAb\ the {\em individual 
classifying algebra\/} for \atype\ $\gg$ and denote it by \clAg. Also, by 
construction the $\Phig_\lambda$ are linearly independent, and hence they 
furnish a basis of \clAg, i.e.\ for every fixed $\g\iN G$ we have
  \be  \clAg = {\rm span}_\complex\{\Phig_\lambda \,|\, S_\lambda\,{\ni}
  \,\gg\} \,.  \ee
Clearly, \clAg\ is again semisimple; its \onedim\ \irrep s are 
obtained by restriction to \clAg\ of those \onedim\ \irrep s $R_\rho$ of 
\clAb\ which satisfy $\gg_\rho\eq\gg$. 

Moreover, the following counting argument shows
that together the elements $\Phig_\lambda$ for all $\lambda$ and all
$\gg\iN G$ span all of \clAb. Namely, associated to every $\Gs$-orbit
$[\lambdab]$ of $\calap$-primaries there are $|\calu_\lambda|$ many
$\cala$-primaries $\Lambdab$, and each of them gives rise to $|S_\lambda|$
many basis elements $\Phig_\lambda$. On the other hand, each such $\Gs$-orbit
contains $|\Gs|/|\cals_\lambda|$ many $\calap$-primaries, each of them 
leading to $|\cals_\lambda|$ many basis elements $\tPhi_{(\lambdab,\phi)}$
of \clAb. With the help of the identities \Erf09 among the sizes of the
various subgroups it thus follows that
  \be  \sum_\lambda |S_\lambda| = \!\sumbo\lambda\!\! |\cals_\lambda|
  \,.  \ee
As a consequence we have indeed -- as \alg s over $\complex$, and with
the distinguished bases related by \erf{Phig} -- an isomorphism
  \be  \clAb \cong \bigoplus_{\gg\in G} \clAg \,.  \Labl op
This decomposition may be regarded as expressing the fact that in the
situation of our interest every boundary condition has an automorphism type.
Put differently, the \alg\ \clAb\ provides a unified description of the
\bc s for the $|G|$ different automorphism types that correspond to the
elements of $G$.

In the special case of trivial \atype, $\gg\eq\bfe$, we can use the result
\I\ that $\tS_{(\lambdab,\psi),\rho}\eq S_{\lambda,\rho}$ for all $\rho$
with $\gg_\rho\eq\bfe$ to see that the ideal \clAe\ is nothing but the
fusion rule \alg\ of the $\cala$-theory, so that we recover the known 
results \cite{card9,prss3} for
\bc s that do not break any of the bulk symmetries. It should also be 
noticed that the precise form of the structure constants of the ideals
\clAg\ does depend on the choice of representatives $\lambdabo$ and $\psio$
(except when $\gg\eq\bfe$, where independence of this choice follows as a
consequence of the simple current relation \Erf sq).
This is, however, perfectly fine, because the twisted boundary block 
depends on the choice of representatives as well, and in fact in a manner so 
as to cancel the overall dependence in all physically meaningful quantities
like one-point correlators for bulk fields on the disk.

\subsection{Independence of $\calc^{(\gg)}$ on the orbifold group}

A remarkable feature of the formula \Erf Ng for the structure constants of 
\clAg\ is that the \atype\ $\gg$ just enters by restricting the range of 
summation. This suggests that the individual \cla\ for \atype\ $g$ in fact
does {\em not\/} depend on the specific orbifold subalgebra $\calap$, i.e.\
on the group $G$ that contains the \auto\ $g$, but rather on $g$ alone. 
Below we will show that this is indeed the case, and actually the
statement already applies to the relevant entries of the diagonalizing
matrix $\tS$. Now
recall from section \ref{t.1} (see the remarks after formula \Erf::) that
a similar statement applies to the twisted boundary blocks $\Bg$.
When combined with the present result, then according to formula
\erf{Bgrho} this independence property of
the {\em blocks\/} $\Bg$, which are chiral quantities, extends to
the (non-chiral) boundary {\em states\/} $\calb_\rho$.

To prove the independence of $\calc^{(\gg)}(\calap)$ on the choice of
orbifold group $G$ containing $g$, it is 
convenient to study the orbifold \wrtt cyclic group 
  \be  G' := \zg \cong \zet_\ordg  \ee
that is generated by $\gg$. We first note that the 
diagonalizing matrix $\tS'$ for the total \cla\ $\calc(\calapp)$ of all
\bc s preserving $\calapp\eq\cala^\zg\,{\subset}\, \cala$ is given by 
the expression \erA tS which contains contributions involving the various
matrices $S^\J$ (though in this special cyclic case the formula simplifies). 
However, in the expression for 
the structure constants of \clAgg\ only those entries $\tS'_{(\lambdabg,\vphi),
[\rhobg,\psu]'}$ appear for which $g\iN S'_\lambdag$, and these entries turn
out to be
particularly simple. Indeed, since $g$ generates $\zg$, the latter property
means that the stabilizer is maximal, $S'_\lambdag\eq G'$; by the duality
relation $\calu'_\lambdag\,{\cong}\,\zgs\!/S'_\lambdag{}^{\!*}$ this implies
$\cals'_\lambdag\eq\calu'_\lambdag\eq1$. Similarly, because of $g_\rhog\eq g$
the monodromy charges of $\rhog$ \wrtt extension from $\calapp$ to $\cala$
have denominator $\ordg$, 
so that $\rhog$ cannot be a fixed point under any simple current in
$G^{\prime *}$, which implies that also $\cals'_\rhog\eq\calu'_\rhog\eq1$. 
Thus we have to deal with full $\zgs$-orbits only; in particular the simple
current summation in the formula for $\tS'$ reduces to the term with $\J\eq\bfe$:
  \be  \tS'_{\lambdabg,[\rhobg,\psu]'}
  = \Frac{\ordg}{\sqrt{|\cals'_\lambdag|\,|\calu'_\lambdag|\,|\cals'_\rhog|\,
  |\calu'_\rhog|}}\, S'_{\lambdabg,\rhobg} = \ordg\, S'_{\lambdabg,\rhobg}
  \,.  \Labl'S

To proceed, we note that the \alg\ $\calapp\eq\cala^\zg$ can be obtained 
 from $\calap\eq\cala^\G$ as a simple current extension by the subgroup
  \be  \Gss := \{ \J\iN\Gs \,|\, g(\J)\eq1 \}  \labl{Gss}
of the simple current group $\Gs\eq G^*$. (The subgroup $\Gss$ has index $\ordg$
in $\Gs$; in fact, the factor group $\Gs/\Gss$ is cyclic of order $\ordg$.)
In particular, the modular S-matrix $S'$ of the $\calapp$-theory can be expressed
through quantities of the $\calap$-theory as
  \be  S'_{\lambdabg,\mubg} = \Frac{|\Gss|}{\sqrt{|\cals''_\lambdag|\,
  |\calu''_\lambdag|\,|\cals''_\mug|\,|\calu''_\mug|}} 
  \sum_{\J\in\calu''_\lambdag\cap\calu''_\mug} \psu^{}_\lambdag(\J)\,
  S^{\J}_{\lambdab,\mub}\, \psu^*_\mug(\J)  \Labl=2
with $\lambdabg\,{\equiv}\,[\lambdab,\psu_\lambdag]''$ and $\mubg\,{\equiv}\,
[\mub,\psu_\mug]''$. 

The $\calapp$-theory has a simple current of the form
  \be  \Jgp = \Jg\cdoT\Gss \equiv [\Jg]'' \,,  \labl{Jgp}
where $\Jg\iN\Gs$ is a simple current of the $\calap$-theory that 
is characterized by the property that $\ordg$ is the smallest positive 
integer $m$ such that $(\Jg)^m$ lies in $\Gss$, so that $\Jgp$ 
has order $\ordg$. By this property of $\Jg$ and the definition \erf{Gss} of
$\Gss$ it follows that
  \be  g_\rho(\Jg^m\JK'') = g(\Jg^m) \ne 1  \ee
for every $\JK''\iN\Gss$ and every $m\eq1,2,...\,,\ordg{-}1$.
This means that none of the monodromy charges $Q_{\Jg^m\JK''}(\rho)$ vanishes,
so that $\rho$ cannot be a fixed point \wrt any of these simple currents
$\Jg^m\JK''$. It follows that the stabilizer $\cals_\rho$ is contained in 
$\Gss$, which in turn implies that $\cals''_\rhog\eq\cals_\rho$ and hence also
$\calu''_\rhog\eq\calu_\rho$.
As a consequence, the boundary labels $[\rho,\psu_\rho]$ of the matrix $\tS$
that satisfy $g_\rho\eq g$ are precisely the same as those appearing in
formula \Erf=2 for $S'$, and the simple current summation in the expression
for the corresponding entries of $\tS$ only runs over elements of $\Gss$.
Using also the fact that $\cals_\lambdag''\eq\cals_\lambda\,{\cap}\,\Gss$,
we can therefore write
  \be  \bearll  \tS_{(\lambdab,\psi_\lambda),\RhoB} \!\!
  &= \Frac{|G|}{\sqrt{|\cals_\lambda|\,|\calu_\lambda|\,|\cals_\rho|\,|\calu_\rho|}}
  \dsty \sum_{\J\in\cals_\lambda\cap\calu_\rho}
  \psi_\lambda(\J)\, S^\J_{\lambdab,\rhob}\, \psu_\rho^*(\J)
  \\{}\\[-.8em]
  &= \Frac{\ordg\,|\Gss|}{\sqrt{|\cals_\lambda|\,|\calu_\lambda|\,|\cals_\rhog''|
  \,|\calu_\rhog''|}} \dsty \sum_{\J\in\cals_\lambdag''\cap\calu_\rhog''}
  \psi^{}_\lambdag(\J)\, S^\J_{\lambdab,\rhob}\, \psu_\rhog^*(\J)
  \,.   \eear  \Labl=3

Furthermore, because of $g_\rho\eq g$ the simple current relation \Erf sq
for $\tS$ implies that these entries of $\tS$ are identical for all $\lambdab$
on one and the same $\Gss$-orbit. We may therefore equate the expression \Erf=3
with its average over $\cals_\lambda''$. The simple current relation for the
matrices $S^\J$ then amounts 
 to a restriction of the summation to $\calu_\lambda''$, so that
  \be   \tS_{(\lambdab,\psi_\lambda),\RhoB}
  = \Frac{\ordg\,|\Gss|}{\sqrt{|\cals_\lambda|\,|\calu_\lambda|\,|\cals_\rhog''|
  \,|\calu_\rhog''|}} \sum_{\J\in\calu_\lambdag''\cap\calu_\rhog''} \!\!\!
  \psi^{}_\lambdag(\J)\, S^\J_{\lambdab,\rhob}\, \psu_\rhog^*(\J)
  = \ordg \sqrt{\Frac{|\cals_\lambdag''|\,|\calu_\lambdag''|}{|\cals_\lambda|\,
  |\calu_\lambda|}}\, S'_{[\lambdab,\psu_\lambdag]''\!,[\rhob,\psu_\rhog]''} 
  \,.  \Labl'7            

To analyze the prefactor appearing here, we first remark that the index 
of the subgroup $\cals''_\lambdag$ in $\cals_\lambda$ is some divisor 
$\nl$ of $\ordg$. There is a simple current $\Jl\iN\cals_\lambda$
which plays an analogous role for the embedding $\cals''_\lambdag
{\subseteq}\,\cals^{}_\lambda$ as $\Jg$ plays for
the embedding of $\Gss$ in $\Gs$, i.e.\ $\nl$ is the smallest power
such that $\Jl^\nl$ is in $\cals''_\lambdag$, and the elements of 
$\cals_\lambda$ are of the form $\Jl^m\JK''$ with $\JK''\iN\cals''_\lambdag$ 
and $m\eq1,2,...\,,\nl{-}1$. Moreover, from the fact that $g\iN S_\lambda$
it follows with the help of duality that $\Jl^m\,{\not\in}\,\calu_\lambda$
for all $m\eq1,2,...\,,\nl{-}1$, and hence we have $\calu^{}_\lambda\,
{\subseteq}\,\cals''_\lambda$. 
 Thus when forming the untwisted stabilizer associated to $\cals''_\lambdag$
one does not lose any elements of $\calu_\lambda$, so that
$\calu^{}_\lambda{\subseteq}\,\calu''_\lambdag$.
Observing that $\calu_\lambda$ is precisely the kernel of the group
homomorphism from $\calu''_\lambdag$ to $\complex$ that maps $\JK''\iN
\calu''_\lambdag$ to the $\nl$th root of unity $F_\lambda(\Jl,\JK'')$,
it follows that the index of $\calu^{}_\lambda$ in $\calu''_\lambdag$ is $\nl$.
Thus we have
  \be  |\cals^{}_\lambda|\,/\,|\cals_\lambdag''| = \nl 
  = |\calu_\lambdag''|\,/\,|\calu^{}_\lambda| \,,  \ee
so that \Erf'7 reduces to
  \be   \tS_{(\lambdab,\psi_\lambda),\RhoB}
  = \ordg\, S'_{[\lambdab,\psu_\lambdag]'',[\rhob,\psu_\rhog]''} \,.  \ee

Comparison with \Erf'S then shows that the relevant matrix elements of
$\tS$ are identical to those of $\tS'$.
We conclude that
  \be  \calc^{(\gg)}(\cala^G) = \calc^{(\gg)}(\cala^\zg) =: \calc^{(\gg)}
  \Labl=4
for every finite abelian orbifold group $G$ with $G\,{\ni}\,\gg$.
It is worth pointing out that the fact that the individual
classifying algebra does not depend on the preserved subalgebra
constitutes another quite non-trivial
check of our ansatz for the diagonalizing matrix $\tS$. We also learn that
the structure constants of the \alg\ $\calc^{(\gg)}$ read
  \be  \Ng\lambdabg\mubg\nubg = \Ng\lambdabg\mubg\nubg\!(\cala^\zg)
  = |G'|^2 \sum_{\ssty [\rhobg]' \atop \ssty \gg_\rhog=\gg}
  \Frac{S'_{\lambdabg,\rhobg}\,S'_{\mubg,\rhobg}\,S'{}^*_{\!\!\!\nubg,\rhobg}}
  {S'_{\vacb',\rhobg}} \,.  \Labl'N
By inserting a suitable projector, they can rewritten in the form
  \be  \Ng\lambdabg\mubg\nubg
  = \sum_{\J'\in\Gs'} \sum_\rhobg \gg(\J')\,\gg_\rhog(\J')^*\,
  \Frac{S'_{\lambdabg,\rhobg}\,S'_{\mubg,\rhobg}\,S'{}^*_{\!\!\!\nubg,\rhobg}}
  {S'_{\vacb',\rhobg}}
  = \sum_{\J'\in\Gs'} \gg(\J')\, {\rm N}'{}^{\ \ \J'\nubg}_{\!\!\!\lambdabg,\mubg}
  \,,  \Labl"N
i.e.\ as a linear combination of fusion coefficients of the $\cala^\zg$-theory.

It is reasonable to expect that similar considerations apply to
orbifold subalgebras \wrt non-abelian groups $G$, too, so that in
particular also in that case every boundary condition possesses
a definite automorphism type. Assuming this to be true, the classifying
algebras for fixed automorphism type studied here should coincide with their
analogues in the non-abelian case. In other words, the set of \bc s will
be exhausted by those \bc s that are already known from the cyclic groups
$\zg$ for all $g\iN G$. (On the other hand, the detailed structure of the
classifying \alg, which involves the distinguished basis $\{\tPhi\}$,
will still be more involved.)

\subsection{Relation with traces on bundles of \cblock s}\label{t33}

The independence on the specific group $G$ finds its natural explanation in
the fact that these numbers are interpretable as traces of appropriate maps 
on bundles of \cblock s \cite{fuSc6}. Namely, 
since for every $\lambda$ we are given the twisted intertwiner maps
$\tg^{(\lambda)}{:}\ \calhl\,{\to}\,\calh_{\gs\!\lambda}$ \Erf tg,
we also have the corresponding tensor product maps 
  \be  \vec\tg \equiv \vec\Tau^{\sss(\lambda_1\,\lambda_2\,...\,\lambda_m)}
  _{\gg,\gg,...,\gg}
  := \tg^{(\lambda_1)}\,{\otimes}\,\tg^{(\lambda_2)}\,{\otimes}\cdots{\otimes}
  \,\tg^{(\lambda_m)}  \ee
on tensor products of $\cala$-modules. In view of the definition \cite{Ueno,beau}
of \cblock s $V_{\lambda_1\lambda_2...\lambda_m}$
as singlets (\wrt a suitable {\em block \alg\/}) in the algebraic dual 
$(\calh_{\lambda_1}{\otimes}\calh_{\lambda_2}{\otimes}\cdots{\otimes}
\calh_{\lambda_m})_{}^*$ of these tensor products, the twisted intertwining 
property together with the fact that the \auto\ $g$ respects the grading
of the chiral \alg\ implies the existence of a linear map 
  \be  \vec\tg^* \equiv \vec\Tau_{\gg,\gg,...,\gg}^{*\,{\sss(\lambda_1\,
  \lambda_2\,...\,\lambda_m)}}:\quad V_{\lambda_1\,\lambda_2\,...\,\lambda_m}
  \to V_{\gs\!\lambda_1\,\gs\!\lambda_2\,...\,\gs\!\lambda_m}  \ee
between spaces of \cblock s. When
$\gs\lambda_i\eq\lambda_i$ for all $i\eq1,2,...\,,m$, this linear map 
$\vec\tg^*$ is an {\em endo\/}morphism so that one can compute its trace;
we will be interested in the traces of three-point blocks. (A concrete 
description of the 
block \alg s and hence of the maps $\vec\tg^*$ is so far only available
for the case of \wzwts, where the situation can be analyzed in terms of
the horizontal subalgebra of the relevant affine \lie.)
Now in the case of our interest, where all simple current stabilizers 
in the $\calapp$-theory are trivial, the fusion rules of the $\cala$-theory 
can be expressed through the modular S-matrix of the $\calapp$-theory as
  \be  \bearll  {\rm N}_{\lambda,\mu}^{\;\ \ \nu} \!\!
  &= |G'|^2
  \dsty\sum_{\ssty [\rhobg]' \atop \ssty Q_{\Gs'}(\rhog)=0}
  \Frac1{|\cals'_\rhog|}\, \Frac{S'_{\lambdabg,\rhobg}\,
  S'_{\mubg,\rhobg}\,S'{}^*_{\!\!\!\nubg,\rhobg}} {S'_{\vacb',\rhobg}} 
  \\{}\\[-.8em]
  &= \dsty\sum_{\J'\iN\Gs'} \sum_\rhobg \eE^{2\pi\ii Q_{\J'}(\rhog)}\,
  \Frac{S'_{\lambdabg,\rhobg}\,S'_{\mubg,\rhobg}\,S'{}^*_{\!\!\!\nubg,\rhobg}}
  {S'_{\vacb',\rhobg}} 
   = \sum_{\J'\iN\Gs'} {\rm N}'{}^{\ \ \J\nug}_{\!\!\!\lambdag,\mug}
  \eear \ee
This indicates that the chiral three-point blocks of our interest can be
decomposed into the direct sum of spaces of \cblock s of the $\calapp$-theory.
Such a decomposition should in fact be expected on general grounds, and the
\cblock s of the $\calapp$-theory should fit together to sub-bundles of the
bundles of \cblock s of the $\cala$-theory. 
Now when restricted to irreducible $\calapp$-modules the maps $\tg$ are
ordinary intertwiners, and as a consequence the map $\vec\tg^*$ acts on the
subspaces of \cblock s as a multiple of the identity. 
On the subspace of dimension $\Ng\lambdabg\mubg{\J\nubg}$ the map
$\vec\Tau_g^*$ should therefore act with eigenvalue $g(\J)$. It thus follows 
that upon choosing representatives of the $\Gs'$-orbits, the trace of this map
is precisely given by the number \Erf"N:
  \be  {\rm tr}^{}_{V_{\lambda\mu\nu}}\,\vec\tg^{*\,{\sss(\lambda\mu\nu)}}
  = \Ng\lambdabg\mubg\nubg\!(\cala^\zg) \,.  \ee
For $g\nE1$ these numbers do depend on the choice of representatives
$\lambdabg$ of the orbits $\lambda\,{\equiv}\,[\lambdabg]$, in agreement
with the fact that the maps $\vec\tg^*$ are defined only up to a phase.

\sect{T-duality}\label{t.4}

In all the considerations so far, we have required that the torus \parfu,
and correspondingly the pairing of the labels $\lambda$ and $\tilde\lambda$ 
of the bulk fields $\pho\lambda$, is given by charge conjugation, which we 
denote by $\ccc$: $\tilde\lambda\eq\ccc(\lambda)\,{\equiv}\,\lambdap$. In this 
section we analyze what happens when a different torus \parfu\ is chosen.\,%
 \futnote{It is by no means necessary that the permutation
that (for maximally extended chiral \alg) characterizes the torus \parfu\
is equal to the permutation $\gs$ defined via the \atype\ 
-- if such an \atype\ exists at all -- of a \bc.
While both mappings are associated to the transition from {\em chiral\/}
\cft\ (i.e., \cft\ on a complex curve) to {\em full\/} \cft\ (\cft\ on a real
\twodim\ surface), they refer to such a transition for two different world
sheets  -- the disk and the torus, \resp\ -- which are not related by any
factorization rules. As a consequence, they can be chosen independently.}
 To this end we first have to state 
what we mean by a \chira\ $\cala$ and its \auto s in more concrete terms than
was done so far. In mathematical terms, a \chira\,%
 \futnote{Not to be confused with the global object for which the term chiral
\alg\ has also been used in the recent mathematical literature \cite{gaiT}, 
which we prefer to call a {\em block \alg\/}.}
 is a {\em vertex operator \alg\/} \cite{FRlm,KAc3}; the relevant data are 
therefore the vector space $\calhv$, the vacuum vector $\vac\iN\calhv$, the 
Virasoro element $\vvir\iN\calhv$, and a `vertex operator map'
$Y$. The latter realizes the state-field correspondence, i.e.\ 
associates to every $v\iN\calhv$ a field operator $Y(v;z)$ (technically, a
linear map from $\calhv$ to ${\rm End}(\calhv)\otimes\complex[[z{,}z^{-1}]]$
with $z$ a formal variable), e.g.\ the energy-momentum tensor $T$ to the Virasoro 
element, $Y(\vvir;z)\eq T(z)$. By an {\em \auto\/} of a vertex operator 
\alg\ $\cala$ we mean an invertible linear map
  \be  \sigm:\quad  \calhv \to \calhv  \ee
that is compatible with state-field correspondence, i.e.\ satisfies
  \be  \sigm^{-1}\, Y(\sigm v;z)\, \sigm = Y(v;z)  \ee
for all $v\iN\calhv$. (Let us stress that -- unlike e.g.\ in \cite{dolm6} --
at this point we do {\em not\/} require that the map $\sigma$ leaves the
vacuum and the
Virasoro element fixed.) As already outlined in section \ref{t.0}, each such
map is accompanied by a permutation $\sigs$ of the label set $I\df\{\lambda\}$
of $\cala$-primaries and by twisted intertwiners $\Tau_\sigm\,{\equiv}\,
\Tau_\sigm^{(\lambda)}{:}\ \calhl\,{\to}\,\calh_{\sigs\!\lambda}$ between
the corresponding irreducible $\cala$-modules.

To proceed, we introduce two particular subgroups of the group \Gmap\ of 
all maps $f$ from $I$ to \auta.
For every $f\iN\Gmap$ and every $\lambda\iN I$, the image
  \be  f_\lambda := f(\lambda) :\quad \calhv \to \calhv  \ee
is an \auto\ in \auta. We denote by \Gtot\ the subgroup of all those elements
$f$ of \Gmap\ for which the map\,%
 \futnote{Note that typically several distinct $f\iN\Gmap$ will give rise to one
and the same permutation $\pi_f^*$.}
  \be  \bearll  \pi_f^*:\ & \;I \to I \\[.2em]
  & \lambda \,\mapsto f_\lambda^*(\lambda)  \eear \ee
preserves conformal weights modulo integers, i.e.\ fulfils
$\Delta_{\pi_f^*(\lambda)}\eq\Delta_\lambda\bmod\zet$ for all $\lambda\iN I$,
and is an \auto\ of the fusion rules, i.e.\ satisfies
  \be  {\rm N}_{\pi_f^*(\lambda),\pi_f^*(\mu),\pi_f^*(\nu)}
  \equiv {\rm N}_{f_\lambda^*(\lambda),f_\mu^*(\mu),f_\nu^*(\nu)}
  = {\rm N}_{\lambda,\mu,\nu}  \ee
for all $\lambda,\mu,\nu\iN I$. (Examples for such \auto s are those induced
by simple currents, see e.g.\ \cite{scya6,gasc}.)
Every element of \Gtot\ gives rise to a modular invariant torus \parfu\ 
  \be  Z_f(\tau) := \sum_{\lambda\in I} \chii_\lambda^{}(\tau)\,
  \chii_{\pi_f^*(\lambdap)}(\tau)^* \,.  \ee

The second subgroup of \Gmap\ of our interest consists of {\em constant\/}
maps $f$ whose image -- to be denoted by $\gf$ -- is an \auto\ of $\cala$ 
that leaves the Virasoro element fixed, $\gf(\vvir)\eq\vvir$ (and hence in
particular obeys $\gfs(\vac)\eq\vac$ and preserves conformal weights exactly,
not only modulo integers).
We denote the subgroup of those maps of this kind that also lie\,%
 \futnote{In fact one should expect that the property of inducing a fusion
rule \auto\ need not be required independently, but is satisfied
automatically as a consequence of the consistency of the relevant orbifold
theory. This has been demonstrated in the case of order-two \auto s in 
\cite{bifs}. Indeed, consistency of the orbifold theory requires that the 
S-matrix of the $\cala$-theory behaves \wrtt permutation $\gs$ that is 
induced by the non-trivial element $\gg$ of $\zet_2$ via the maps $\tg$ as 
$S_{\lambda,\gs\!\mu}\eq S_{\gs\!\lambda,\mu}$. When combined with the 
Verlinde formula, this implies that $\gs$ furnishes an \auto\ of the \furu s 
of the $\cala$-theory.}
 in \Gtot\ by \GT. Every \auto\ $\gf$ of this subgroup of $\Gtot$ with 
$f\iN\GT$ can be used to define conformally invariant \bc s. Furthermore, 
for every $f\iN\GT$ and every $f'\iN\Gtot$ the modular invariant \parfu s 
$Z_{f'}$ and $Z_{f f'}$ are physically indistinguishable, i.e.\ upon a 
suitable relabelling of the fields all correlation functions in the associated
\cfts\ coincide. Accordingly, it is appropriate to refer to \GT\ as the 
T-{\em duality group\/} of the theory. For every \chira\ $\cala$ the 
T-duality group \GT\ contains in particular the map $\fc$ whose image 
is the charge conjugation \auto\ $\cc$, i.e.\ $\fc(\lambda)\eq\cc$ for all
$\lambda\iN I$, and $\pi_{\!\fc}^*(\lambda)\eq\cc(\lambda)\eq\lambdap$.

The two theories with \parfu s $Z_{f'}$ and $Z_{f f'}$ being
indistinguishable, in particular the respective sets of all conformally
invariant \bc s must be the same. It is worth investigating this
correspondence in some detail. Let us denote for any $f\iN\GT$ and any
$f'\iN\Gtot$ by
  \be  \cg\gf{f'} \equiv \cg\gf{f'}(\cala^\zg)  \Labl cg
the \cla\ for \bc s of \atype\ $\gf$ for a \cft\ with torus \parfu\ $Z_{f'}$.
Recall from section \ref{t.3} that this \alg\ (as well as its distinguished
basis) can be constructed by starting with any arbitrary finite abelian group
$G$ containing $g$. As also discussed there, for the case where $f'\eq\fc$ 
corresponds to charge conjugation, the structure constants of $\cg\gf{f'}$ 
are given by the traces over the linear maps induced by $\tf$ on the 
three-point \cblock s:
  \be  {\rm N}^{(\gf;\fc)}_{\lambda,\mu,\nu} = {\rm tr}^{}_{V_{\lambda,\mu,\nu}}
  \llb \vec\Tau_{\!\gf,\gf,\gf}^{*\,{\sss(\lambda\,\mu\,\nu)}} \lrb \,.  \ee
In the general case, what should matter are not the individual maps $f$ and $f'$
by themselves, but rather only the information on how the pairing described by
the \bc s relates to the pairing in the torus \parfu. In other words, a
simultaneous action of an element of \GT\ on both labels of the \cla\ $\cg\gf{f'}$
will not change the situation in an observable manner. A well-known example of
this effect is seen in the theory of a free boson compactified at radius $R$,
where the T-duality map $R\,{\mapsto}\,2/R$ (which corresponds to $f\,{\mapsto}\,
f\fc$) amounts to exchanging Dirichlet and Neumann conditions (compare 
subsection \ref{t.81}). (For orbifolds of free bosons, T-duality has been 
studied recently e.g.\ in \cite{brer}.) 

In short, for every $f''\iN\GT$ there is an isomorphism
  \be  \cg{\gg_{f f''}}{f'f''} \,\cong\, \cg\gf{f'}  \ee
of \cla s. With the help of these isomorphisms we deduce that the
structure constants of the \cla\ $\cg\gf{\fc f'}$ are given by the traces 
  \be  {\rm N}^{(\gf;\fc f')}_{\lambda,\mu,\nu} = {\rm tr}^{}_{V_{\lambda,\mu,\nu}}
  \llb \vec\Tau_{\!\gF g_{f'}^{-1}\!,\gF g_{f'}^{-1}\!,\gF g_{f'}^{-1}}
  ^{*\,{\sss(\lambda\,\mu\,\nu)}} \lrb \,.  \labl{Nff}
In this formula it is manifest that simultaneous application of a T-duality
transformation $f\iN\GT$ on the bulk modular invariant and on the \atype\
of the \bc s does yield isomorphic \cla s.

Building on this result, we expect that formula \erf{Nff} extends naturally to 
arbitrary elements $f$ of \Gtot, i.e.\ to the case that $f$ is not necessarily
constant on $I$. We are thus led to conjecture that in this case the relation 
\erf{Nff} gets generalized to\,%
 \futnote{One might have expected that here products of $\Tau$-maps rather than
the $\Tau$-map for the product of \auto s appears. But when the group \Gtot\ is
realized projectively, this would lead to inconsistencies. When \Gtot\ is 
realized genuinely, then the two descriptions are equivalent.}
  \be  {\rm N}^{(\gf;\fc f')}_{\lambda,\mu,\nu} = {\rm tr}^{}_{V_{\lambda,\mu,\nu}}
  \llb \vec\Tau_{\!\gf f_\lambda^{\prime -1}\!,\gf f_\mu^{\prime -1}\!,\gf
  f_\nu^{\prime -1}}^{*\,{\sss(\lambda\,\mu\,\nu)}} \lrb \,.  \Labl,x
In the special case of $D_{\rm odd}$-type modular invariants (which correspond
to a simple current \auto\ for an order-two simple current of half-integral 
conformal weight), the relation \Erf,x is already known \cite{prss2,fuSc5} to hold.

Our observations imply in particular that the \atype\ $g_\rho$ of a \bc\ is 
{\em not\/} an {\em observable\/} concept. What {\em is\/} observable is the 
product $g_\rho^*\pi_f^{*-1}$, i.e.\ the `difference'
between an automorphism in the torus partition function and on the boundary.
(In other words, one should regard \atype s as
elements of the `affinum' -- see section \ref{t.5} below -- of the group
\GT\ rather than as elements of \GT\ itself.)
Similarly, also the difference $g(g')^{-1}$ of two \atype s is observable; 
e.g.\ the annulus partition function will be different in the situation where
one deals with two boundary conditions of distinct automorphism type as compared
to the situation where the automorphism type of both \bc s is the same.

While in the considerations above the T-duality transformations had to be
applied to the torus \parfu\ and to the \bc s simultaneously, there is 
also a slightly different notion of T-duality for boundary conditions,
to which we now turn our attention. Namely, we keep the bulk theory fixed, 
and ask whether boundary conditions of different automorphism type can be 
associated to each other through a suitable element of the T-duality group. 
While in many cases this question turns out 
to have an affirmative answer, the relationship in question is between 
{\em families\/} of boundary conditions rather than between individual 
\bc s. In short, T-duality on \bc s is not a one-to-one map, in general.
Closer inspection shows that the relevant families of \bc s can be understood 
as orbits of \bc s \wrt some suitable symmetry. Here by the term symmetry we 
mean a bijection $\gamma$ of the space of boundary conditions of a given 
automorphism type $g$ such that the annulus amplitudes coincide, i.e.
  \be  \A_{\gamma(\rho)\,\gamma(\rho')}(t) = \A_{\rho\,\rho'}(t)  \labl{bsym}
for all $\rho,\rho'$ with automorphism type $g_\rho\eq g\eq g_{\rho'}$. 

Let us study the presence of such a symmetry first in the example of the 
critical three-state Potts model. It is known \cite{afos} that the duality 
symmetry of this model maps the free \bc\ to any of the three fixed \bc s, 
and indeed this is a specific case of the general duality \cite{drwa} 
between free and `configurational' \bc s of lattice spin models. 
Similarly, the new boundary condition discovered in \cite{afos} gets mapped 
to any of the three mixed boundary conditions. In this case the symmetry 
group $\GGG\eq\zet_3$ of the fixed or mixed \bc s is directly inherited
from the lattice realization of the Potts model. We also observe that precisely
those boundary conditions for which this $\zet_3$-symmetry is spontaneously 
broken are grouped in non-trivial orbits. Furthermore, one readily 
checks that for orbits that are related by the T-duality 
$\pi_{\rm T}^*\eq\ccc$, the sum rule 
  \be  N_{\pi_{\rm T}^*(\rho)}^{-1}\, N_{\pi_{\rm T}^*(\rho')}^{-1}
  \sum_{\gamma,\gamma'\in\GGG}\!
  \A_{\gamma\pi_{\rm T}^*(\rho)\,\gamma'\pi_{\rm T}^*(\rho')} 
  = N_\rho^{-1}\, N_{\rho'}^{-1}
  \sum_{\gamma,\gamma'\in\GGG}\! \A_{\gamma(\rho)\,\gamma'(\rho')}
  \labl{srule}
for the annulus coefficients holds, where $N_\rho$ is the order of the 
stabilizer of the $\GGG$-action on $\rho$. Roughly speaking, the sum 
rule \erf{srule} tells us that T-dual {\em orbits\/} give rise to an 
equal number of open string states on the boundary.
  
This pattern can be detected in various other examples as well. For instance,
in the theory of a single uncompactified free boson, there is a single 
Neumann boundary condition, whereas the Dirichlet
boundary conditions are labelled by a position in $\dl R$, which should
be interpreted as the affine space over the group $\GGG\eq{\dl R}$ of translations.
This group is spontaneously broken for Dirichlet boundary conditions,
and it is straightforward to check that relations \erf{bsym} and \erf{srule}
are satisfied in this case as well.
Another class of examples is provided by boundary conditions in \wzwts\ 
that break the bulk symmetry via an {\em inner\/} automorphism of the
underlying simple \lie\ \gb. In these cases the boundary conditions of each
automorphism type are labelled by the same set, namely by the primary 
labels of the original theory (see \cite{bifs} for the case of 
automorphisms of order two). The group $\GGG$ is in this case realized by the
action of simple currents of the original theory, which account for the
different possible choices of the shift vector that characterizes the inner
\auto\ of \gb. Again the validity of relations \erf{bsym} and
\erf{srule} is easily verified. 

We would like to emphasize, though, that the existence of such T-duality
relations is in fact a quite special feature of an individual model. In 
general we do {\em not\/} expect that boundary conditions of different 
automorphism type are related in such a manner. For instance, in the case
of boundary conditions of \wzwts\ that
break the bulk symmetry via an {\em outer\/} automorphism of \gb, no relations
of the form \erf{bsym} or \erf{srule} are known to us.

\sect{Boundary homogeneity}\label{t.5} 

In this section we exhibit another general aspect of the space of conformally 
invariant boundary conditions that preserve only a subalgebra $\calap$ of 
$\cala$. Namely, we show that the orbifold group $G$ is realized as a group
of symmetries on the space of boundary conditions. These symmetries permute 
the boundary conditions within each set $\{\RhoB\,|\,\psu_\rho\iN\calu_\rho^*\}$
with fixed $\Gs$-orbit $\rhoB$; the permutation is the same for all 
$\Gs$-orbits. This behavior, to which we refer as {\em boundary 
homogeneity\/}, is similar to the so-called {\em fixed point homogeneity\/} 
that is present in simple current extensions, and indeed the arguments closely 
resemble the ones needed in the latter context \cite{fusS6}.

We start from the observation that the orbifold group $G$ can be identified 
with the dual $\Gs^*_{}$ of the simple current group $\Gs\eq G^*_{}$, and 
consider some character $\Psi$ of $\Gs$. For every field $\rhob$ of the 
$\calap$-theory and every character $\psu\iN\calu_\rho^*$ we then define a 
new character $\PSI\psu\iN\calu_\rho^*$ by
  \be  \PSI\psu(\J) := \Psi(\J)\, \psu(\J)  \ee
for all $\J\iN\calu_\rho$.  This is indeed again a character of $\calu_\rho$. 
Moreover, manifestly the group law of $G$ is reproduced for different choices 
of $\Psi$, so that our construction provides an action of $G$ on the group 
$\calu_\rho^*$. Typically $G$ does not act freely, but it does act transitively.

The next step is to realize that this prescription supplies us with
a well defined action on the space of boundary conditions. This is not
entirely trivial, because the labels for the \bc s
are not pairs of orbits, but rather are obtained by 
the equivalence relation \ErF eq, i.e.\ $(\lambdab,\psu_\lambda)\,{\sim}\,
\J\,(\lambdab,\psu_\lambda)\,(\J{\star}\lambdab,\JJ\psu_\lambda)$,
which involves a non-trivial manipulation of the characters. 
However, this complication does not do any harm, because just as for the
action of $\Psi\iN\Gs^*$ it consists of a multiplication,
so that the two operations commute. We write
  \be  \PSI\!\rho := [\rhob,\!\PSI\psu] \quad{\rm for}\quad
  \rho = [\rhob,\psu] \,.  \ee
Similarly, when we extend by some smaller group $\Hp\,{\subset}\,\Gs$, we can 
define the analogous object $\PSI[\rhob,\psup]'\,{:=}\,[\rhob,\PSI\psup]'$. 

Let us now explain in which sense the elements of $G$ are to be regarded as
symmetries. Using the explicit expression \ErF8s for the annulus coefficients,
we can establish the identity 
  \be \A_{\PSI\RhoBe\,\PSI\RhoBz}^{\PSI\SigmaBp} 
  = \A_{\RhoBe\,\RhoBz}^\SigmaBp \ee
between annulus coefficients. Thus when we also act on the corresponding 
chiral labels of the open string states in the annulus amplitude, we can absorb 
the transformation into a relabelling of the summation variables, so as to 
conclude that the annulus partition function is invariant. We expect that 
this extends indeed to a full-fledged symmetry at the level of correlation 
functions, where again one has to act on the insertions on the boundary. 

As a consequence, the boundary conditions that correspond to one and the 
same $\Gs$-orbit $\rhoB$ should in fact better be labelled by the 
elements of what may be called the {\em affinum\/} over 
the character group $\calu_\rho^*$ rather than by $\calu_\rho^*$ itself.
Here by the term affinum over a group \G\ we refer to the elementary structure 
of a {\em set\/} $\AG$ that carries a free and transitive action of \G.\,%
 \futnote{Thus there exists a map $A\,{\times}\,\G\,{\to}\,A$ acting as 
$(p,g)\,{\mapsto}\,p^g$ such that $p^g\eq p$ if and only if $g\eq e$, 
$p^{gh}\eq (p^g)^h_{}$, and such that for each pair $(p,q)\iN A\,{\times}\,A$ 
there is a unique $g\iN\G$ with $p^g\eq q$. When \G\ is non-abelian, one must 
distinguish between left and right actions, and hence left and right affina.}
 Conversely, the group \G\ can be identified with the quotient of 
$\AG\,{\times}\,\AG$ by the equivalence relation $(p,q)\,{\sim}\,(p^h,q^h)$ for 
all $h\iN\G$. For every $p\iN\AG$ we are given an identification 
$g\,{\leftrightarrow}\,p^g$ between the group \G\ and its affinum $\AG$, but 
there does not exist any {\em canonical\/} identification.
Roughly speaking, in the structure of the affinum one ignores the special
role played by the identity element; thus an affinum and its group are
related in much the same way as an affine space $A_V$ is related to the 
corresponding vector space $V$, which can be identified with its group 
of translations. Indeed, the group of automorphisms (that is, bijections 
intertwining the action of \G) of an affinum $\AG$ is precisely \G. It 
follows that any two affina over \G\ are isomorphic; but the isomorphism is 
never canonical; it is always determined only up to the isomorphism group \G.

We should admit that, even though we avoided using the term, the concept of
an affinum is already implicit at several other places of our discussion of
\bc s. For instance, the quantities $\psi_\lambda$ and $\psu_\rho$ that appear
in the definition \erA tS of the matrix $\tS$ are best regarded as elements of the
affina of the respective character groups, because \cite{bant6} the matrices
$S^\J$ are only defined up to certain changes of basis in the space of one-point
blocks on the torus and because such a change amounts to a relabelling of the
characters.

We finally add a comment on how this symmetry property of the \bc s
looks like in the case of the three-state Potts model. In this case 
it exchanges simultaneously two fixed and two mixed boundary conditions,
while it leaves the third fixed and mixed boundary condition invariant.
In the Potts model we actually have yet another symmetry on the space 
of boundary conditions, the action of a $\zet_3$-group. These two symmetries 
combine to the symmetric group $S_3$.

\sect{The universal classifying algebra}\label{t.6}\label{s.72}

The decomposition \Erf op of the \cla\ \clAb\ into individual \cla s 
for fixed \atype s implies that for every subgroup $H$ of the
orbifold group $G$ one has $\calc(\cala^H)\,{\subseteq}\,\calc(\cala^G)$.
Indeed, $\calc(\cala^H)$ is just the direct sum of the ideals
$\clAg\,{\subseteq}\,\calc(\cala^G)$ for all $\gg\iN H$.
It is also known \cite{doMa2} that the mapping $H\,{\mapsto}\,\cala^H$
provides a bijection between the set of subgroups of $G$ and the set of 
consistent chiral sub\alg s of $\cala$ that contain $\cala^G$.\,%
 \futnote{A similar Galois correspondence has been established in the
context of braided monoidal *-categories in \cite{muge6}.}
 In our situation the orbifold group $G$ is abelian, but the latter result
continues to hold for all nilpotent finite groups \cite{doMa3} and is 
expected to be true for arbitrary finite orbifold groups.\,%
 \futnote{Assuming that the statement holds for every group within a
certain class $X$, it
follows in particular that whenever there exists at least one \bc\ that
does not possess an \atype, then the Virasoro \alg\ by itself cannot be 
the orbifold sub\alg\ of $\cala$ \wrt any group that belongs to $X$.}

In this section we would like to address the issue of consecutive
breakings of bulk symmetries in more generality, which leads us in 
particular to introduce the notion of a {\em universal classifying algebra\/}.
While our detailed studies have so far been restricted to cases where
the preserved subalgebra $\calap$ of the bulk symmetries satisfies
$\calap\eq\cala^G$ for some finite abelian group $G$, it is reasonable to
expect that several features of our analysis will persist for general
$\calap$. In particular, it should again be possible to determine a \cla,
provided that the following two pieces of information are available:
\nxt
the decomposition of $\cala$-modules into direct sums of irreducible 
$\calap$-modules;
\nxt concrete expressions relating the bundles of \cblock s of the
  $\cala$-theory with those of the $\calap$-theory.
\\[.1em] 
We also expect that the statements about inclusions of \cla s valid for
the case of finite abelian orbifold groups generalize as follows.
To every inclusion $\calap\,{\hookrightarrow}\,\cala$ of preserved bulk 
symmetry algebras there is associated a projection of the corresponding 
classifying algebras; the classifying algebra for $\cala$ is a suitable
quotient of the one for $\calap$. More generally, for every chain of inclusions
  \be \calap \,\hookrightarrow\, \cala'\,\hookrightarrow\, \cala \ee
of symmetry algebras, there should exist a corresponding chain of projections 
  \be \calc(\calap) \,\stackrel\pi{\tto}\, \calc(\cala') \,\stackrel{\pi'}{\tto}\,
  \calc(\cala)  \labl{s78}
for the associated classifying algebras.
As a consequence, every \irrep\ of $\calc(\cala')$ gives rise to an \irrep\ of
$\calc(\calap)$. This makes sense indeed: an \irrep\ of $\calc(\cala')$
corresponds to a boundary condition that preserves $\cala'$ and thus, a 
fortiori, also preserves the smaller algebra $\calap$; it should therefore
correspond to some \irrep\ of $\calc(\calap)$. Relation \erf{s78} clearly
holds when both $\calap\eq\cala^G$ and $\cala'\eq\cala^H$ are orbifold 
subalgebras for abelian orbifold groups with $H\,{\subseteq}\,G$.
(Moreover, the projections are compatible with the distinguished bases
of the \alg s, compare e.g.\ the arguments leading to formula \Erf=4
for the case $H\eq\zg$). More generally, one can hope to obtain this way also
quantitative information on solvable orbifold groups.

This way the following picture emerges. The set $\calm$ of all consistent 
subalgebras of a given chiral algebra $\cala$ that possess the same 
Virasoro element as $\cala$ is partially ordered by 
inclusion. It is reasonable to expect that $\calm$ is even an inductive 
system; that is, given any two consistent subalgebras $\calap_1$ and
$\calap_2$ of $\cala$, one can find another consistent subalgebra $\calap_3$ 
that is contained in their intersection, 
  \be  \calap_3\subseteq\calap_1\,{\cap}\,\calap_2 \,.  \ee
Note that this implies in particular that we do not have to make the assumption
that the intersection of all consistent subalgebras of $\cala$ contains a 
consistent subalgebra; rather, one only needs to deal with intersections of
{\em finitely many\/} subalgebras.

Assuming that also in general the inclusion $\calap_1\,{\subset}\,\calap_2$ 
implies that the classifying algebra for $\calap_2$ is a quotient of the 
one for $\calap_1$, one arrives at a projective system $(\calap_i)$ of 
classifying algebras. Taking the {\em projective limit\/} over this system,
we then arrive at a {\em universal classifying algebra\/} 
  \be  \univ := \lim_{\dsty\leftarrow}\,\calap_i \,.  \ee
(The projective limit of the closely related structure of a fusion ring
has been studied in \cite{fuSc3}.)
By construction, this \alg\ $\univ$ governs {\em all\/} conformally
invariant boundary conditions. In other words, it is the \cla\ $\calc(\vir)$
for the case where the preserved sub\alg\ just consists of the Virasoro \alg. 
The universal classifying algebra can be found explicitly in simple models,
e.g.\ for the free boson compactified on a circle or for the 
$\zet_2$-orbifold of these theories (see subsection \ref{t.91}).

The construction of a surjective homomorphism $\pi{:}\ \clAb\,{\to}\,
\calc(\cala')$ that maps the elements of the distinguished basis 
$\rp{\cal B}$ of \clAb\ to elements of the distinguished basis ${\cal B}'$
of $\calc(\cala')$ is in fact straightforward as far as the $\lambdab$-part
of the labels $(\lambdab,\varphi)$ for $\rp{\cal B}$ are concerned. In contrast,
concerning the $\varphi$-part one faces complications which stem from the
absence of a simple relationship between $\calup_\lambda$ and $\calu_\lambda$
(this fact had also to be taken into account in the manipulations that
were necessary to establish integrality of the annulus coefficients,
see subsection \ref{s.64} of \I). As a matter of fact, for a complete
discussion of this issue even in the case of abelian orbifold subalgebras 
additional simple current technology is required that goes beyond the 
results of \cite{fusS6}. In particular it will be necessary to implement 
the powerful results that have recently been obtained in \cite{sche10}.

We also note that for consistency, along with the projection $\pi$ there
should come an injection $\iota$ from the set of \bc s that preserve
$\cala'$ to those that preserve $\calap$, in such a way that the 
diagonalizing matrices for \clAb\ and for $\calc(\cala')$ are related as
  \be  \tS_{(\lambdab,\varphi),\iota([\rhob,\psu]\oei)}^{} \,\propto\,
  \tS'_{\pi((\lambdab,\varphi)),[\rhob,\psu]\oei} \,.  \ee
Again the explicit construction of $\iota$ proves to be difficult; it will
not be pursued here.

\sect{Involutary \bc s} \label{t.7}

In this section we focus our attention to the situation where the symmetries
$\calap\eq\cala^G$ that are preserved by the \bc s form a sub\alg\ that is
fixed by an involution $\omm$. In other words, for such {\em involutary 
\bc s\/} the orbifold group $G$ is just the $\zet_2$-group consisting of 
$\omm$ and the identity. For the associated $\zet_2$-orbifolds, a lot of
information is available (see e.g.\ \cite{dvvv,bifs}. The vacuum sector 
of the $\cala$-theory decomposes into subspaces as
  \be  \calh_\vac \cong \calhb_\vacb \oplus \calhb_\J  \ee
with $\J$ a simple current of order two,
and the \auto\ $\omm$ acts as
  \be  \omm|^{}_{\calhb_\vacb} = \id^{}_{\calhb_\vacb} \,, \qquad
  \omm|^{}_{\calhb_\J} = - \id^{}_{\calhb_\J} \,.  \ee
Involutary sub\alg s are very
special indeed. Several of the structures that required a detailed discussion
in the general case are realized rather trivially here. For instance, as 
cyclic groups have trivial second cohomology, all
untwisted stabilizers are equal to the full stabilizers; this already
simplifies various equations considerably.

The reason why we nevertheless study this simple situation in quite some
detail is that it is realized in various interesting systems. We will
soon display several of these examples, but first we summarize
some generic features of the $\zet_2$ case; in particular we will study 
the individual \cla\ for automorphism type $\omm$.

\subsection{Even and odd \bc s}

In the case at hand, the orbifold group $G$ and the simple current group 
$\Gs\eq G^*$ are both isomorphic to $\zet_2$. Thus in particular the 
exponentiated monodromy charge $\gg$ takes its value in $\zet_2$, so 
that there are just two automorphism
types of boundary conditions. We refer to those boundary conditions whose
automorphism type is given by the identity as {\em even\/} boundary conditions
while those with \atype\ $\omm$ will be called {\em odd\/}.

The \ctype s for the even \bc s are labelled by the primary fields $\lambda$
of the $\cala$-theory, while those for the odd \bc s are labelled by the
orbits of primary fields of the orbifold theory (whose chiral \alg\ is 
$\calap\eq\cala^{\zet_2}$) with monodromy charge $Q_\J\eq1/2$.
The simple current group is $\Gs\eq\{\vacb,\J\}$, so in particular its 
orbits $\lambdaB$ either have length two (i.e.\ have stabilizer
$\cals_\lambda\eq\ID$) or length one (i.e.\ are
fixed points, with $\cals_\lambda\eq\zet_2$). 
Only fields $\lambdab$ with vanishing monodromy charge can be fixed points;
therefore fixed points cannot give rise to odd boundary conditions.

The even boundary conditions preserve, of course, all bulk symmetries.
The relevant boundary blocks $\bbN\lambda$ can therefore be 
expressed in terms of the boundary blocks of the $\calap$-theory as
  \be  \bbN\lambda = \Frac1{|\cals_\lambda|}\,
  \llb \bbb\lambda \oplus \bbbJ\lambda \lrb \,,  \Labl13
and the Ward identities that come from a field $Y$ in the \chira\ $\cala$
look like
  \be  \bbN\lambda \circ \llb Y_n \oT\bfe + \zetay\, \bfe \oT Y_{-n}
  \lrb = 0  \Labl66
with $\zetay\eq(-1)^{\Delta_Y-1}$.
As for any \bc s that preserve all of $\cala$, the \cla\ \clA\
for the even \bc s is the fusion \alg\ of the $\cala$-theory, with structure
constants expressible via the Verlinde formula in terms of the modular
transformation matrix $S$ of the $\cala$-theory.

As established in subsection \ref{t.2},
in the case of odd boundary conditions, the Ward identities \Erf66 get
replaced by those for twisted boundary blocks; here they read
  \be  \bbD\lambda \circ \llb Y_n \oT\bfe + \zetay\, \bfe \oT
  \omm(Y_{-n}) \lrb = 0 \,.  \Labl15
The odd boundary blocks $\bbD\lambda$ satisfying these constraints
are `differences'
  \be  \bbD\lambda = \bbb\lambda \oplus \llb {-}\bbbJ\lambda \lrb  \Labl25
of the boundary blocks of the $\calap$-theory. They are thus related to the 
ordinary boundary blocks as in formula \Erf Bg, i.e.\ we have
  \be  \bbD\lambda = \bbN\lambda \circ \llb \tomm\oT\id \lrb \,,  \Labl ob
where the maps $\tomm$ satisfy the $\omm$-twisted intertwining property
$\tomm\,Y\eq\omm(Y)\,\tomm$ (which together with their action on the 
highest weight vector determines them uniquely).

\subsection{The \cla}

Let us now display the total \cla\ \clAb\ which governs even and
odd \bc s simultaneously. We already know what the labels for the basis
of \clAb\ and for its \onedim\ \irrep s look like. Moreover,
in the formula \erA tS for the diagonalizing matrix $\tS$, in the 
$\zet_2$ case only two different matrices appear, namely the modular 
S-matrix $\bS\,{\equiv}\,S^\vacb$ of the orbifold theory and the matrix
$\breve S\,{:=}\SJ$ for the simple current $\J$. The length 
$\L\lambda$ of an orbit of $\J$ can be either one or two; it will be convenient
to use different symbols for labels for fixed points and those for
length-two orbits; we choose roman letters $f,g,...$ for the former and
greek letters $\alpha,\beta,...$ from the beginning of the alphabet for the
latter. Also, for simplicity we will use one and the same symbol $\psi$ to
refer to a $\Gs$-character and to its value on the non-trivial element
$\J\iN\Gs$, which can be either of $\pm1$. For the subsets of the whole 
set $\M\eq\{\lambda\}$ of primary labels of the $\cala$-theory that consist 
of the labels for full orbits and for fixed points we write $\Mo$ and $\Mf$,
\resp, i.e. 
  \be  \Mo:=\{ \mu\iN\M \,|\, \L\mu\eq2 \} \,,\qquad
  \Mf:=\{ \mu\iN\M \,|\, \L\mu\eq1 \} \,.  \ee
We also introduce the corresponding subsets
  \be  \Mob := \{\rpa\iN\Mb \,|\, \alpha{=}\{\rpa,\J\rpa\}\iN\Mo\}\,,
  \qquad \Mfb := \{\rp f \,|\, \feta\iN\Mf\}  \ee
of the label set $\Mb$ of the orbifold theory.
With these notations the dimension of \clAb\ reads
  $    {\rm dim}\,\clAb 
  \eq |\Mob| + 2\,|\Mfb| $.
By the sum rule \erA sr this must equal the number of \bc s, i.e.\ the number of
$\zet_2$-orbits of the $\calap$-theory, counted with their stabilizer:
  \be  {\rm dim}\,\clAb = \Frac12\,|\Mob|
  + \Frac12\,|\Meb| + 2\,|\Mfb| \,.  \ee
This tells us that for every $\zet_2$-orbifold the number $|\Mo|$ of length-two
$Q\eq0$ orbits coincides with the total number $|\Me|$ of $Q\eq1/2$ orbits.
On the other hand the number $|\Mf|$ of fixed points, which necessarily have
$Q\eq0$, can be arbitrary. Also note that ${\rm dim}\,\clA\,{\equiv}\,
|\M|\eq|\Mo|\,{+}\,|\Mf|$, i.e.\ there are always at least as many even as 
there are odd \bc s. The numbers of odd and even boundary conditions are equal 
precisely in those cases where there are no fixed points, which happens 
precisely when the associated automorphism of the fusion rules is the identity.

Further, the entries of the diagonalizing matrix $\tS$ read explicitly
  \be  \bearll  \YS\rpa\rhoB = 2\, \Sb\alpha\rho \,,
  & \YS\fetab\rhoB = \Sb f\rho \qquad{\rm for}\,\ \L\rho\eq2 \,,
  \\{}\\[-.7em]
  \YS\rpa\gepSb = \Sb\alpha g\,, \ \
  & \YS\fetab\gepSb = \Frac12\, (\Sb fg + \psi\eps\,\oS fg) \,.  \eear \ee
This leads to the formul\ae\
  \be  \bearl  \tNl\rpa\rpb\rpc = 2\, \Nn\alpha\beta\gamma \,, \qquad\qquad
  \tNl\rpa\rpb\fetab = \Nn\alpha\beta f \,,
  \\{}\\[-.5em]
  \tNl\rpa\fetab\gepsb = \frac12\, (\Nn\alpha fg + \psi\eps\,\oNn\alpha fg)
  = \NDn\alpha\feta\geps \,,
  \\{}\\[-.5em]
  \tNl\fetab\gepsb\heppb = \NDn\feta\geps\hepp  \eear \labl{z21}
for the structure constants with three lower indices. 
Here $\rp{\rm N}$ and $\evtyp{\rm N}$ denote the fusion coefficients
of the $\calap$-theory and of the $\cala$-theory, \resp. Moreover, we have 
introduced {\em twined\/} fusion coefficients,
defined with the help of the twined S-matrix $\breve S$, according to
  \be
  \pNn f\rho g := \sumfb h \frac {\oS fh^{}\Sb \rho h^{}\oS gh^*} {\Sb\vac h} 
  \,.  \ee
Recall from subsection \ref{t33} that these twined fusion coefficients are 
the traces of the action of the outer automorphisms associated to $\J$ on 
bundles of chiral blocks.

Furthermore, from equation \erf{z21} we read off that the matrix
$\tC\eq\tS\tS^{\rm t}$ furnishes a conjugation on the basis labels (not just
an involution as in the generic case). We have 
  \be  \CY\rpa\rpb = 2\,\delta_{\rpa,\rpb^+}^{}\,, \qquad
  \CY\rpa\feta = 0 \,,  \qquad
  \CY\feta\geps = \delta_{\rp f,\rp g^+}^{}\, \Cl f_{\eta,\eps} \,;  \ee
in particular, conjugation on the fixed points is exactly as in the 
$\cala$-theory. It follows that the structure constants 
of the classifying algebra read
  \be  \bearll  \tNm\rpa\rpb\rpc = \N\alpha\beta\gamma \,, &
  \tNm\rpa\rpb\fetab = \N\alpha\beta f \,, \\{}\\[-.5em]
  \tNm\rpa\fetab{\;\ \ \ \rpc} = \Frac12\,\ND\alpha\feta{\,\ \ \ \gamma} \,,&
  \tNm\rpa\fetab{\ \gepsb} = \ND\alpha\feta{\ \geps} \,,  \\{}\\[-.5em]
  \tNm\fetab\gepsb{\ \ \ \ \ \ \ \rpc}
  = \Frac12\,\ND\feta\geps{\,\ \ \ \ \ \ \ \gamma} \,, \ \ &
  \tNm\fetab\gepsb{\ \ \ \ \heppb} = \ND\feta\geps{\ \ \ \ \hepp}
  \,.  \eear \Labl67

\subsection{The individual \cla\ for odd \bc s}

The basis $\{\tPhi_{(\lambdab,\psi)}\,|\,\lambdab\iN\Mob,\,\psi\iN
\cals_\lambda\}$ of the classifying algebra \clAb\ 
is mapped by \erf{Phig} to the distinguished basis of the fusion \alg\ of 
the $\cala$-theory. In the present situation we find
  \be  \evtyp\Phi_\alpha = \Frac12\,(\tPhi_\rpa+\tPhi_{\J\rpa}) \quad
  {\rm and} \quad \evtyp\Phi_\feta = \tPhi_\fetab \,.  \Labl3E
As we have seen, this provides an algebra homomorphism $\varphi$ from the 
classifying algebra \clAb\ to the fusion \alg\ of $\cala$, which is just 
the classifying algebra \clA\ for the even boundary conditions, i.e.\ 
$\clA\eq\evtyp\calc(\calap)\eq{\rm span}_\complex \{\evtyp\Phi_\lambda\}$.
On the other hand the kernel of $\varphi$ is an ideal of \clAb\ which 
provides us with a classifying algebra 
$\clAm\,{\equiv}\,\calc^{(\omm)}(\calap)$ for the odd boundary conditions.

Let us analyze this classifying algebra \clAm\ in more detail. Only 
$\Gs$-orbits of length two contribute. On each such orbit we choose 
one distinguished representative $\rpa$. 
We can and will assume that these choices are made in such a manner
that, first, for the vacuum orbit the orbifold vacuum $\vacb$ is taken  
as the representative, and second, on conjugate orbits one chooses conjugate 
representatives. Then the set of elements 
  \be  \odtyp\Phi_\alpha:= \Frac12\,(\tPhi_\rpa-\tPhi_{\J\rpa}) \,,  \Labl3F
with $\rpa$ the distinguished representative, form a basis for \clAm.
In this basis the structure constants of \clAm\ are given by the 
following traces on spaces of \cblock s (which are integers):
  \be  \dN\alpha\beta\gamma
  := \N\alpha\beta\gamma - \NJJJ\alpha\beta\gamma \,.  \Labl dN
As an ideal of \clA, the \alg\ \clAm\ inherits several properties
of \clA: it is semisimple, commutative and associative; it
is unital, the unit element being $\odtyp\Phi_\vac\eq\tPhi_\vacb-\tPhi_\J$;
and it has a conjugation which is evaluation at the unit element.
The dimension of \clAm\ is
  \be  {\rm dim}\,\clAm \equiv |\Mo| = |\Me| \,.  \Labl oe

Since \clAm\ is semisimple, it must possess a diagonalizing matrix. 
Indeed, by the Verlinde formula for the orbifold fusion rules
$\N\alpha\beta\gamma$ we can write the structure constants \Erf dN as
  \be  \dN\alpha\beta\gamma = \sumb m\Frac{\Sb\alpha m\,\Sb\beta m}{\Sb\vac m}
  \, \llb \Sb\gamma m^* - \Sj\gamma m^* \lrb \,.  \ee
Here a priori the summation is over all sectors of the orbifold theory.
But only the twisted fields give a non-vanishing contribution, and in that
case the two terms with $\rp\gamma$ and
$\J\rp\gamma$ are equal. Labelling these twisted fields by roman letters 
$\rp a,\rp b,...$ and calling the corresponding index set $\Meb$, we thus obtain
  $    \dN\alpha\beta\gamma \eq 2 \sumeb d
  \Sb\alpha d^{}\,\Sb\beta d^{}\,\Sb\gamma d^* / \Sb\vac d $.
Moreover, elements $\rp d$ on the same $\zet_2$-orbit
give identical results, hence we may rewrite this formula as a sum over orbits
$d\df\{\rp d,\J\rp d\}$. Denoting the set of these orbits by $\Me$, so that
  \be  \Meb := \{\rp a\iN\Mb \,|\, a{=}\{\rp a,\J\rp a\}\iN\Me\}\,,  \ee
we arrive at
  \be  \dN\alpha\beta\gamma= 4 \sume d
  \frac{\Sb\alpha d^{}\,\Sb\beta d^{}\,\Sb\gamma d^*}{\Sb\vac d} \,.  \Labl3z
We can interpret this result as stating that the structure constants
$\odtyp{\rm N}$ are governed by the matrix $\dS$ with entries
  \be  \dS_{\alpha,b}:= 2\,\Sb\alpha b \qquad\mbox{with }\
  \alpha\iN\Mo,\; b\iN\Me \,;  \Labl ts
this is the diagonalizing matrix for \clAm.
Also note that owing to $\dS_{\vac,b}^{}\eq2\,\Sb\vac b\,{>}\,0$ for all 
$b\iN\Me$, this matrix shares the positivity property of modular S-matrices
(this does {\em not\/} already follow from the commutativity and semisimplicity 
of \clAm). Moreover, by combining unitarity of $\rp S$ with the simple current 
symmetry $\Sj mn\eq(-1)^{2 Q_\J(n)}\Sb mn$ one obtains
  \be  \sumeb a \Sb a l^*\,\Sb a m^{}
  = \Frac12\, (\delta_{\rp l,\rp m}^{}-\delta_{\rp l,\J\rp m}^{}) \,, \qquad
  \sumofb n \Sb n l^*\,\Sb n m^{}
  = \Frac12\, (\delta_{\rp l,\rp m}^{}+\delta_{\rp l,\J\rp m}^{}) \,, \Labl1z
with the help of which one can show that the matrix \Erf ts is unitary, which
in turn tells us once again that $\dS$ is a square matrix.

The structure constants $\dN\alpha\beta\gamma$ as defined by \Erf dN
do depend on the choice of representatives that has been made. Indeed, upon
replacing $\rpc$ by $\J\rpc$, $\dN\alpha\beta\gamma$ goes to minus itself. 
But this does not change the \cla\ \clAm, since the choice of a sign 
constitutes a one-cocycle on \clAm, which in turn can be absorbed by
choosing a correlated sign for the boundary blocks. More concretely,
for the full $\zet_2$-orbits we have the isomorphism $\calh_\alpha\,{\cong}\,
\calhb_\rpa\opluS\calhbJ\gka\,{\cong}\,\calh_{\ommdual\alpha}$
of $\calap$-modules, so that the associated even boundary blocks are precisely 
$\bbN\alpha\eq\bbb\alpha\,{\oplus}\,\bbbJ\alpha$, as given by formula \Erf13, 
where $\Betab_{\rp m}{:}\ \calhb_{\rp m}{\otimes}\calhb_{\rp m^+}{\to}\,\complex$
are the orbifold boundary blocks. In contrast,
there are two possible choices of odd boundary blocks, namely 
$\bbD\alpha\eq\bbb\alpha\,{\oplus}\,(-\bbbJ\alpha)$ as in \Erf25 as well as
  \be  \bbD{\ommdual\alpha} = \bbbJ\alpha\,{\oplus}\,(-\bbb\alpha)
  = -\,\bbD\alpha \,.  \Labl1B
By construction both of these linear forms $\bbD\alpha$ and $\bbD{\ommdual\alpha}$
on $\calh_\alpha{\otimes}\calh_{\alpha^+_{}}$ satisfy the appropriate Ward 
identities, but of course we must keep just one out of the two. The right 
prescription is to keep $\odtyp\Beta_\beta$ when the label $\rpb\iN\Mob$
is the chosen representative for the orbit $\beta\iN\Mo$. 
(Also, positivity of mixed annulus amplitudes is guaranteed only with this 
choice, see below.) In short, the label of the orbifold boundary state
that appears with a positive sign in the boundary block \Erf25
is the one that is to be chosen as the representative of the orbit. 

\subsection{Annulus amplitudes}

Using our general results from section 6 of \I\ it is also straightforward to 
calculate all annulus amplitudes. For the case of two even \bc s we obtain
  \be \bearll \A_{\alpha\,\beta}
  &= \dsty\sumofb m (\N\gkb m\gka+\NJJ\gkb m\gka)\,\rchi_{\rp m}(\ii t/2) \,,
  \\{}\\[-.8em]
  \A_{\alpha\,\feta} &= \dsty\sumofb m \N fm\gka\,\rchi_{\rp m}(\ii t/2) \,,
  \\{}\\[-.8em]
  \A_{\feta\,\geps} \!\!\!
  &= \frac12 \dsty \sumofb m (\N gmf + \psi\eps\,\pN gmf)\,\rchi_{\rp m}(\ii t/2)
  \,. \eear \labl{rrr}
Similarly, for two odd \bc s the annulus amplitudes are
  \be  \A_{a\,b}^{}(t)
  = \Frac12\,\sumofb m \L m\,(\N bm a+\NJJ bm a) \, \rchi_{\rp m}(\ii t/2)
  \,.  \ee
Finally, for {\em mixed\/} annuli, i.e.\ annuli with one even and one odd
boundary, we find
  \be  \A_{a\,\mu}^{}(t)
  = \Frac12\, \sumeb c \L\mu\,(\N\mu ca+\NJJ\mu ca)\,
  \rchi_{\rp c}(\ii t/2)  \,.  \ee
Thus in particular we correctly establish the $\zet_2$ selection rule
  \be  \A_{\rhoe\,\rhoz}^\sigma = 0 \qquad{\rm whenever}\ \
  Q_\J(\rhoe)+Q_\J(\rhoz)+Q_\J(\sigma) \in \zet{+}1/2 \,,  \Labl2A
which is analogous to the selection rule that is valid for the orbifold 
fusion rules.
 
For every triple $\rhoe\,\rhoz,\sigma$ the annulus coefficient
$\A_{\rhoe\,\rhoz}^\sigma$ is manifestly a {\em non-negative integer\/}, in 
agreement with the physical meaning of the annulus amplitude as a partition 
function. We would like to stress that this coefficient is not only a 
non-negative integer, but in addition has a natural representation 
theoretic interpretation, namely as the dimension of a space of \cblock s. 
The expressions $\N gmf\,{\pm}\,\pN gmf$, for example, that 
appear in the last line of \erf{rrr}, are equal \cite{fuSc8} 
to the ranks of two invariant subbundles of bundles of \cblock s.
This observation also seems to be relevant for a better understanding of the
multiplicities of boundary fields which are counted by these dimensions
(that is, there is a separate boundary field $\Psi_\sigma^{\rhoe,\rhoz;\ell}$
for every $\ell\eq1,2,...\,,\A_{\rhoe\,\rhoz}^\sigma$)
and for which a satisfactory understanding is unfortunately so far still 
lacking. Inspecting the calculations more closely, one also observes that the 
coefficients $\A_{a\,\mu}^c$ check the correctness of the
sign convention in the definition of the odd 
boundary blocks that was discussed after formula \Erf1B; with a different 
prescription, some of these coefficients would become negative.
 
With the results above, it is easily verified that the annulus coefficients
$\A_{\rhoe\,\rhoz}^\sigma$ satisfy the relations that are expected on the 
basis of factorization arguments,\,%
 \futnote{As already pointed out in subsection 6.5 of \I\ for the case of 
general orbifold group $G$, a rigorous derivation of these relations is,
however, not yet available.}
 i.e.\ that they are `associative' in the sense that
  \be  \sum_\sigma \A_{\rhoe\,\rhoz}^\sigma \A_{\rhod\,\rhov}^{\sigma^+_{}}
  = \sum_\sigma \A_{\rhoe\,\rho_3^+}^\sigma
  \A_{\rho_2^+\,\rhov}^{\sigma^+_{}}  \Labl as
for all possible values of the $\rho_i$,
and that they are `complete' in the sense that
  \be  \A^{\rho_1}\, \A^{\rho_2}
  = \sum_{\rho_3} \Ntot{\rho_1}{\rho_2}{\ \rho_3}\, \A^{\rho_3} \,, \Labl1M
where the $\A$'s are regarded as matrices in their two lower indices.
The non-vanishing coefficients $\NM$ are
  \be \bearl
  \Ntot\lambda\mu\nu = \ND\lambda\mu\nu\,, \\[.5em]
  \Ntot ab\nu  = \Frac12\,\L\nu\,(\N ab\nu+\NJJ ab\nu)\,, \\[.5em]
  \Ntot a\mu c = \Frac14\,(\L\mu)^2\,(\N a\mu c+\NJ a\mu c)\,.  \eear \ee
Note that the matrix $\NM_\vac$ is the identity matrix.

\sect{Examples for involutory \bc s}\label{t.8}

We now present several classes of examples with involutory \bc s.
We do not intend to be exhaustive, but concentrate on particularly 
interesting \cfts. 

\subsection{Dirichlet and Neumann conditions for the free boson}\label{t.81}

A simple and well-known realization of involutory \bc s is encountered for
the $c\eq1$ theory of a single free boson, compactified on a
circle of radius $R\eq\sqrt{2\caln}$, where $\caln$ is a positive integer
-- \resp, at the T-dual radius ${\rm T}(R)\eq2/R\eq\sqrt{2/\caln}$.
This theory has $2\caln$ primary fields, which at the chiral level we label by
integers $\lambda\bmod2\caln$; their \uone\ charge modulo $\sqrt{2\caln}$ is 
$q_\lambda\eq\lambda/\sqrt{2\caln}$. At radius $\sqrt{2\caln}$ the theory has 
diagonal torus \parfu, i.e.\ the primaries are of the form $(\lambda,\lambda)$,
while for radius $\sqrt{2/\caln}$ one deals with the charge conjugation 
invariant, i.e.\ with primaries $(\lambda,-\lambda)$. We will follow our 
general convention to describe the \bc s for the case of the charge 
conjugation invariant; the corresponding results for the diagonal 
invariant can be deduced via T-duality, as explained in section \ref{t.4}.

The chiral algebra $\cala$ of the free boson
theory consists of operators of the form
  \be  P(\partial X) \exp(\ii n\sqrt{2\caln}\,X) \,, \labl{ex1}
where $n\iN\zet$ and $P(\partial X)$ is a (normally ordered) polynomial in the
\uone\ current $j\eq\ii\partial X/\sqrt{2\caln}$. This \alg\ has an obvious 
involutory \auto, which in terms of the 
Fubini\hy Veneziano field $X$ is expressed as
  \be  X \mapsto -X  \Labl-x
and has the physical meaning of charge conjugation.
It maps the \uone\ current to minus itself and exchanges the fields
$\exp(\pm\ii n\sqrt{2\caln}\,X)$; thus its fixed point \alg\ 
consists of the even polynomials in $j$ combined with the operators
$\cos(n\sqrt{2\caln}X)$ with $n\iN\zet$ and odd polynomials in $j$ 
combined with $\sin(n\sqrt{2\caln}X)$. This \alg\ $\calap$ is just 
the chiral \alg\ of the $\zet_2$-orbifold of the free boson.

We recall \cite{dvvv} that this $\zet_2$-orbifold has $\caln{+}7$ primary 
fields.  First, one has the vacuum $\vacb$ and a simple current $\J$ of 
conformal weight $\Delta_\J\eq1$, which comes from the \uone\ current of the 
$\cala$-theory. Besides these two fields there is one other length-two orbit
$\{\psi^1,\psi^2\}$ of monodromy charge zero (coming from the self-conjugate
field with $\lambda\eq\caln$ of the $\cala$-theory), as well as 
$\caln{-}1$ fixed points $\varphi_\lambdab^{}$ with 
$\lambdab\eq1,2,...\,,\caln{-}1$, and finally there is
one pair of twisted fields for each of the two self-conjugate fields of the
$\cala$-theory, which are denoted by $\{\sigma,\sigma'\}$ and $\{\tau,\tau'\}$.

The number of even \bc s is equal to the number $2\caln$ of primary fields of
the boson theory. According to our general prescription they are labelled by 
the orbits of the orbifold and the
characters of their stabilizers, i.e.\ there is one even \bc\ for each of 
$\{\vacb,\J\}$ and $\{\psi^1,\psi^2\}$ and two for each of the fixed points 
$\varphi_\lambdab$. (In the language of the circle theory, the latter
correspond to the two primary fields of opposite charge $\pm\lambdab$.)
There are just two odd \bc s, corresponding to the two 
$\J$-orbits of twisted fields. The \cla\ \clAm\ for the odd \bc s turns out to
be just the group \alg\ of $\zet_2$, and the total \cla\ \clAb\ is isomorphic
to the direct sum of this $\complex\zet_2$ and the fusion rule \alg\
$\complex\zet_{2\caln}$ of the boson theory.

As is already apparent from \Erf-x, the odd
\bc s are nothing but Neumann conditions for the free boson $X$ (with
definite rational values of the Wilson line), while the
odd ones are Dirichlet conditions, with definite rational values of the
position of the D0-brane, namely at $\xi R$ with $\xi$ any of the 
$2\caln$th roots of unity. (Recall that this formulation refers to the
case where the torus \parfu\ is the charge conjugation invariant;
for the true diagonal invariant, in agreement with T-duality the
role of Dirichlet and Neumann conditions get interchanged.)
The fact that the Wilson line (for Neumann conditions) \resp\ the position
of the brane (for Dirichlet conditions) are restricted to a discrete set
of values is of course a consequence of the fact that in the situation
considered here the preserved bulk symmetries correspond to a rational \cft.
By breaking more bulk symmetries one arrives at more general possibilities.
In particular, as will be seen in subsection \ref{t.91} below, one may 
consider orbifolds that correspond to a change of the radius of the circle 
and thereby arrive at Wilson lines and brane positions at arbitrary points 
of the circle.

\subsection{\sutwo\ \wzwts}

Another example is given by \sutwo\ \wzwts\ at levels $k\iN4\zet$. 
In this case the full \cft\ based on the diagonal modular invariant for the 
orbifold theory can be realized as a sigma model on the group manifold 
\SU(2), while the diagonal modular invariant for the $\cala$-theory 
corresponds to a sigma model on the non-simply connected group manifold 
\SO(3). In the usual notation \cite{caiz} these are the theories
labelled $A_{k+1}$ and $D_{k/2+2}$, \resp.

The non-trivial simple current $\J\iN\Gs$ has conformal
weight $\Delta_\J\eq k/4$. Labelling the sectors of the $A_{k+1}$ model by their
highest \sutwo-weights and taking as representatives of $\zet_2$-orbits
those with smaller weight, the various label sets look like follows.
  \be  \bearlll
  \Mob= \{ 0,2,4, ...\,, \ekz{-}2, \ekz{+}2, \ekz{+}4, ...\,,k \} \,,\;  &
  \Mfb= \{ \ekz \} \,,\;  &
  \Meb= \{ 1,3,5, ...\,, k{-}1 \} \,, \\{}\\[-.6em] 
  \Mo = \{ 0,2,4, ...\,, \ekz{-}2 \} \,, & \hsp{-6.24}
  \Mf = \{ (\ekz,\psi) \,|\, \psi\eq{\pm}1 \} \,, &
  \Me = \{ 1,3,5, ...\,, \ekz{-}1 \} \,.  \eear \ee
Thus $|\Mo|\eq|\Me|\eq k/4$ and $|\Mf|\eq2$.

For the modular matrix $S\,{\equiv}\,\SDo$ of the $D_{k/2+2}$-model, a natural 
ordering of the labels in $\M\,{\equiv}\,\Mo{\cup}\Mf$ is
  \be  \mu = 0\,,\,2\,,\,4\,,\, ...\,,\, \ekz{-}2\,,\, (\ekz{,}+)\,,\,
  (\ekz{,}-) \,,  \ee
both for rows and for columns. With this labelling the general formula
for the modular S-matrix of a simple current extension 
(established in \cite{fusS6} and displayed also in appendix A of \I) yields
  \be  \SD\mu\nu = \left\{ \bearll
  2\,\Sb\mu\nu & {\rm for}\,\ \mu,\nu\eq0,2,...\,,\ekz{-}2 \,, \\{}\\[-.8em]
  \Sb\mu\nu &{\rm for}\,\ \mu\eq0,2,...\,,\ekz{-}2,\ \nu\eq(\ekz{,}\psi)\\[.3em]
    &\hsp{.3} {\rm or}\,\ \mu\eq(\ekz{,}\psi), \ \nu\eq0,2,...\,,\ekz{-}2 \,,
  \\{}\\[-.8em]
  \Frac12\,(\Frac1{\sqrt{k/2+1}}+\psi\eps\,\II_{})
   & {\rm for}\,\ \mu\eq(\ekz{,}\psi),\ \nu\eq(\ekz{,}\eps) 
  \,. \eear \right.   \Labl3t
Here $\rp S$ is the S-matrix of the $A_{k+1}$-model, i.e.\ $\Sb\mu\nu\eq\sqrt
{2/(k{+}2)}\sin((\rp\mu{+}1)(\rp\nu{+}1)\pi/(k{+}2))$, and we introduced the number
  \be  \II := \left\{ \bearll 1 & {\rm for}\ k\iN8\zet \,, \\[.3em]
  \ii\equiv\sqrt{-1} & {\rm for}\ k\iN8\zet+4 \,,\eear \right. \Labl II
which is nothing but the (one-by-one) matrix $\SJ$.
In contrast, for the diagonalizing matrix $\dS$ of \clAm\ there will typically 
not exist any preferred ordering of the labels $\alpha\iN\Mo$ (for the rows 
of $\dS$) and $a\iN\Me$ (for the columns); amusingly,
in the present case it is possible to order rows and 
columns in such a fashion that $\dS$ is symmetric. We have
  \be  \dS_{\alpha,b} = 2\,\Sb\alpha b= \sqrt{\Frac8{k+2}}\, 
  \sin\llb \Frac{(\alpha+1)(b+1)\,\pi}{k+2} \lrb \,.  \Labl4t
Now ordering again $\Mo$ according to the value of the weight, i.e.\ as
$\alpha\eq0,2,4,...\,,k/2{-}2$, symmetry is achieved when $\Me$
is ordered by taking first the weights of the form $4j{+}1$
in ascending order and afterwards the weights of the form $4j{+}3$
in descending order, i.e.\ $a\eq1,5,9,13,....,\,15,11,7,3$.
Doing so, the matrix \Erf4t simply becomes the matrix with entries
  \be  \widetilde{\Odtyp S}_{p,q} = \sqrt{\Frac8{k+2}}\, \sin\llb 
  \Frac{(2p-1)(2q-1)\,\pi}{k/2+1} \lrb\,,  \ee
where the integers $p$ and $q$ run from 1 to $k/4$. In particular, inspection
shows 
that $\clAm$ coincides with the fusion \alg\ \cite{wang'}
of the $(\ekz{+}1,2)$ Virasoro minimal models.

Via the simple current symmetries of the matrices $\bS$ and $\SJ$, the
diagonalizing matrix $\tS$ of the \cla\ is already completely determined by
$\SDo$ and $\dS$. Concretely, when we choose the ordering of the rows of
$\tS$ as
  \be  (\lambdab,\varphi) = 0\,,\,2\,,\,4\,,\, ...\,,\, \ekz{-}2\,,\, 
  (\ekz{,}+)\,,\,(\ekz{,}-)\,,\,k\,,\,k{-}2\,,\, ...\,,\, \ekz{+}2   \ee
and the ordering of the columns according to
  \be  (\rhob,\psi) = 0\,,\,2\,,\,4\,,\, ...\,,\, \ekz{-}2\,,\, 
  (\ekz{,}+)\,,\,(\ekz{,}-)\,,\,1\,,\,3\,,\, ...\,,\, \ekz{-}1 \,, \hsp{.6}
  \ee
then $\tS$ is block diagonal of the form
  \be  \tS = \left( \begin{array}{cc} \SDo & \evodtyp S \\[.3em]
  \odevtyp S & \dS \eear \right) ,  \Labl2t
and the off-diagonal blocks are related to the diagonal ones by
  \be  \evodtyp S|^{}_{\rm full} = - \dS \,,\quad
  \evodtyp S|^{}_{\rm fixed} = 0 \,,\quad
  \odevtyp S = \SDo|^{}_{\rm full} \,,  \ee
where the symbols $|^{}_{\rm full}$ and $|^{}_{\rm fixed}$ stand for 
restriction of the
rows to those corresponding to full orbits and to fixed points, \resp.

\subsection{Relation with incidence matrices of graphs}

In the \sutwo\ case under consideration, it is not too difficult to establish
that -- just like the \alg\ $\clAm$ -- also the total \cla\ \clAb\ constitutes 
a structure that has been encountered in \cft\ before. Indeed, we will
construct an isomorphism between \clAb\ and an algebra that appears in the 
work of Pasquier et al. To simplify some
of the formul\ae\ below, let us write $k\eq4\ell$ with $\ell\iN\zet$. Then
a $2\ell{+}2$-dimensional associative \alg\ with structure constants
  \be  \hN\hr\hs\hT :=
  \sum_\hu \hS_{\hr,\hu}^{} \hS_{\hs,\hu}^{} \hS_{\hT,\hu}^* / \hS_{1,\hu}^{}
  \labl{16'}
for $\hr,\hs,\hT\iN\{1,2,...\,,2\ell{+}2\}$ has been considered in
\cite{pasq4,dizu,pezu} and been called the {\em Pasquier \alg\/}
associated to the situation of our interest. In formula \erf{16'},
$\hS$ is the matrix with entries
  \be  \hS_{\hr,\hs}^{} := \left\{ \bearll
  \Frac12\,(\Frac1{\sqrt{2\ell+1}}+\II)
   & {\rm for}\,\ \hr\eq2\ell{+}1,\ \hs\eq\ell{+}1\,\
     {\rm or}\,\ \hr\eq\hs\eq2\ell{+}2\,, \\{}\\[-.8em]
  \Frac12\,(\Frac1{\sqrt{2\ell+1}}-\II)
   & {\rm for}\,\ \hr\eq2\ell{+}1,\ \hs\eq2\ell{+}2\,\
     {\rm or}\,\ \hr\eq2\ell{+}2,\ \hs\eq\ell{+}1\,, \\{}\\[-.8em]
  \Frac1{\sqrt{2\ell+1}}\,(-1)^{(\hr-1)/2}_{}\,(1-(-1)^\hr_{})
   &{\rm for}\,\ \hr\eq1,2,...\,,2\ell,\ \hs\eq\ell{+}1,2\ell{+}2\,,
  \\{}\\[-.8em]
  \Frac1{\sqrt{4\ell+2}}
   &{\rm for}\,\ \hr\eq2\ell{+}1,2\ell{+}2,\ \hs\nE\ell{+}1,2\ell{+}2\,,
  \\{}\\[-.8em]
  \Frac1{\sqrt{4\ell+2}} \cdot2\,\cos\llb \Frac{(2\ell-\hr+1)(2\hs-1)\,\pi}
   {4\ell+2} \lrb & {\rm else}\,, \eear \right.  \Labl hS
with $\II$ as in \Erf II.
Note that the matrix $\hat N_2$ with entries $(\hat N_2)_\hs^{\ \hT}\eq\hN2
\hs\hT$ is nothing but the {\em incidence matrix} of the graph $D_{2\ell+2}$.
Conversely, up to a phase the columns of $\hS$ are uniquely determined by the
two requirements that $\hS$ is unitary and
diagonalizes $\hat N_2$ -- with the exception, however, of the columns numbered
$\ell{+}1$ and $2\ell{+}2$, which both are eigenvectors to the
same eigenvalue zero. For the latter two columns, in formula \Erf hS
(unlike in table 2 of \cite{pasq4}) we have chosen specific linear combinations
that are singled out by the property that all structure constants $\hN\hr\hs\hT$ are
non-negative integers.
 We also remark that the matrix \Erf hS is unitary, but it is not symmetric, 
nor can it be made symmetric by any re-ordering of rows and columns,\,%
 \futnote{However, as follows from the identifications below, when combining
suitable re-orderings with rescalings, certain submatrices of $\hS$ become
symmetric.}
 and finally that
  \be  \hS_{\hr,2\ell+2-\hs}^{} = (-1)^{\hr+1}\, \hS_{\hr,\hs}^{} 
  \qquad{\rm for}\quad \hr\nE2\ell{+}1,2\ell{+}2\,,\ \hs\nE\ell{+}1,2\ell{+}2
  \,.  \ee
  
By inspecting the formul\ae\ \Erf hS and \Erf2t, we observe the relation 
  \be  \hS_{\ro([\rhob,\psi]),\si((\lambdab,\varphi))}^{} 
  = \Frac{\varepsilon^{}_\lambda}{\sqrt{N_\lambda}}\,
  \tS_{(\lambdab,\varphi),[\rhob,\psi]}  \Labl ht
between the matrices $\hS$ and $\tS$. Here $N_\lambda$ is the length of the
$\Gs$-orbit through $\lambdab$ (i.e.\ $N_\lambda\eq2$ except for 
$N_{2\ell}\eq1$), $\varepsilon$ is the sign factor
  \be  \varepsilon_\lambda^{} := \left\{ \begin{array}{cl}
  1\,\  & {\rm for}\,\ \lambda\eq2\ell
  \,, \\ (-1)^{\lambda/2}& {\rm else}\,,  \eear \right.  \ee
and we introduced bijections $\ro$ 
and $\si$ between the respective index sets of $\hS$ and $\tS$ which act as
  \be  \ro([\rhob{,}\psi]) := \left\{ \bearll
  \rhob+1 & {\rm for}\,\ \rhob\nE2\ell\,, \\{}\\[-.9em]
  2\ell+1 & {\rm for}\,\ (\rhob{,}\psi)\eq(2\ell{,}+)\,, \\{}\\[-.9em]
  2\ell+2 & {\rm for}\,\ (\rhob{,}\psi)\eq(2\ell{,}-)
  \eear \right. \hsp{.5}  \Labl ro
and
  \be  \si((\lambdab{,}\varphi)) := \left\{ \bearll
  \lambdab/2+1 & {\rm for}\,\ \lambdab\nE2\ell\,, \\{}\\[-.9em]
  \ell+1  & {\rm for}\,\ (\lambdab{,}\varphi)\eq(2\ell{,}+)\,, \\{}\\[-.9em]
  2\ell+2 & {\rm for}\,\ (\lambdab{,}\varphi)\eq(2\ell{,}-) \,,
  \eear \right.  \Labl si
\resp. In particular, the diagonalizing matrices \Erf3t
and \Erf4t for the even and odd \bc s obey
  \be  \SD\mu\nu = \sqrt{N_\mu}\, (-1)^{\mu/N_\mu}\,
  \hS_{\ro(\nu),\si(\mu)}^{}  \Labl NP
and
  \be  \dS_{\alpha,b} = \sqrt2\, (-1)^{\alpha/2}\,
  \hS_{b+1,\si(\alpha)} \,, \Labl DP
\resp.

Using the fact that $S$ is symmetric, the result \Erf ht tells us that
up to normalizations of the rows of $\tS$, transposition, and reordering of 
the rows and columns, the two matrices $\tS$ and $\hS$ coincide. As all these 
manipulations can be absorbed into a basis transformation, it follows that
the two \alg s that via these matrices are associated to the simple current
extension of $\sutwo_\ell$, i.e.\ the \cla\ \clAb\
and Pasquier's \alg\ are {\em isomorphic\/}.

For concreteness, let us also display explicitly these matrices in the 
simplest case, i.e.\ for $\ell\eq1$:
  \be  \bearl \hS = \Frac1{\sqrt6} \left(\!\! \begin{array}{cccc}
  1 & \sqrt2 &   1 & \sqrt2   \\[.3em]
  \sqrt3 & 0 & -\sqrt3 & 0    \\[.3em]
  1 & \frac1{\sqrt2}\,(1{+}\ii\sqrt3) & 1 & \frac1{\sqrt2}\,(1{-}\ii\sqrt3)
  \\[.3em]
  1 & \frac1{\sqrt2}\,(1{-}\ii\sqrt3) & 1 & \frac1{\sqrt2}\,(1{+}\ii\sqrt3)
  \eear \!\right) ,
  \\{}\\[-.5em]
  \tS = \Frac1{\sqrt3} \left(\!\! \begin{array}{cccc}
  \ 1 & 1 & 1 & -\sqrt3\,    \\[.3em]
  \ 1 & \,\frac12\,(1{+}\ii\sqrt3)\, & \,\frac12\,(1{-}\ii\sqrt3)\, & 0 \\[.3em]
  \ 1 & \,\frac12\,(1{-}\ii\sqrt3)\, & \,\frac12\,(1{+}\ii\sqrt3)\, & 0 \\[.3em]
  \ 1 & 1 & 1 &  \sqrt3\,  
  \eear \!\right) .  \eear\ee

\subsection{Virasoro minimal models}

The unitary minimal models of the Virasoro \alg\ are labelled by $m=2,3,...\,$;
they have conformal central charge $c\eq c_m\df1{-}6/(m{+}1)(m{+}2)$. Via their
realization as a coset theory $\sutwo_{m-1}{\oplus}\sutwo1/\sutwo_m$, the \sltwo\
WZW situation of the previous subsection gives rise to similar effects in
these minimal models. The requirement that the level must be divisible by four
translates to the condition $m\iN4\zet\,{\cup}\,(4\zet{+}1)$ on the label $m$.
In these cases the chiral algebra of the $\cala$-theory is obtained from 
$\calap$, which is just the Virasoro \alg, via extension by the field $\J$ with
label $\lambdab_\J\eq(m,1)$ (Kac table notation), which has conformal weight
$\Delta_J\eq m(m{-}1)/4$.

In these cases we even know that the Virasoro \alg\ $\calap$ is the only
consistent sub\alg\ of $\cala$, simply because no other unitary
\cfts\ exist at the same value of $c$. Thus our methods supply us\,%
 \futnote{At least modulo what goes \cite{sasT2} under the 
name of `complex charges'.}
with {\em all\/} conformally invariant \bc s of the $\cala$-theory. In
particular there are precisely two \atype s, the even \bc s which preserve
the full bulk symmetry $\cala$, and the odd \bc s which preserve only
the Virasoro sub\alg.

The primary fields of the $\calap$-theory with central charge $c_m$ are 
labelled by $\lambdab\,{\equiv}\,(r,\Br)$ with
  \be  1 \le r \le m \,, \qquad 1 \le \Br \le m+1 \,,  \ee
modulo the identification $(r,\Br)\,{\sim}\,(m{+}1{-}r,m{+}2{-}\Br)$,
so that there is a total of $m(m{+}1)/2$ sectors. We first look at
the cases with $m\eq4\ell$ for some $\ell\iN\zet_{>0}$.
Then the $\calap$- and $\cala$-theory are commonly \cite{caiz} denoted by
$(A_{4\ell},A_{4\ell+1})$ and $(A_{4\ell},D_{2\ell+2})$, \resp.
(In the simplest of these, obtained for $\ell\eq1$, the $\calap$-theory is the
tetracritical Ising model $(A_4,A_5)$ while the $\cala$-theory is the 
three-sta\-te Potts model $(A_4,D_4)$.) We have
  \be  \bearl
  \Mob = \{(r,\Br)\,|\, r\eq1,3,5,...\,,4\ell{-}1,\;
         \Br\eq1,3,5,...\,,2\ell{-}1,2\ell{+}3,...\,,4\ell{+}1 \} \,, \\[.4em]
  \Mfb = \{(r,2\ell{+}1)\,|\, r\eq1,3,5,...\,,4\ell{-}1 \} \,, \\[.4em]
  \Meb = \{(r,\Br)\,|\, r\eq2,4,6,...\,,4\ell,\;
         \Br\eq2,4,6,...\,,4\ell \} \,, \eear \ee
and hence $|\Mob|\eq|\Meb|\eq4\ell^2$, $|\Mfb|\eq2\ell$.
Thus we obtain $2\ell(\ell{+}2)$ even and $2\ell^2$ odd \bc s, and hence a
total of $4\ell(\ell{+}1)$ conformally invariant \bc s.

The similar series with $m\eq4\ell{+}1$ for $\ell\iN\zet_{>0}$
can be treated analogously. The $\calap$- and $\cala$-theory are now known
under the names $(A_{4\ell+1},A_{4\ell+2})$ and $(D_{2\ell+2},A_{4\ell+2})$,
\resp. We have
  \be  \bearl
  \Mob = \{(r,\Br)\,|\, r\eq1,3,5, ...\,,2\ell{-}1,2\ell{+}3,...\,,
           4\ell{+}1\,,\,\; \Br\eq1,3,5,...\,,4\ell{+}1 \,, \\[.4em]
  \Mfb = \{(2\ell{+}1,\Br)\,|\, \Br\eq1,3,5,...\,,4\ell{+}1 \,, \\[.4em]
  \Meb = \{(r,\Br)\,|\, r\eq2,4,6,...\,,4\ell
           \,,\,\; \Br\eq2,4,6,...\,,4\ell{+}2 \,, \eear \ee
so that $|\Mob|\eq|\Meb|\eq2\ell(2\ell{+}1)$, $|\Mfb|\eq2\ell{+}1$, and the
number of even and odd \bc s is $(2\ell{+}1)(\ell{+}2)$ and $\ell(2\ell{+}1)$,
\resp.

Via the coset construction, it is possible to express all the ingredients 
in the formula for $\tS$ through quantities of the underlying \sltwo\ \wzwm s,
so that again the \cla\ \clAb\ can easily be obtained explicitly.
We refrain from displaying any details, which are not too illuminating.
We would like to mention, however, that these results are in perfect
agreement with the findings of \cite{bePz,bppz}.
In the latter papers, various statements were encoded in the language of
graphs; the following remarks allow to make contact to that point of view.

The total number of conformally invariant \bc s is
  \be  |\Mob|+2\,|\Mfb| = \left\{ \bearll
  4\ell(\ell{+}1) = \Frac12 \cdoT \rank{A_{4\ell}} \cdoT \rank{D_{2\ell+2}}
  & {\rm for}\,\ m\eq4\ell \,, \\[.21em]
  2(\ell{+}1)(2\ell{+}1) = \Frac12 \cdoT \rank{A_{4\ell+2}} \cdoT \rank{D_{2\ell+2}}
  & {\rm for}\,\ m\eq4\ell{+}1 \,. \eear\right.  \labl{nbc}
Regarding the graphs $A_{4\ell}$ and $D_{2\ell+2}$ (i.e.\ the Dynkin diagrams
of the respective simple \lie s) as bi-colored, starting with (say) a 
black node, this can be understood as follows.
The even \bc s are in one-to-one correspondence with pairs of black nodes from
the `product' of the two graphs, while odd \bc s are in one-to-one
correspondence with pairs of white nodes.
Mixed pairs of nodes do not correspond to \bc s, which accounts for the
factor of $1/2$ in \erf{nbc}. The latter selection rule may be implemented by 
a suitable folding of the $A_{4\ell}$ graph. The resulting graph has a loop,
hence in particular it is no longer bi-colorable; pairs of nodes from 
$D_{2\ell+2}$ and the folded graph are then in one-to-one correspondence
with {\em all\/} conformally invariant \bc s, including both even and odd ones.

Further, let us denote by $\e\Hr$ the $\Hr$th {\em exponent\/} of the \lie\ 
$D_{2\ell+2}$. For every $\Hr\eq1,2,...\,,2\ell{+}2$, the integer $\e\Hr$ 
lies in the label set of the $A_{4\ell+1}$ graph; indeed, the exponents
correspond precisely to the black nodes of
$A_{4\ell+1}$, with the middle node appearing twice.
We can therefore define for every $\bs\eq 1,2,...\,,4\ell{+}1$ a
matrix $V_\bs$ through
  \be  (V_\bR)_\Hs^{\;\Ht} := \sum_\Hu \AS_{\bR,\e\Hu}\, \DS_{\Hs,\Hu}\,
  \DSs_{\Ht,\Hu} / \AS_{1,\e\Hu} \,,  \Labl35
where $\AS$ and $\DS$ are the unitary diagonalizing matrices for the graphs 
$A_{4\ell+1}$ and $D_{2\ell+2}$. Thus $\AS$ is nothing but the modular S-matrix 
of the $\sutwo_{4\ell}$ \wzwm.\,%
 \futnote{On the other hand, the matrix $\DS$ defined this way is by far 
not unique; in particular, the incidence matrix of $D_{2\ell+2}$ has an 
eigenvalue with multiplicity $2$. For a generic choice of diagonalizing
matrix the numbers $\HN\Hr\Hs\Ht$ \Erf16 will not be integral. But \cite{bePz} 
there is a unique choice such that these numbers {\em are\/} integral.}
 By direct calculation one checks that the matrices \Erf35 furnish a \rep\ 
of the fusion ring of $\sutwo_{4\ell}$. Further, it can be shown that $V_2$ 
coincides with the incidence matrix of $D_{2\ell+2}$. By the explicit form 
of the $\sutwo_{4\ell}$ fusion rules, this implies that one may equivalently 
define the matrices $V_\bs$ inductively via $V_1\df\one$, 
$V_2\df\,$incidence matrix of $D_{2\ell+2}$ and
  \be  V_\bs:= V_2V_{\bs-1}-V_{\bs-2}\quad\
  {\rm for}\;\ \bs\eq 3,4,...\,,4\ell{+}1 \,.  \ee
One also has the matrix equation
  \be  V_\bq\,\HNe\Hr = \sum_\Hs (V_\bq)_\Hr^{\;\ \Hs}\,\HNe\Hs  \Labl19
for all $\bq\eq1,2,...\,,4\ell{+}1$ and all $\Hr\eq1,2,...\,,2\ell{+}2$,
where $\HNe\Hs$ are the `graph fusion matrices' associated to $\DS$, i.e.\ the
matrices with entries
  \be  \HN\Hr\Hs\Ht :=
  \sum_\Hu \DS_{\Hr,\Hu} \DS_{\Hs,\Hu} \DSs_{\Ht,\Hu} / \DS_{1,\Hu}  \ee
for $\Hr,\Hs,\Ht\eq1,2,...\,,2\ell{+}2$.

In \cite{bePz,bppz} also the $E$-type series of modular invariants for
\mimo s were discussed in the framework of graphs, and 
cyclic groups larger than $\zet_2$ have been addressed in \cite{pezu3}. It
will certainly be interesting to compare the results obtained there
with the ones that can be derived by the methods of the present paper,
and in particular to study how the graph oriented approach deals with the
case of non-cyclic abelian groups, where non-trivial two-cocycles appear.

\sect{More examples}\label{t.9}

In this section we present a few further examples, in which the orbifold
group $G$ is larger than $\zet_2$. We first discuss two examples of 
direct relevance to string theory.
Afterwards we turn to some specific examples in which the effects of
non-trivial two-cocycles can be analyzed in detail.

\subsection{General cyclic groups}\label{t.91}

Let us study a situation of immediate interest in which the orbifold group 
is cyclic and hence has trivial second cohomology, so that the untwisted 
and full stabilizers
coincide. We start with the $c\eq1$ theory of a free boson, compactified on a 
circle of radius $R\eq\sqrt{2\caln}$ with $\caln\iN\zetplus$. The 
chiral algebra $\cala$ of this theory consists of operators of the form 
\erf{ex1}. Inspection shows that for every $m\iN\zetplus$ there is a subalgebra 
$\calap^{(m)}$ of $\cala$ which is obtained by restricting the value of $n$ in 
\erf{ex1} to be a multiple of $m$. The algebra $\calap^{(m)}$ is
nothing else but the chiral algebra of another free boson theory, with the
free boson $X$ compactified on a circle of radius
$m\sqrt{2\caln}$. Therefore it is a consistent subalgebra in the sense 
that it allows for the construction of \cblock s which obey factorization 
rules and have a \kzc. The orbifold group is $G\eq\zet_m$; its generator 
acts on an operator of the form \erf{ex1} by multiplication with the phase 
$\exp(2\pi\ii n/m)$; in terms of the Fubini\hy Veneziano field $X$ this means
  \be  X \mapsto X + 2\pi/(m\sqrt{2\caln})  \,.  \Labl mx
This leaves the \uone\ current $j$ invariant and multiplies 
$\exp(\ii n\sqrt{2\caln}\,X)$ by the phase $\exp(2\pi\ii n/m)$.
(Together with the \auto\ $X\,{\mapsto}\,{-}X$ \Erf-x, the transformation
\Erf mx generates the dihedral group $D_m$.)

In this example no fixed points are present, so that it is most 
straightforward to write down the classifying algebra. The 
$\calap^{(m)}$-theory has $2m^2\caln$ primary fields, which may be 
labelled by integers $\lambdab\bmod 2m^2\caln$;
their \uone\ charge is $q_\lambdab\eq\lambdab/m\sqrt{2\caln}$. The twist 
sector is determined by the value of $\lambdab\bmod m$; in particular, those
$\lambdab\eq m\lb$ which are multiples of $m$ are in the untwisted sector and
label a basis of the classifying algebra $\calc(\calap^{(m)})$. The 
multiplication in $\calc(\calap^{(m)})$ is just
given by the restriction of the fusion product of the $\calap^{(m)}$-theory
to the untwisted sector, i.e.\
  \be  \tNn\lb{\lb'}{\lb''} = \delta_{\lb+\lb'+\lb''\bmod2m\caln}^{} \,.  \ee
Thus $\calc(\calap^{(m)})$ is the group algebra of the cyclic group 
$\zet_{m\caln}$. The reflection coefficients $\Rc a\mub\vacb$ must 
therefore obey the relation $\Rc a{m\lb_1}\vacb \Rc a{m\lb_2}\vacb\eq 
\Rc a{m(\lb_1+\lb_2)}\vacb$,
where the addition is modulo $2m\caln$. The solutions to this requirement are
  \be  \Rc a{m\lb}\vacb = \exp(2\pi\ii m\lb a/2m^2\caln)
  = \exp(\pi\ii\lb a/m\caln) \ee
with $a\iN\zet$. Moreover, $a$ must be taken only modulo
$2m\caln$, and hence the possible values of $a$ are in one-to-one 
correspondence with the $\zet_m$-orbits of the 
$\calap^{(m)}$-theory, in accordance with the general theory.

This result allows for the following geometric interpretation. After
performing a suitable T-duality transformation, we can assume that 
we are dealing with a Dirichlet boundary condition, so that we can
characterize the boundary state by the position $a$ of a
point-like defect on the circle. Breaking the bulk symmetry to the
subalgebra $\calap^{(m)}$ correspond to defects located at $2m\caln$-th roots
of unity on the unit circle. In general, we expect that boundary conditions 
breaking more chiral symmetries of the bulk correspond to more generic locations 
in the space of boundary conditions.

In the case at hand we can also study explicitly the projective limit
of classifying algebras that was used in the definition of the universal 
classifying algebra in section \ref{t.6}. Manifestly, when
$m'$ divides $m$, then $\calap^{(m')}$ is a subalgebra of $\calap^{(m)}$.
In more fancy terms this can be expressed as follows. Divisibility 
introduces a partial ordering on the set $I$
non-negative integers; the subalgebras $\calap^{(m)}$ form an inductive
system over $I$. The inductive limit consists just of the subalgebra
$\calap^{(\infty)}$ of uncharged elements of $\cala$. Moreover,
there is a natural projection relating the classifying algebras:
  \be \calc(\calap^{(m)}) \cong \complex \zet_{m\caln} \,\to\, 
  \complex \zet_{m'\caln} \cong \calc(\calap^{(m')}) \,.  \ee
There then exists a projective limit 
  \be  \hat\zet_{\caln} := \lim_{\dsty\leftarrow} \zet_{m\caln}  \ee
of the classifying algebras, from which
every classifying algebra $\calc(\calap^{(m)})$ can be obtained as a quotient.
The group algebra of the infinite group $\hat\zet_{\caln}$ is the
projective limit of the classifying algebra. (For $\caln\eq1$ this group
is a well-known object; it appears as the absolute Galois group of any finite 
field, or as the Galois group of the infinite extension 
$\rationals(W)/\rationals$, where $W$ is the group of all roots of unity.)

Up to this point we were exclusively considering subalgebras that are rational
in the sense that the number of primary fields of the orbifold is finite. When
we allow also for boundary conditions that preserve non-rational subalgebras,
then the situation simplifies considerably: $\calap^{(\infty)}$ is then
an allowed subalgebra, and the projective limit becomes isomorphic to
the group algebra of $\zet$.\,%
 \futnote{This might look confusing at first sight; but indeed the projective 
limit does depend on the category in which it is taken, i.e.\ on the 
selection of the objects -- here boundary conditions -- one considers.}
 In this case the irreducible representations of the classifying algebra,
and thus the corresponding boundary conditions, are labelled by the group 
U(1), the dual group of $\zet$. Recall that depending on whether the trivial
or the charge conjugation modular invariant has been chosen in the bulk,
the elements of this U(1) group correspond either to values of the Wilson 
lines or to positions of D0-branes.

\subsection{Simple current symmetries in string theory}

Our second example concerns the construction of perturbative superstring 
theories. Such theories contain fermionic degrees of freedom, and one must 
impose several projections to obtain consistency. All these projections can be 
formulated in terms of simple currents (see e.g.\ \cite{scya6,schW2}). First, 
one has to impose alignment of the fermionic degrees of freedom; the space-time
fermions, the fermionic degrees of the internal theory and the superghosts 
have to be either all in Neveu\hy Schwarz or all in the Ramond sector.\,%
 \futnote{For a discussion of boundary states in the (super-)ghost
sector, see e.g.\ \cite{dfpslr}.} 
 This is accomplished by enlarging the chiral algebra by all bilinears 
of the supercurrents of the world sheet theory. We assume that the theory has
a space-time sector containing $D$ free bosons and $D$ free fermions with
a supercurrent $J_{{\rm st}}$, which is a simple current of conformal
weight $\Delta\eq3/2$ and of order two. Similarly, there are supercurrents
$J_{{\rm int}}$ for the inner sector and $J_{{\rm sgh}}$ for the superghosts.
The relevant simple current group $\Gs$ is $\zet_2\timeS\zet_2$,
consisting of the identity and the three non-trivial currents
  \be  (J_{{\rm st}}, J_{{\rm int}},\bfe)\,,\quad
  (J_{{\rm st}},\bfe,J_{{\rm sgh}}) \quad{\rm and}\quad
  (\bfe,J_{{\rm int}},J_{{\rm sgh}}) \,.  \ee
This extension ensures supersymmetry of the world sheet theory. Boundary 
conditions that break this symmetry seem to be unacceptable, since they would 
spoil the consistency of the theory; they lead to a tachyonic spectrum and 
other undesired features. These remarks refer to {\em world sheet 
supersymmetry\/} and hence apply to all string theories built from $N\eq1$ 
superconformal field theories. 

When we have superconformal theories with extended $N\eq2$
superconformal symmetry on the world sheet, another simple current extension
allows to build space-time supersymmetric theories. To this end one
imposes the GSO projection, which can be achieved by enlarging the 
chiral algebra by another integer spin simple current \jgso. This simple
current \jgso\ implements the spectral flow in the world sheet theory 
(see e.g.\ \cite{scya6,schW2}); it has non-trivial components 
both in the space-time and in the superghost sector. The order $M$ of
\jgso\ is essentially the common denominator of the \uone\ charges in the
Ramond sector; thus it depends on the specific $N\eq2$ model under 
consideration.

\jgso\ can be described more explicitly as follows. The \uone\ current 
  \be  J(z) = J_{{\rm st}}(z)+J_{{\rm int}}(z)  \ee
of the $N\eq2$ algebra is the sum
of a component $J_{{\rm st}}$ in the space-time sector and $J_{{\rm int}}$ 
in the inner sector; it can be expressed in terms of a standard free boson 
as $J\eq\sqrt{c/3}\,\ii\partial X\,{\equiv}\,\sqrt5\,\ii\partial X$. 
The spectral flow simple current is then realized as the Ramond ground
state $\exp(\ii(\sqrt5/2)X)$ of conformal weight $\Delta\eq5/8$. It has 
to be combined with its counterpart in the superghost sector. Expressing 
the superghost in terms of a free boson $\Phi$, the spectral flow operator 
is $\exp(\ii\Phi/2)$ which has conformal weight $-5/8$, so that the total 
simple current
  \be  \jgso = \exp(\ii\,\Frac{\sqrt5}2\,X(z))\, \exp(\ii\,\Frac12\,\Phi(z))
  \ee
has integral conformal weight. In total, the projections in the construction of 
a superstring theory require a simple current extension by the abelian group
  \be  \zet_2\timeS\zet_2\timeS\zet_M \,.  \ee
In this extension typically fixed points do occur, as well as untwisted
stabilizers that differ from the full stabilizers.

In contrast to the situation with supersymmetry
on the world sheet, it is definitely of interest to study boundary conditions 
that do not preserve all {\em space-time\/} supersymmetries. A particular
example is given by BPS conditions, but our formalism allows to describe 
also boundary conditions in which all space-time supersymmetries are broken.
For reviews of non-BPS states and their
conformal field theory description, we refer to \cite{sen18,Leru}.

Let us mention that the formalism developed in the present work has 
another application in string theory. Namely, every superconformal field 
theory with \nn supersymmetry
has an automorphism of order two that reverses the sign of the \uone\
current $J$ of the \nn algebra and exchanges the two supercurrents 
$G^{\pm}$ of charge $\pm1$. Accordingly, when studying \bc s that
correspond to this \auto\ we get two automorphism types;
they are usually called `A-type' and `B-type' (see e.g.\ \cite{oooy}).
According to the general results in section \ref{t.4}, T-duality interchanges 
these two automorphism types. Notice that both types of boundary
conditions are encompassed by a single classifying algebra.

\subsection{The $\zet_2$ orbifold of the free boson and fractional branes}
\label{t.84}

Another illustrative example for our formalism is provided by the
$\zet_2$-orbifold of a free boson, compactified at a rational radius
squared. For concreteness, we again restrict our attention to the case when
$R^2\eq2\caln$ with $\caln\iN\zetplus$, which corresponds to the diagonal
modular invariant. The boundary conditions that preserve all bulk symmetries 
are in one-to-one correspondence to the labels of primary fields, for which
we will use the convention of \cite{dvvv}. These boundary conditions can
be given the following interpretation. 

In the untwisted sector of the orbifold
there are $\caln{-}1$ primaries $\Phi_q$, $q\eq1,2,...\,,\caln{-}1$, of 
conformal weight $\Delta_q\eq q^2/4\caln$, as well as two pairs $1,J$ (of 
conformal weight 0 and 1) and $\psi^1,\,\psi^2$ (with $\Delta_\psi\eq\caln/4$)
which each combine to a single primary field of the underlying circle theory.
The \bc s labelled by the $\Phi_q$ constitute D0-branes sitting at $\xi R$ 
with $\xi$ one of the $2\caln$th roots $\eE^{\pi\ii q/\caln}$,
$0\,{<}\,q\,{<}\,\caln$, of unity; those labelled by $1$ and $J$ describe 
D0-branes which are both localized at one orbifold point, and those labelled
by $\psi^1$ and $\psi^2$ are D0-branes localized at the other orbifold point.
The latter boundary conditions deserve particular attention, as in these cases
the position in target space is not sufficient to describe uniquely the boundary
conditions. Rather, an additional discrete label is needed. Now
the primary fields $1,\,J,\,\psi^1$ and $\psi^2$ have quantum dimension 1,
while the primaries $\Phi_q$ all have quantum dimension 2. It is also known that
in a string compactification the Ramond\hy Ramond charge is proportional to
certain (generalized) quantum dimensions. Thus we can conclude that the
D-branes sitting at orbifold points have a Ramond\hy Ramond charge that
is only half the one of D-branes sitting at smooth points; accordingly
\cite{didg} they are referred to as {\em fractional branes\/}.

In terms of boundary states (or, equivalently, reflection coefficients) 
this behavior is explained as follows. For \bc s that preserve the full bulk
symmetry, the reflection coefficients are (ratios of) elements of
the modular matrix $S$. The pairs $1,J$ and $\psi^1,\psi^2$ form full
orbits of the order-two simple current $J$, while the fields $\Phi_q$
are fixed points of $J$. Since the monodromy charge \wrt $J$ is
0 for fields in the untwisted sector and $1/2$ in the twisted sector,
the standard simple current relation
$S_{J\lambda,\mu}\eq \eE^{2\pi\ii Q_J(\mu)}S_{\lambda,\mu}$
implies that in the boundary conditions labelled by $\Phi_q$
boundary blocks of the twisted sector do not appear. Indeed, because of
$J\Phi_q\eq\Phi_q$ we have $S_{q,\mu}\,{\equiv}\,S_{Jq,\mu}\eq{-}S_{q,\mu}$
for all $\mu$ in the twisted sector. In contrast, orbits of full length 
do appear, but with opposite sign for the two primaries $\lambda$ and $J\lambda$
on the orbit. Briefly, the ambiguity for boundary conditions that are
localized at the orbifold points reflects the fact that on a disk with
such a boundary condition bulk fields in the twisted sector can acquire
a non-vanishing one-point function and that two values with
opposite sign are possible for this \corfu.

The twisted sector contributes four more primaries $\sigma^{1,2}$ and
$\tau^{1,2}$ of conformal weight $1/16$ and $9/16$, respectively. It turns
out that the corresponding boundary conditions are not localized and are
thus Neumann-like. It is therefore tempting to identify them with the four
different types of $\zet_2$-equivariant line bundles over the circle. 

To gain insight in symmetry breaking boundary conditions, we need consistent
subalgebras $\calap$ of the chiral algebra $\cala$ of the $\zet_2$-orbifold.
Examples are easily obtained by observing that the vector space that underlies 
the chiral algebra $\cala$ can be decomposed according to the absolute value 
of the \uone\ charge in the circle theory. This quantity is well-defined because
we only have to deal with fields in the untwisted sector and only states 
with opposite charge are identified. In the decomposition
all multiples of $\sqrt\caln$ appear:
  \be \calh_\vac = \bigoplus_{n\in\zetpluso} \calh^\uonE_{n\sqrt\caln}
  \,.  \ee
This decomposition does not constitute a grading of $\cala$ over
the additive group $\zetpluso$, because the fusion structure within the chiral 
algebra reads $[q_1]\,{\star}\,[q_2]\eq [q_1{+}q_2]\,{+}\,[|q_1{-}q_2|]$. 
Still these fusions imply that for every integer $\ell\,{\ge}\,2$ the subspace
  \be  \bar\calh^{(\ell)}_\vac := \bigoplus_{n\in\zetpluso}
  \calh^\uonE_{n\ell\sqrt\caln}  \ee
provides a subalgebra of the \chira, and it is in fact a consistent subalgebra 
because it is nothing but the chiral algebra of the $\zet_2$-orbifold of a
free boson at radius $\bar R^{(\ell)}\eq\ell R$.

Moreover, inspection shows that, except for $\ell\eq2$, we are {\em not\/} 
dealing with an orbifold subalgebra $\cala^G$ of $\cala$. Indeed, for 
$\ell\,{\ge}\,3$ there does not exist any automorphism of $\cala$
for which the fixed point set is $\bar\calh^{(\ell)}_\vac$.
(Note that for generic $\caln$ the orbifold theory also has very few
automorphisms of the fusion rules that preserve the conformal weights.)
In particular, this situation is not covered by our formalism. However,
the specific \cft\ in question is simple enough to allow for a direct
construction of the corresponding boundary states \cite{osaf}. They are not 
related to automorphisms
of the chiral algebra, and thus it is in general not possible to associate an
automorphism type to such boundary conditions. Hence they provide a simple 
counter example to the common misconception that every boundary condition 
should possess a definite automorphism type.

In the particular case $\ell\eq2$, our techniques can be applied to obtain 
still more boundary conditions. We denote primary fields in
the orbifold theory at radius $\bar R\df2R$ by analogous labels as above,
with an additional bar. The boundary conditions then correspond to orbits 
of primaries \wrtt simple current $\bar\psi^1$, which has order two. We 
only describe the orbits of non-trivial
automorphism type. The orbits $\{\bar\Phi_q,\bar\Phi_{4\caln-q}\}$ for
$q\eq1,3,...\,,2\caln{-}1$ give D0-branes localized at $R$ times a
$4\caln$th root of unity that is not a $2\caln$th root of unity. {}From
the other orbits in the untwisted sector of the $\calap$-theory we recover
the other boundary conditions in the untwisted sector that were described
earlier. In the twisted sector of the $\calap$-theory we find two orbits
that are fixed points, $\bar\sigma^1$ and $\bar\tau^1$; they give rise to
four boundary conditions that preserve all bulk symmetries, corresponding
to $\sigma^{1,2}$ and $\tau^{1,2}$ in the $\cala$-theory. Finally,
the remaining orbit $\{\bar\sigma^2,\bar\tau^2\}$ gives rise to a 
Neumann-like boundary condition 
which breaks the symmetries down to $\bar\calh^{(2)}_\vac$.

We conclude our discussion with the remark that, unlike for the free 
boson case, the automorphism type of a boundary condition in the
$\zet_2$-orbifold of a free boson does not allow any longer to
distinguish Dirichlet and Neumann boundary conditions. Still, all 
Dirichlet boundary conditions come from the untwisted sector while all 
Neumann boundary conditions come from the twisted sector.

\subsection{Examples with genuine \ustab}

We now consider in detail an example where an untwisted stabilizer occurs 
that is a proper subgroup of the full stabilizer. It is worth emphasizing 
that the situation with genuine untwisted stabilizers arises rather 
naturally in string compactifications. This is a consequence of the 
following elementary fact that applies to any tensor product of three or 
more subtheories with three simple currents $\hat\J_i$ ($i\eq1,2,3$) 
of half-integral conformal weight in three distinct subtheories.
In the Gepner construction of superstring vacua such subtheories can be
factors in the inner sector, but also the \cfts\ describing space-time fermions
or the (super-)ghosts. Generic examples for simple currents with 
$\Delta\iN\zet{+}1/2$ are the various components of the supercurrent on the
world sheet, but typically in such compactifications other simple
currents of this type are present as well.

It has been shown in \cite{fusS6} that in \wzwts\ the commutator cocycle
obeys
  \be F_\lambda(\J,\JK)= \exp(2\pi\ii (\Delta_\J^{}\,{-}\,\Delta_\J^{(\JK)}))
  \,,  \labl{simpler}
where $\Delta_\J$ is the conformal weight of $\J$, while $\Delta_\J^{(\JK)}$ is
the conformal weight of the projection of $\J$ in what is called the fixed 
point theory \cite{scya6} with respect to $\JK$. This formula also applies 
to those simple currents of coset theories which come from simple 
currents of the underlying \wzwm s.\,%
 \futnote{In coset theories there can, however, exist additional simple 
currents which arise from resolving fixed points with integral quantum 
dimension.}
 In the situation at hand, each of the simple currents $\hat\J_i$ is
projected to the identity primary field of the fixed point theory, so that 
\erf{simpler} implies that $F_\lambda(\hat\J_i,\hat\J_i)\eq{-}1$. On the
other hand, for $i\nE j$ the two currents are fields in distinct subtheories,
so that $F_\lambda(\hat\J_i,\hat\J_j)\eq1$.
Now out of the three currents $\hat\J_i$ we can form three simple currents
$J_i$ of the tensor product theory by setting $\J_1\df\hat\J_2\hat\J_3$ and
cyclic. These have integral conformal weight and hence -- unlike the original
currents $\hat\J_i$ themselves -- can be used to extend the \chira, by a
$\zet_2{\times}\zet_2$ group. By the bihomomorphism property of $F_\lambda$
we then see that $F_\lambda$ is non-trivial on this group:
  \be  F_\lambda(\J_i,\J_j) = \left\{ \begin{array}{rl} 1 & {\rm for}\ i\eq j
  \,, \\[.15em] -1 & {\rm for}\ i\nE j \,. \eear\right.  \ee
As a consequence, the untwisted stabilizer 
is a proper subgroup of the full stabilizer.

We will now study the effect of a non-trivial untwisted stabilizer
in the situation that the $\cala$-theory is a \wzwt, based on
some affine Lie algebra $\g\eq\gb\untw$ and that the orbifold theory is a
\wzwt\ as well, now based on $\g'\eq\gb'{}\untw$ with $\gb'\,{\subset}\,\gb$.
Such WZW orbifolds have been studied in \cite {kaTo2,bifs}. Here we are 
interested in conformally invariant boundary conditions. Thus the Virasoro
algebras of $\g$ and $\g'$ must coincide; this is precisely the case when
$\g'\,{\hookrightarrow}\,\g$ is a conformal embedding \cite{scwa,babo}.

Many, though not all, conformal embeddings can be understood in terms of
simple currents. An example where this is possible is the following. For 
any three odd positive integers $d_1,d_2,d_3$ there is a conformal embedding
  \be  \g':=\so(d_1)_1\opluS \so(d_2)_1\opluS \so(d_3)_1 \,\hookrightarrow\,
  \so(d_1{+}d_2{+}d_3)_1 =: \g \,.  \ee
The theory based on \g\ can be obtained as an extension of the
$\g'$-theory by a simple current group $\Gs\eq\zet_2\timeS\zet_2$, consisting 
of the fields 
  \be  (\Om{,}\Om{,}\Om)\,, \quad (\Om{,}\vv{,}\vv)\,, \quad (\vv{,}\Om{,}\vv)
  \quad{\rm and }\quad (\vv{,}\vv{,}\Om)\,,  \Labl lx
where $\Om$ and $\vv$ refer to the singlet and vector representation of 
$\so(d_i)$, \resp. There is only a single fixed point, namely the tensor 
product $(\ss{,}\ss{,}\ss)$ of three $\so(d_i)$ spinor representations;
it has stabilizer $\cals\eq\Gs$.
The conformal weight of the vector simple current $\vv$ at level 1 is
$1/2$; according to the general arguments presented at the beginning of
this subsection, the untwisted stabilizer is therefore trivial, 
$\calu\eq\{(\Om{,}\Om{,}\Om)\}$. (The value $[\cals\,{:}\,\calu]\eq2^2$
of the index is in agreement with the fact that the ground state 
degeneracy of the irreducible spinor \rep\ of the $\so(d_1{+}d_2{+}d_3)$ 
theory is twice as large as the one of $(\ss{,}\ss{,}\ss)$, namely 
$2\cdoT2^{(d_1-1)/2} 2^{(d_2-1)/2} 2^{(d_3-1)/2}$.)

As a side remark, we mention that these theories can be realized in terms 
of free fermions. Thus the effect of a genuine untwisted stabilizer can
occur even in {\em free\/} \cfts. This is in fact not too surprising. As 
we have explained, the presence of \ustab s
is related to the fact that the orbifold group acts only projectively
on certain sectors of the theory; this is also known \cite{dvvv} to be true for
the action of the three polyhedral groups on the free boson (compactified at 
the self-dual radius) that gives rise to the exceptional $c\eq1$ theories.
     
In the case at hand, the relevant \auto s of the chiral \alg\ $\cala$ 
can be understood in terms of the \findim\ Lie groups $\GG$ and $\GG'$
associated to \g\ and $\g'$. Namely, for every boundary block
${\rm B}_\lambda{:}\ \calh_\lambda^{}{\otimes}\calh_\lambdap$ that preserves 
the full bulk symmetry and every element $\ga\iN\GG$ the combination
${\rm B}_\lambda^{(\ga)} \df {\rm B}_\lambda^{}\,{\circ}\,(\ga\oT\id)$ 
provides us with a twisted 
boundary block. The corresponding automorphism on the affine \lie\ \g\ is 
the inner automorphism that acts on the modes $J^a_n$ of \g\ as 
$J^a_n\,{\mapsto}\,(\ga J^a\ga^{-1})_n$. This \auto\ preserves 
the symmetries in the subalgebra $\g'$ if and only if $\ga$ is in the 
centralizer $C_\GG(\GG')$ of $\GG'$ in $\GG$, and it acts trivially
if and only if $\ga$ is even in the center $Z(\GG)$
of $\GG$. Thus the non-trivial twists are those by elements in the group
$C_\GG(\GG')/Z(\GG)$.

In the case of our interest, the relevant embedding on the level of Lie 
groups reads
  \be  \GG':= (\Spin(d_1)\timeS\Spin(d_2)\timeS\Spin(d_3))/(\zet_2\timeS\zet_2)
  \,\hookrightarrow\, \Spin(d_1{+}d_2{+}d_3) =: \GG \,;  \Labl GG
note that the subgroup $\GG'$ is not simply connected. For determining
$C_\GG(\GG')/Z(\GG)$ it is instructive to consider first the embedding 
  \be \tilde\GG':=\SO(d_1)\timeS \SO(d_2)\timeS \SO(d_3)
  \,\hookrightarrow\, \SO(d_1{+}d_2{+}d_3)=:\tilde\GG \ee
that is obtained from \Erf GG by dividing out the center of 
$\Spin(d_1{+}d_2{+}d_3)$. Then the center of the $\tilde\GG$ is trivial; 
moreover, using the matrix realization of these groups, one shows that the 
centralizer $C_{\tilde\GG}(\tilde\GG')$ is a $\zet_2\timeS\zet_2$ group 
consisting of the unit matrix $\one_{d_1+d_2+d_3}$ and the diagonal matrices
  \be  \ \tilde M_{23}:=\! \Dmatrix{}-- ,\ \
  \tilde M_{13}:=\! \Dmatrix-{}-,\ \  \tilde M_{12}:=\! \Dmatrix--{} \,.  \ee
Already at this stage we can conclude that the centralizer $C_{\GG'}(\GG)$ is 
an extension of this $\zet_2\timeS\zet_2$ group by the center $\zet_2$
of $\Spin(d_1{+}d_2{+}d_3)$. To decide which extension we are dealing with,
we introduce gamma matrices $\gamma^i$ that satisfy the Clifford relations
$\{\gamma^i,\gamma^j\}\eq2\,\delta^{ij}$. We can then determine the two lifts
$M^\pm_{ij}$ of the matrices $\tilde M_{ij}$ to $\Spin(d_1{+}d_2{+}d_3)$ 
from the requirement that $M_{12}\gamma^i(M_{12})^{-1}\eq{-}\gamma^i$ for 
$1\,{\le}\,i\,{\le}\,d_1{+}d_2$ \,and\, $M_{12}\gamma^i(M_{12})^{-1}_{}
\eq\gamma^i$ for $i\,{>}\,d_1+d_2$. It is easy to verify that
  \be \bearl
  M^\pm_{12} = \pm\, \gamma^1\gamma^2 \cdots \gamma^{d_1+d_2}\,, \qquad\
  M^\pm_{23} = \pm\, \gamma^{d_1+1}\gamma^{d_1+2} \cdots
  \gamma^{d_1+d_2+d_3} \,, \\[.8em]
  M^\pm_{13} = \pm\, \gamma^1\gamma^2 \cdots \gamma^{d_1}\,
  \gamma^{d_1+d_2+1} \cdots \gamma^{d_1+d_2+d_3}
  \,.  \end{array} \ee
It can be checked that these matrices commute with all elements in $G'$.
They form a group of order $8$; the structure of this group depends
on the values of $d_1,d_2$ and $d_3$. 
First, when all $d_i$ leave the same rest modulo $4\zet$, then the group is
isomorphic to the eight-element generalized {\em quaternion\/} group; 
this group has one \twodim\ and four \onedim\ \irrep s. Otherwise, i.e.\
when only two of the $d_i$ leave the same rest modulo $4\zet$, the group is
isomorphic to the {\em dihedral\/} group $D_4$ of 8 elements. This group has
four \onedim\ and one \twodim\ \irrep, too. 
In both cases the centralizer is non-abelian. 

Let us present some more details. The boundary blocks are labelled by
the fields of monodromy charge zero and characters of their (full) stabilizers;
thus there are 8 boundary blocks coming from full orbits; we label them 
lexicographically,
  \be \begin{array}{llll}
  \Beta_1:=\Beta_{\Om\Om\Om}\,,\;\ & \Beta_2:=\Beta_{\Om\Om\vv}\,,\;\ &
  \Beta_3:=\Beta_{\Om\vv\Om}\,,\;\ & \Beta_4:=\Beta_{\Om\vv\vv}\,, \\[.4em]
  \Beta_5:=\Beta_{\vv\Om\Om}\,,\;\ & \Beta_6:=\Beta_{\vv\Om\vv}\,,\;\ &
  \Beta_7:=\Beta_{\vv\vv\Om}\,,\;\ & \Beta_8:=\Beta_{\vv\vv\vv}\,. \eear\Labl77
Note that the blocks numbered as $1,4,6,7$ come from the vacuum of the
$\cala$-theory, while the others come from the field that carries the
vector \rep\ of $\so(d_1{+}d_2{+}d_3)$. In addition we have 4 boundary
blocks coming from the fixed point $(\ss{,}\ss{,}\ss)$. They correspond to the 
four irreducible characters $\psi$ of $\zet_2\timeS\zet_2$; we label them as
  \be \Beta_9:=\Beta_{\sss++++}\,,\quad\; \Beta_{10}:= \Beta_{\sss++--}\,,
  \quad\;
  \Beta_{11}:= \Beta_{\sss+-+-}\,,\quad\; \Beta_{12}:= \Beta_{\sss+--+}\,, \ee
where the $\pm$ labels indicate the values $\pm1$ of $\psi$ on the four 
elements of $\zet_2\timeS\zet_2$, in the lexicographic order chosen in 
formula \Erf lx.

The \bc s are labelled by the orbits and characters of their \ustab s. Thus in
addition to the three \bc s that preserve all of $\cala$, there are three
conditions from the length-4 orbits and two conditions for each of the three
length-2 orbits which have stabilizer $\zet_2$. We label them according to
  \be  \begin{array}{llll}
  \calb_1  \;\hat=& \{(\Om\Om\Om)\,,(\Om\vv\vv)\,,(\vv\Om\vv)\,,(\vv\vv\Om)\}\,,
  \ \ \ &
  \calb_7  \;\hat=& \{(\Om\ss\ss)\,,(\vv\ss\ss)\} \ {\rm with}\ \psi\eq1 \,,
  \\[.3em]
  \calb_2  \;\hat=& \{(\Om\Om\vv)\,,(\Om\vv\Om)\,,(\vv\Om\Om)\,,(\vv\vv\vv)\}\,,
  &
  \calb_8  \;\hat=& \{(\Om\ss\ss)\,,(\vv\ss\ss)\} \ {\rm with}\ \psi\eq{-1} \,,
  \\[.3em]
  \calb_3  \;\hat=& \{(\ss\ss\ss)\}\,,
  &
  \calb_9  \;\hat=& \{(\ss\Om\ss)\,,(\ss\vv\ss)\} \ {\rm with}\ \psi\eq1 \,,
  \\[.3em]
  \calb_4  \;\hat=& \{(\ss\Om\Om)\,,(\ss\Om\vv)\,,(\ss\vv\Om)\,,(\ss\vv\vv)\}\,,
  &
  \calb_{10}\;\hat=&\{(\ss\Om\ss)\,,(\ss\vv\ss)\} \ {\rm with}\ \psi\eq{-1} \,,
  \\[.3em]
  \calb_5  \;\hat=& \{(\Om\ss\Om)\,,(\Om\ss\vv)\,,(\vv\ss\Om)\,,(\vv\ss\vv)\}\,,
  &
  \calb_{11}\;\hat=&\{(\ss\ss\Om)\,,(\ss\ss\vv)\} \ {\rm with}\ \psi\eq1 \,,
  \\[.3em]
  \calb_6  \;\hat=& \{(\Om\Om\ss)\,,(\Om\vv\ss)\,,(\vv\Om\ss)\,,(\vv\vv\ss)\}\,,
  &
  \calb_{12}\;\hat=&\{(\ss\ss\Om)\,,(\ss\ss\vv)\} \ {\rm with}\ \psi\eq{-1} \,.
  \eear \Labl78
With this numbering, the diagonalizing matrix $\tS$ looks as follows:
  \be  \tS = \frac12  \left(\begin{array}{rrrrrrrrrrrr}
  1& 1& \nb  & \nb  & \nb  & \nb  &  1&  1&  1&    1&  1&  1  \\
  1& 1& \nbm & \nb  & \nb  & \nbm & -1& -1& -1&   -1&  1&  1  \\
  1& 1& \nbm & \nb  & \nbm & \nb  & -1& -1&  1&    1& -1& -1  \\
  1& 1& \nb  & \nb  & \nbm & \nbm &  1&  1& -1&   -1& -1& -1  \\
  1& 1& \nbm & \nbm & \nb  & \nb  &  1&  1& -1&   -1& -1& -1  \\
  1& 1& \nb  & \nbm & \nb  & \nbm & -1& -1&  1&    1& -1& -1  \\
  1& 1& \nb  & \nbm & \nbm & \nb  & -1& -1& -1&   -1&  1&  1  \\
  1& 1& \nbm & \nbm & \nbm & \nbm &  1&  1&  1&    1&  1&  1  \\
  \nb&\nbm& 0&  0& 0& 0&  b_{23}&\!-b_{23}& b_{13}&\!-b_{13}& b_{12}&\!-b_{12} \\
  \nb&\nbm& 0&  0& 0& 0&  b_{23}&\!-b_{23}&\!-b_{13}& b_{13}&\!-b_{12}& b_{12} \\
  \nb&\nbm& 0&  0& 0& 0& \!-b_{23}& b_{23}& b_{13}&\!-b_{13}&\!-b_{12}& b_{12} \\
  \nb&\nbm& 0&  0& 0& 0& \!-b_{23}& b_{23}&\!-b_{13}& b_{13}& b_{12}&\!-b_{12}
  \eear\right) \,. \ee
Here we have put
  \be  b_{ij} := \sqrt2\, \ii^{-(d_i+d_j)/2}_{}  \ee
for $i,j\iN\{1,2,3\}$.
We note that
  \be  \tS \tS^\dagger = 4\cdot\one = \tS^\dagger\tS  \quad{\rm and}\quad
  (\tS \tS^{\rm t})^2 = 16\cdot\one \,,  \ee
but $\tS \tS^{\rm t}$ is a permutation if and only if precisely one out of the
expressions
  \be  2 \pl b_{23}^2 \pl b_{13}^2 \pl b_{12}^2 \,,\quad
  2 \pl b_{23}^2 \mi b_{13}^2 \mi b_{12}^2 \,,\quad
  2 \mi b_{23}^2 \pl b_{13}^2 \mi b_{12}^2 \,,\quad
  2 \mi b_{23}^2 \mi b_{13}^2 \pl b_{12}^2  \ee
is non-vanishing, which are precisely those cases where after division by 2
these numbers furnish a character of $\zet_2{\times}\zet_2$.

Let us now add some general remarks. Every $\GG$-module can be
decomposed into irreducible modules of the product $C_{\GG}(\GG')\timeS 
\GG'$. The fact that the group $C_{\GG}(\GG')$ is non-abelian implies that it 
has higher-dimensional \irrep s, which in turn means that higher-dimensional
degeneracy spaces appear; this way we recover a generic feature of genuine
untwisted stabilizers. Conversely, we are led to the following conjecture.
Let $\GG'\,{\hookrightarrow}\,\GG$ be an embedding of reductive compact Lie 
groups such that the associated embedding of affine \lie s is a
simple current extension. Then the centralizer $C_\GG(\GG')$ is non-abelian
if and only if at some value of the
level a genuine untwisted stabilizer appears. For instance,
there is a conformal embedding $(D_4)_2\,{\hookrightarrow}\,(E_7)_1$ which
again is a simple current extension by a $\zet_2{\times}\zet_2$ group.
As $E_7$ is an exceptional \lie, the relevant centralizer, namely
the one of $\SO(8)/\zet_2$ in $E_7$, is now difficult to compute,
but in any case the analysis of the \bc s indicates that this 
centralizer is non-abelian. 
For completeness, we mention that in the $E_7$ case there are eight
\bc s preserving the affine $D_4$ subalgebra, and the matrix $\tS$ reads
  \be  \tilde S = \frac1{\sqrt2}\, \left(\begin{array}{rrrrrrrr}
  1&1&1&1    &1&1&1&1 \\ 1&1&1&1    &-1&-1&-1&-1 \\
  1&1&-1&-1    &1&1&-1&-1 \\ 1&1&-1&-1    &-1&-1&1&1 \\
  1&-1&\ib&-\ib    &\ib&-\ib&\ib&-\ib \\ 1&-1&\ib&-\ib    &-\ib&\ib&-\ib&\ib \\
  1&-1&-\ib&\ib    &\ib&-\ib&-\ib&\ib \\ 1&-1&-\ib&\ib    &-\ib&\ib&\ib&-\ib
  \eear\right)\,.  \ee

On the other hand, the automorphisms themselves are classified by the group 
$C_\GG(\GG')/Z(\GG)$, which in both cases considered above
is $\zet_2\timeS\zet_2$. Thus they precisely correspond to the automorphism
types that are predicted by the general analysis.

\vskip3em\noindent{\small {\bf Acknowledgement} \\
We would like to thank Valya Petkova and Patrick Dorey for discussions,
and Peter Bantay and Bert Schellekens for helpful correspondence.}
 
     \newpage

\appendix

\sect{Collection of formulae}

Here we collect a few basic formulae from \I\ that are used in this paper. 
The equation numbers are the same as in \I.
\nxt
The boundary blocks $\tBeta_{(\lambdab,\varphi)}$ are the linear forms
  \def\theequation{4.23}
  \be  \tBeta^{}_{(\lambdab,\psi)}
  := \Norm\mu\psu\,d_\lambda^{-1/2}\,\Bet_\psi \oT \bbb\lambda  \labl{tBeta}
on
  $$ \calhl^{}{\otimes}\calhlp \equiv
  \bigoplus_{\J\in\Gs/\cals_\lambda}\V_\psu{\otimes}\calhb_{\J\lambdabo}
  \;{\otimes}\; \bigoplus_{\J\in\Gs/\cals_\lambda}
  \V_\psup{\otimes}\calhb_{\J\lambdabop} \,.  $$
Here $\Bet_\psi^{}{:}\;\Vpsu\otimeS\Vpsup\,{\to}\,\complex$ is a linear form
on the degeneracy spaces, $\Betab_\lambdab{:}\; \calh_\lambdab^{}\otimeS\calh_\lambdab
\,{\to}\,\complex$ is an ordinary boundary block of the $\calap$-theory,
$d_\lambda\eq\sqrt{|\cals_\lambda|/|\calu_\lambda|}$, and $\Norm\lambda\psu$
is a phase that is left undetermined (in the first place, this normalization
is introduced as $\norm\lambda\psi$, but as shown in \I\ it depends only on the
primary label $\lambda\eq\Lambdab$ of the $\cala$-theory).
The linear form $\Bet_\psi$ can be written as
  \def\theequation{4.20}
  \be  \Bet_\psi = \n \circ (\clo_\psi\tims\id)  \Labl nO
with $\n$ defined by $\n(v\ot w)\eq{\BB(v\ot\po\ot w\ot\qo)}/{\BBB(\po\ot\qo)}$
(where $\po\iN\calhb_\lambdab$ and $\qo\iN\calhb_\lambdabp$ are any vectors
such that $\BBB(\po\ot\qo)\nE0$) and 
  \def\theequation{4.11}
  \be  \clo_\psi := d_\lambda^{-3/2}\!
  \sum_{\J\in\cals_\lambda/\calu_\lambda}\! \psi(\J)^* R_{\psu}(\J)
  \,, \labl{clo}
where $R_\psu$ denotes the \irrep\ of the twisted group \alg\ that is
labelled by $\psu\,{\prec}\,\psi$. The $\clo_\psi$ with $\psi\,{\gt}\,\psu$
(that is, $\psi_{|\calu_\lambda}{=}\,\psu$) form a partition of unity:
  \def\theequation{4.17}
  \be  \Sumpsipsu\lambda\! \clo_\psi = d_\lambda^{1/2}\, \one_{d_\lambda}^{} \,.
  \Labl2d
\nxt
The operator product expansion that describes the excitation on the boundary
caused by a bulk field approaching it reads
  \def\theequation{5.4}
  \be  \phi_{(\lambdab,\psi),(\lambdabp,\psi^+)}(r\eE^{\ii\sigma})
  = \sum_\mub (1{-}r^2)^{-2\Delta_\lambdab+\Delta_\mub}_{}\,
  \Rc a{(\lambdab,\psi)}\mub\,\Psi^{aa}_\mub(\eE^{\ii\sigma})
  + \mbox{descendants} \quad\; {\rm for}\;\ r\,{\to}\,1 \,.  \labl{oben2}
\nxt
The diagonalizing matrix $\tS$ of \clAb\ can be written as
  \def\theequation{5.9}
  \be  \tS_{(\lambdab,\psi_\lambda),\RhoB}
  := \Frac{|G|}{[\cals_\lambda\calu_\lambda\cals_\rho\calu_\rho]^{1/2}}
  \sum_{\J\in\cals_\lambda\cap\calu_\rho} \psi^{}_\lambda(\J)\,
  \psu_\rho(\J)^*\, S^\J_{\lambdab,\rhob} \,,  \Labl tS
where the matrices $S^\J$ represent the modular S-transformation on 
the one-point \cblock s with insertion $\J$ on the torus.
The result that $\tS$ is a square matrix is equivalent to the sum rule
  \def\theequation{5.22}
  \be  \sumbo\lambda |\cals_\lambda| = \sum_\rhoB |\calu_\rho| \,.  \Labl sr

 \vskip3em
\small
 \newcommand\wb{\,\linebreak[0]} \def\wB {$\,$\wb}
 \newcommand\Bi[1]    {\bibitem{#1}}
 \renewcommand\J[5]   {{\sl #5}, {#1} {#2} ({#3}) {#4} }
 \newcommand\PhD[2]   {{\sl #2}, Ph.D.\ thesis (#1)}
 \newcommand\Prep[2]  {{\sl #2}, preprint {#1}}
 \newcommand\BOOK[4]  {{\em #1\/} ({#2}, {#3} {#4})}
 \def\jf    {J.\ Fuchs}
 \newcommand\inBO[7]  {in:\ {\em #1}, {#2}\ ({#3}, {#4} {#5}), p.\ {#6}}
 \newcommand\inBOx[7] {in:\ {\em #1}, {#2}\ ({#3}, {#4} {#5})}
 \newcommand\gxxI[2] {\inBO{GROUP21 Physical Applications and Mathematical
              Aspects of Geometry, Groups, and \A s{\rm, Vol.\,2}}
              {H.-D.\ Doebner, W.\ Scherer, and C.\ Schulte, eds.}
              \WS\Si{1997} {{#1}}{{#2}}}
 \def\anop  {Ann.\wb Phys.}
 \def\comp  {Com\-mun.\wb Math.\wb Phys.}
 \def\duke  {Duke\wB Math.\wb J.}
 \def\duki  {Duke\wB Math.\wb J.\ (Int.\wb Math.\wb Res.\wb Notes)}
 \def\foph  {Fortschr.\wb Phys.}
 \def\ijmp  {Int.\wb J.\wb Mod.\wb Phys.\ A}
 \def\imrn  {Int.\wb Math.\wb Res.\wb Notices}
 \def\jams  {J.\wb Amer.\wb Math.\wb Soc.}
 \def\jopa  {J.\wb Phys.\ A}
 \def\jhep  {J.\wb High\wB Energy\wB Phys.}
 \def\nupb  {Nucl.\wb Phys.\ B}
 \def\phlb  {Phys.\wb Lett.\ B}
 \def\phrd  {Phys.\wb Rev.\ D}
 \def\phrl  {Phys.\wb Rev.\wb Lett.}
 \def\thmp  {Theor.\wb Math.\wb Phys.}
 \newcommand\geap[2] {\inBO{Physics and Geometry} {J.E.\ Andersen, H.\
            Pedersen, and A.\ Swann, eds.} \MD\NY{1997} {{#1}}{{#2}} }
 \def\AMS    {{American Mathematical Society}}
 \def\AP     {{Academic Press}}
 \def\CUP    {{Cambridge University Press}}
 \def\MD     {{Marcel Dekker}}
 \def\NH     {{North Holland Publishing Company}}
 \def\SV     {{Sprin\-ger Ver\-lag}}
 \def\WS     {{World Scientific}}
 \def\Ad     {{Amsterdam}}
 \def\Be     {{Berlin}}
 \def\Ca     {{Cambridge}}
 \def\NY     {{New York}}
 \def\PR     {{Providence}}
 \def\Si     {{Singapore}}

\end{document}